\newif\iffigs\figstrue
\documentclass[12pt]{article}
\usepackage{latexsym,amssymb}
\usepackage{latexsym,amssymb}
\iffigs
            \input{epsf}
\else
\fi
\def\IC{\relax\,\hbox{$\inbar\kern-.3em{\rm C}$}}
\def\IG{\relax\,\hbox{$\inbar\kern-.3em{\rm G}$}}
\def\IB{\relax{\rm I\kern-.18em B}}
\def\ID{\relax{\rm I\kern-.18em D}}
\def\IL{\relax{\rm I\kern-.18em L}}
\def\IF{\relax{\rm I\kern-.18em F}}
\def\IH{\relax{\rm I\kern-.18em H}}
\def\II{\relax{\rm I\kern-.17em I}}
\def\IN{\relax{\rm I\kern-.18em N}}
\def\IP{\relax{\rm I\kern-.18em P}}
\def\IQ{\relax\,\hbox{$\inbar\kern-.3em{\rm Q}$}}
\def\bfzero{\relax\,\hbox{$\inbar\kern-.3em{\rm 0}$}}
\def\IK{\relax{\rm I\kern-.18em K}}
\def\IG{\relax\,\hbox{$\inbar\kern-.3em{\rm G}$}}
 \font\cmss=cmss10 \font\cmsss=cmss10 at 7pt
\def\IR{\relax{\rm I\kern-.18em R}}
\def\ZZ{\relax\ifmmode\mathchoice
{\hbox{\cmss Z\kern-.4em Z}}{\hbox{\cmss Z\kern-.4em Z}}
{\lower.9pt\hbox{\cmsss Z\kern-.4em Z}}
{\lower1.2pt\hbox{\cmsss Z\kern-.4em Z}}\else{\cmss Z\kern-.4em
Z}\fi}
\def\bfone{\relax{\rm 1\kern-.35em 1}}

\def\inbar{\vrule height1.5ex width.4pt depth0pt}
\def\bfzero{\relax{\rm I\kern-.18em 0}}
\def\bfone{\relax{\rm 1\kern-.35em 1}}

\newcommand{\ft}[2]{{\textstyle\frac{#1}{#2}}}
\def\tilde{\widetilde}

\def\1bar{1\hskip -.275cm -}
\def\2bar{2\hskip -.275cm -}
\def\3bar{3\hskip -.275cm -}

\newsavebox{\uuunit}
\sbox{\uuunit}
             {\setlength{\unitlength}{0.825em}
                     \begin{picture}(0.6,0.7)
                                 \thinlines
                                 \put(0,0){\line(1,0){0.5}}
                                 \put(0.15,0){\line(0,1){0.7}}
                                 \put(0.35,0){\line(0,1){0.8}}
                                \multiput(0.3,0.8)(-0.04,-0.02){10}{\rule{0.5pt}{0.5pt}}
                     \end {picture}}

\makeatletter \@addtoreset{equation}{section} \makeatother

\newcommand{\be}{\begin{equation}}
\newcommand{\ee}{\end{equation}}
\newcommand{\ba}{\begin{eqnarray}}
\newcommand{\ea}{\end{eqnarray}}
\def\bfone{\relax{\rm 1\kern-.35em 1}}

\def\bfone{\relax{\rm 1\kern-.35em 1}}
\font\cmss=cmss10 \font\cmsss=cmss10 at 7pt
\begin{document}
\begin{titlepage}
\begin{flushright}
ITP-UU-03/49 \\SPIN-03/30
\end{flushright}
\vskip 1.5cm
\begin{center}
{\LARGE \bf   Cosmological backgrounds
of superstring theory and Solvable Algebras: Oxidation and Branes$^ \dagger
$}\nonumber \\
\vfill {\large
 P. Fr\'e$^1$, V. Gili$^2$, F. Gargiulo$^1$, \nonumber \\
 A. Sorin$^3$,  K. Rulik$^1$, M. Trigiante$^4$} \nonumber \\
\vfill {
$^1$ Dipartimento di Fisica Teorica, Universit\'a di Torino,
$\&$ INFN -
Sezione di Torino\nonumber \\
via P. Giuria 1, I-10125 Torino, Italy  }\nonumber \\
\vskip 0.3cm
{
$^2$ Dipartimento di Fisica Nucleare e Teorica, Universit\'a di Pavia,
$\&$ INFN
Sezione di Pavia\nonumber \\
via A. Bassi 6, I-27100 Pavia, Italy  }\nonumber \\
\vskip 0.3cm { $^3$ Bogoliubov Laboratory of Theoretical Physics,
JINR, 141980 Dubna, Moscow Region, Russian Federation  }\nonumber \\
\vskip 0.3cm {$^4$ Spinoza Institute, Leuvenlaan 4, Utrecht, The
Netherlands }
\end{center}
\vfill
\begin{abstract}
We develop a systematic algorithm to construct, classify and study
exact solutions of type II A/B supergravity which are
time--dependent and homogeneous and hence represent candidate
cosmological backgrounds. Using the formalism of solvable Lie
algebras to represent the geometry of non--compact coset manifolds
$\mathrm{U/H}$ we are able to reduce the supergravity field
equations to the geodesic equations in $\mathrm{U/H}$ and rephrase
these latter in a completely algebraic setup by means of the so
called Nomizu operator representation of covariant derivatives in
solvable group manifolds. In this way a systematic method of
integration of supergravity equations is provided. We show how the
possible $D=3$ solutions are classified by non--compact
subalgebras $ \mathbf{G} \subset \mathrm{E_{8(8)}}$ and their
ten--dimensional physical interpretation (oxidation) depends on
the classification of the different embeddings $ \mathbf{G}
\hookrightarrow \mathrm{E_{8(8)}}$. We give some preliminary
examples of explicit solutions based on the simplest choice
$\mathbf{G}=\mathrm{A_2}$. We also show how, upon oxidation, these
solutions provide a smooth and exact realization of the bouncing
phenomenon on Weyl chamber walls envisaged by the cosmological
billiards of Damour et al. We also show how this physical
phenomenon is triggered by the presence of euclidean $D$--branes
possibly interpretable at the microscopic level as S--branes. We
outline how our analysis could be extended to a wider setup where,
by further reducing to $D=2,1$, more general backgrounds could be
constructed applying our method to the infinite algebras
$E_{9,10}$.
\end{abstract}
\vspace{2mm} \vfill
\hrule width 3.cm {\footnotesize
$^ \dagger $
This work is supported in part by
the European Union RTN contracts
HPRN-CT-2000-00122 and HPRN-CT-2000-00131. The work of M. T. is supported by an
European Community Marie Curie Fellowship under contract HPMF-CT-2001-01276.}
\end{titlepage}
\newpage
\section{Introduction}
In view of the new observational data in cosmology that appear to
confirm the inflationary scenario and provide evidence for a small
but positive cosmological constant \cite{experiment,linde90}, there has been wide
interest in the context of M--theory/string theory and extended supergravities
for the search of
de Sitter like vacua (see for instance \cite{Kachru:2003sx}--\cite{Fre:2002pd} and
references therein)  and more generally for the
analysis of time--dependent backgrounds \cite{Gutperle:2002ai}--\cite{ mart}.
This has been done in various approaches and at different levels,
namely both from the microscopic viewpoint, considering
time--dependent boundary states and boundary CFTs (see for instance
\cite{Sen:2002vv,Sen:2002nu} and references therein) and
from the macroscopic viewpoint studying supergravity solutions. In
this latter context, great attention has been devoted to the
classification of \textit{gaugings}
\cite{QXX1f,deWit:2003ja,Fischbacher:2003yw,Nicolai:2001sv,Nicolai:2000sc}
their relation to
compactifications with fluxes \cite{QXX5} and the ensuing
cosmological solutions  \cite{Kachru:2003sx,Kachru:2003aw,Burgess:2003ic}. Indeed de
Sitter like or anti
de Sitter like backgrounds require an effective cosmological
constant, or better a scalar potential that is typically produced by
the gauging procedure.
\par
As it is well known, gauged supergravities apparently break the large
symmetry groups of ungauged supergravities encoding those perturbative
and non perturbative dualities which are responsible for  knitting together
the five consistent perturbative
superstrings into  a single non--perturbative theory. Yet the
interpretation of gaugings as compactifications with suitable fluxes
and branes restores the apparently lost symmetries.
\par
Notwithstanding this fact it is, to begin with, very much interesting
to study cosmological backgrounds of superstring theory in the context of
pure ungauged supergravity where the role of duality symmetries is
more direct and evident.
\par
In this setup a very much appealing and intriguing scenario has
been proposed in a series of papers
\cite{bill99}--\cite{Demaret:sg}: that of cosmological billiards.
Studying the asymptotic behaviour of supergravity field equations
near time (space--like) singularities, these authors have
envisaged the possibility that the nine cosmological scale factors
relative to the different  space dimensions of string theory plus
the dilaton could be assimilated to the lagrangian coordinates of
a fictitious ball moving in a ten--dimensional space. This space
is actually the Weyl chamber associated with the $E_{10}$ Dynkin
diagram and the cosmological ball scatters on the Weyl chamber
walls in a chaotic motion. \iffigs
\begin{figure}
\begin{center}
\epsfxsize =12cm
{\epsffile{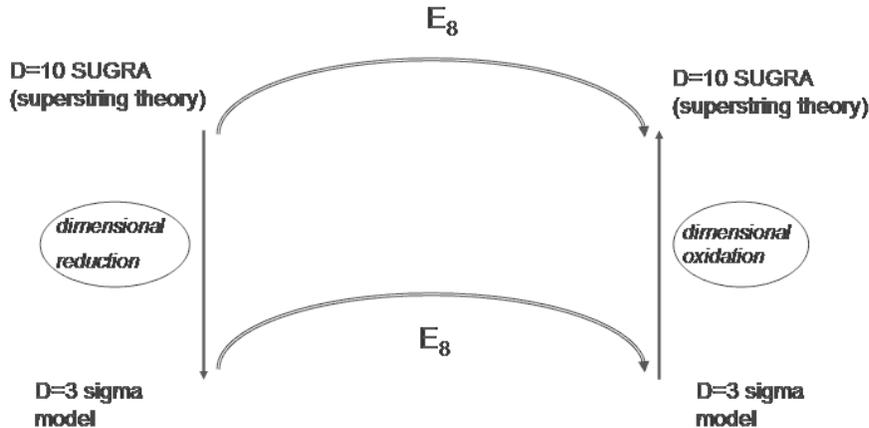}}
\caption{Time--dependent homogeneous supergravity backgrounds in
$D=10$ can be obtained by first dimensionally reducing to $D=3$,
solving the differential equations of the sigma model and then
oxiding back the result to $D=10$. This procedure defines also the
action of the hidden symmetry group $E_{8(8)}$ on the ten
dimensional configurations. \label{disegnino}} \hskip 2cm
\unitlength=1.1mm
\end{center}
\end{figure}
\fi There is a clear relation between this picture and the duality
groups of superstring theories. Indeed it is well known that
compactifying type II A or type II B supergravity on a $T^{r-1}$
torus, the massless scalars which emerge from the Kaluza-Klein
mechanism in dimension $D=10-r+1$ just parametrize the maximally
non--compact coset manifold
\begin{equation}
  \mathcal{M}_{scalar} \, = \, \frac{\mathrm{E_{r(r)}}}{\mathrm{H_r}}
\label{eseries}
\end{equation}
where $H_r$ is the maximally compact subgroup of the simple Lie group
$\mathrm{E_{r(r)}}$ \cite{julia80}. Furthermore the restriction of
$\mathrm{E_{r(r)}}$ to integers is believed to be an exact
non--perturbative symmetry of superstring theory compactified on such
a torus \cite{Hull:1994ys}. Since compactification and truncation to the massless
modes is
an alternative way of saying that we just focus on field configurations \textit{that
depend only
the remaining}:
$$
  10-r+1 \, \mbox{coordinates}
$$
it follows that cosmological backgrounds, where the only non trivial
dependence is just on \textit{one coordinate}, namely \textit{time},
should be related to compactifications on a $T^9$ torus and hence
linked to the $E_{10}$ algebra \cite{Julia:1982gx}--\cite{e10}. Furthermore the
Cartan generators of
the $\mathrm{E_{r(r)}}$ algebra are dual to the \textit{radii} of the
$T^{r-1}$ torus plus the dilaton. So it is no surprise that
the evolution of the cosmological scale factor should indeed
represent some kind of motion in the dual of the Cartan subalgebra of
$E_{10}$. Although naturally motivated, the $E_{10}$ billiard picture was so
far considered only in the framework of an approximated asymptotic
analysis and no exact solution with such a behaviour was actually
constructed. This originates from two main difficulties. Firstly,
while up to $r=8$, which corresponds $D=3$ dimensions, the Lie
algebras $E_{r(r)}$ are normal finite dimensional simple algebras,
for $r=9,10$ they become infinite dimensional algebras
whose structure is much more difficult to deal with and the
corresponding coset manifolds need new insight in order to be
defined. Secondly, the very billiard phenomenon, namely the
scattering of the fictitious ball on the Weyl chamber walls requires the
presence of such potential walls. Physically they are created
by the other bosonic fields present in the supergravity theory, namely the
\textit{non diagonal} coefficients of the metric and the various
\textit{$p$--form field strengths}.
\par
In this paper we focus on three--dimensional maximal supergravity
\cite{Marcus:1983hb}--\cite{Nicolai:2001sv},
namely on the dimensional reduction of type II theories on a $\mathrm{T^7}$
torus, instead of going all the way down to reduction to
one--dimension, by compactifying on $\mathrm{T^9}$. The advantage of
this choice is that all the bosonic fields are already scalar fields,
described by a non--linear sigma model without, however, the need of
considering Ka\v c--Moody algebras which arise as isometry algebras of scalar
manifolds in $D<3$
space--times. In this way we are able to utilize the \textit{solvable Lie
algebra approach} to the description of the whole bosonic sector which
enables us to give a completely algebraic characterization of the
microscopic origin of the various degrees of freedom
\cite{Andrianopoli:1996bq,Andrianopoli:1996zg}.
Within this framework the supergravity field equations for bosonic
fields restricted to only time dependence reduce simply to the
geodesic equations in the target manifold $\mathrm{E_{8(8)}/SO(16)}$.
These latter can be further simplified to a set of differential
equations whose structure is completely determined in Lie algebra
terms. This is done through the use of the so
called \textit{Nomizu operator}. The concept of Nomizu operator
coincides with the concept of covariant derivative for solvable group
manifolds and the possibility of writing covariant derivatives in
this algebraic way as linear operators on solvable algebras relies on
the theorem that states that a non--compact coset manifold with a
transitive solvable group of isometries is isometrical to the
solvable group itself.
\par
The underlying idea for our approach is rooted in the concept of
hidden symmetries. Cosmological backgrounds of superstring theory,
being effectively one--dimensional fill orbits under the action of a
very large symmetry group, possibly $\mathrm{E_{10}}$ that necessarily
contains $\mathrm{E_{8(8)}}$, as the manifest subgroup in three dimensions.
Neither $\mathrm{E_{10}}$ nor $\mathrm{E_{8(8)}}$ are manifest in
$10$--dimensions but become manifest in lower dimension. So an
efficient approach to finding spatially homogeneous solutions in ten
dimensions consists of the process schematically described in
fig.\ref{disegnino}. First one reduces to $D=3$, then solves the
geodesic equations in the algebraic setup provided by the Nomizu--operator--formalism
and then oxides back the result to a full fledged $D=10$
configuration. Each possible $D=3$ solution is characterized by a
non--compact subalgebra
\begin{equation}
  \mathbf{G} \, \subset \,\mathrm{ {E}_{8(8)}}
\label{Gsubalg}
\end{equation}
which defines the smallest consistent truncation of the full
supergravity theory within which the considered solution can be
described. The inverse process of oxidation is not unique but
leads to as many physically different ten dimensional solutions as
there are algebraically inequivalent ways of embedding
$\mathbf{G}$ into $\mathrm{E_{8(8)}}$. In this paper we will
illustrate this procedure by choosing for $\mathbf{G}$ the
smallest non abelian rank two algebra, namely
$\mathbf{G}=\mathrm{A_2}$ and we will see that the non abelian
structure of this algebra reflects interaction terms that are
present in the ten dimensional theory like, for instance, the
Chern--Simons term. The solvable Lie algebra formalism allows us
to control, through the choice of the $\mathbf{G}$--embedding, the
physical ten--dimensional interpretation of any given
$\sigma$--model solution. In this paper we choose a particular
embedding for the subalgebra $\mathrm{A_2}$ which leads to a type
II B time dependent background generated by a system of two
euclidean D-branes or S-branes
\cite{Gutperle:2002ai,otherSbranes}: a D3 and a D1, whose world
volumes are respectively four and two dimensional. This physical
system contains also an essential non trivial B--field reflecting
the three positive root structure of the $\mathrm{A_2}$ Lie
algebra, one root being associated with the RR $2$--form
$C^{[2]}$, a second with the RR $4$--form $C^{[4]}$ and the last
with the NS $2$--form $B^{[2]}$. In the time evolution of this
exact solution of type II B supergravity we retrieve a smooth
realization of the bouncing phenomenon envisaged by the
\textit{cosmic billiards} of \cite{bill99}--\cite{Demaret:sg}.
Indeed the scale factors corresponding to the dimensions parallel
to the S--branes first expand and then, after reaching a maximum,
contract. The reverse happens to the dimensions transverse to the
$S$--branes. They display a minimum approximately at the same time
when the parallel ones are maximal. Transformed to the dual CSA
space this is the bouncing of the cosmic ball on a Weyl chamber
wall. This is not yet the full cosmic billiard, but it illustrates
the essential physical phenomena underlying its implementation. We
shall argue that in order to obtain a repeated bouncing we need to
consider larger subalgebras and in particular extend our analysis
to the Ka\v c--Moody case where the dual CSA becomes a space with
lorentzian signature. Such an extension is postponed to future
publications, yet we stress that the main features of the key
ingredients for this analysis have been laid down here. Moreover
it is worth emphasizing that the same $\mathrm{A_2}$ solution
presented in this paper can be oxided to different
ten--dimensional configurations corresponding to quite different
physical systems. In particular, as we explain in later sections,
it can be lifted to a purely gravitational background describing
some sort of gravitational waves. In the present paper we give the
general scheme, but the detailed study of these alternative
oxidations is also postponed to future publications.
\par
It is also worth mentioning that our approach to cosmological
backgrounds makes it clear how, at least on the subspace of time
dependent homogeneous configurations, the hidden symmetry $\mathrm{E_{8(8)}}$ or
its further Ka\v c--Moody extensions, can be made manifest directly
in ten dimensions. Indeed it suffices to follow the diagram of
fig.\ref{disegnino}. Reducing first to $D=3$, acting with the group
and then oxiding back the result to ten dimensions defines the group action
in ten dimensions.
\par
Our paper is organized as follows.
In section \ref{geodesinomi} after showing how three dimensional
gravity can be decoupled from the sigma model, we recast the equations of
motion of the latter into the geodesic equations for the manifold
$\mathrm{E_{8(8)}/SO(16)}$ by using the Nomizu operator formalism.
This leads to a system of first order non linear differential
equations whose structure is completely encoded in the $\mathrm{E_{8(8)}}$
positive root system. In the same section we also outline an
algorithm, valid for any maximally non compact homogeneous manifold
$\mathrm{U/H}$, which allows to find the general solution of the
geodesic differential system by means of compensating
$\mathrm{H}$--transformations. Actually the original differential system is
transformed into a new one for the parameters of the compensating
$\mathrm{H}$--rotations which has the advantage of being integrable in an
iterative way, namely by substituting at each step the solution of
one differential equation into the next one.
\par
In section \ref{exampsolv} we apply the general method to an abstract
$\mathrm{A_2}$--model, namely to the manifold
$\mathrm{SL(3,\mathbb{R})/SO(3)}$. We derive explicit solutions of
the differential equations which provide a paradigma to illustrate our method
but also interesting examples which in a later section we oxide to
ten dimensions.
\par
In section \ref{genoxide} we construct the mathematical framework,
based on the solvable Lie algebra formalism, which allows us to oxide
any given solution of the three--dimensional theory to ten dimensions
by choosing an embedding of the Lie algebra $\mathbf{G}$ into $\mathbf{E_{8(8)}}$.
\par
In section \ref{occidoa2} we consider the explicit oxidation of
the $\mathrm{A_2}$ solutions previously found. First we classify
the different available embeddings, which are of eight different
types. Then, choosing the fourth type of embedding in our
classification list, we show how it leads to a type II B
supergravity solution which describes the already mentioned
system of interacting  $S3$ and $S1$ branes. We illustrate the
physical properties of this solution also by plotting the time
evolution of relevant physical quantities like the scale factors,
the energy densities and the pressure eigenvalues. In this plots
the reader can see the bouncing phenomenon described before.
\par
Finally section \ref{inconcludo} contains our conclusions and
perspectives.
\section{Geodesics on maximally non
compact cosets $\mathrm{U}/\mathrm{H}$ and differential equations}
\label{geodesinomi}
We have recalled how both maximally extended supergravities (of type A and
B) reduce, stepping down from $D=10$ to $D=3$ to the following non
linear sigma model coupled to $D=3$ gravity:
\begin{equation}
\label{lag_sm} \mathcal{L}^{\sigma-model} \,=\,
\sqrt{-\mbox{det} \, g} \, \left[ \, 2 \, R [g] \,+ \,
\ft 12 \, h_{IJ}\left(\phi\right) \partial_{\mu} \phi^I
\partial_{\nu} \phi^J \, g^{\mu\nu} \right]
\end{equation}
where $h_{IJ}$, $I, \, J \,=\, 1,\,\ldots,\,128$, is the metric of
the homogeneous $128$-dimensional coset manifold
\begin{equation}
  \mathcal{M}_{128}=\frac{\mathrm{E_{8(8)}}}{\mathrm{SO(16)}}
\label{m128}
\end{equation}
\par
The above manifold falls in the general category of manifolds
$\mathrm{U/H}$ such that $\mathbb{U}$ (the Lie algebra of
$\mathrm{U}$) is the maximally non-compact real section of a
simple Lie algebra $\mathbb{U}_C$ and the subgroup $\mathrm{H}$ is
generated by the maximal compact subalgebra $\mathbb{H}\subset
\mathbb{U}$. In this case the solvable Lie algebra description of
the target manifold $\mathrm{U/H}$ is universal. The manifold
$\mathrm{U/H}$ is isometrical to the solvable group manifold:
\begin{equation}
\mathcal{M}_{d} =\exp \, \left [  Solv\left( \mathrm{U/H}\right)
\, \right ] \label{geneUH}
\end{equation}
where the solvable algebra $Solv\left( \mathrm{U/H}\right)$ is
spanned by all the Cartan generators $\mathcal{H}_i$ and by the
step operators $E^\alpha$ associated with all positive roots
$\alpha >0$ (on the solvable Lie algebra parametrization of supergravity scalar
manifolds see \cite{Andrianopoli:1996bq,Andrianopoli:1996zg}). On the other hand the
maximal compact subalgebra
$\mathbb{H}$ is spanned by all operators of the form $E_\alpha -
E_{-\alpha}$ for all positive roots $\alpha >0$. So the dimension
of the coset $d$, the rank $r$ of $\mathbb{U}$ and the number of
positive roots $p$ are generally related as follows:
\begin{equation}
 \mbox{dim} \left[ U/H\right] \equiv d \, = \, r +p \quad ; \quad p
 \,
 \equiv \, \# \mbox{positive roots}=  \mbox{dim} \,
 \mathbb{H} \quad ; \quad r \, \equiv \, \mbox{rank} \,
\mathbb{U}
\label{relazionibus}
\end{equation}
\par
In the present section we concentrate on studying solutions of a
bosonic field theory of type (\ref{lag_sm}) that are only
time-dependent. In so doing we consider the  case of a generic
manifold $\mathrm{U/H}$ and we show how the previously recalled
algebraic structure allows to retrieve a complete generating
solution of the field equations depending on as many essential
parameters as the rank of the Lie algebra $\mathbb{U}$. These
parameters label the orbits of solutions with respect to the
action of the two symmetries present in (\ref{lag_sm}), namely
$\mathrm{U}$ global symmetry and $\mathrm{H}$ local symmetry.
\par
The essential observation is that, as long as we are interested in
solutions depending only on time, the field equations of
(\ref{lag_sm}) can be organized as follows. First we write the
field equations of the matter fields $\phi^I$ which supposedly
depend only on time. At this level the coupling of the sigma-model
to three dimensional gravity can be disregarded. Indeed the effect
of the metric $g_{00}$ is simply that the  field equations for the
scalars $\phi^I$ have the same form as they would have in a rigid
sigma model with just the following proviso. The parameter we use
is proper time rather than coordinate time. Next  in the variation
with respect to the metric we can use the essential feature of
three--dimensional gravity, namely the fact that the Ricci tensor
completely determines also the Riemann tensor. This means that
from the stress energy tensor of the sigma model solution we
reconstruct, via Einstein field equations, also the corresponding
three dimensional metric.
\par
\subsection{Decoupling the sigma model from gravity}
Since we are just interested  in configurations where the fields
depend only  on time, we take the following ansatz for the three
dimensional metric:
\begin{equation}
  ds^2_{3D} = A^2(t)\, dt^2 - B^2(t)\left(  dr^2 + r^2 d\phi^2 \right)
\label{confpiatto}
\end{equation}
where $A(t)$  and $B(t)$ are undetermined  functions of time. Then
we observe that one of these functions can always be reabsorbed
into a redefinition of the time variable. We fix such a coordinate
gauge by requiring that the matter field equations for the sigma
model should be decoupled from gravity, namely should have the
same form as in a flat metric. This will occur for a special
choice of the time variable. Let us see how.
\par
In general, the sigma model equations, coupled to gravity, have
the following form:
\begin{equation}
  \Box_{cov} \, \phi^I + \Gamma^I_{JK} \partial_\mu \phi^J \,
  \partial_\nu \phi^K \, g^{\mu \nu }= 0
\label{genersigma}
\end{equation}
In the case we restrict dependence only on time the above
equations reduce to:
\begin{equation}
  \frac{1}{\sqrt{-\mbox{det}g}} \,\frac{d}{dt} \left(
  \sqrt{-\mbox{det}g} \, g^{00} \, \frac{d}{dt} \, \phi^I \right)  +
  \Gamma^I_{JK} \frac{d}{dt} \phi^J \,
  \frac{d}{dt}\phi^K \, g^{00 } = 0
\label{sigmamodtime}
\end{equation}
We want to choose a new time $\tau =\tau(t)$ such that  with
respect to this new variable equations (\ref{sigmamodtime}) take
the same form as they would have in a sigma model in flat space,
namely:
\begin{equation}
\ddot{\phi}^I \,+\, \Gamma^{I}_{JK}\, \dot{\phi}^J \, \dot{
\phi}^K \,=\,0 \qquad I, \, J, \, K\,=\, 1,\,\ldots,\,\mbox{dim}
\, \mathcal{M} \label{geodesiaque}
\end{equation}
where $\Gamma^{I}_{JK}$ are the Christoffel symbols for the metric
$h_{IJ}$. The last equations are immediately interpreted as
geodesic equations in the target scalar manifold.
\par
In order for equations (\ref{sigmamodtime}) to reduce to
(\ref{geodesiaque}) the following condition must be imposed:
\begin{equation}
  \sqrt{-\mbox{det}g} \, g^{00} \, \frac{d}{dt} = \frac{d}{d\tau} \,
  \Rightarrow\, dt = \sqrt{-\mbox{det}g} \, g^{00} \, d\tau
\label{timegauge}
\end{equation}
Inserting the metric (\ref{confpiatto}) into the above condition
we obtain an equation for the coefficient $A(t)$ in terms of the
coefficient $B(t)$. Indeed in the new coordinate $\tau$ the metric
(\ref{confpiatto}) becomes:
\begin{equation}
  ds^2_{3D} = B^4(\tau)\, d\tau^2 - B^2(\tau)\left(  dr^2 + r^2 d\phi^2 \right)
\label{confpiatto2}
\end{equation}
 The choice (\ref{confpiatto2}) corresponds
to the following choice of the \textit{dreibein}:
\begin{equation}
  e^0 = B^2(\tau) \, d\tau \quad ; \quad e^1 = B(\tau) dr \quad ; \quad e^2 =
  B(\tau) \, r \, d\phi
\label{dreibeinconfpiatto}
\end{equation}
For such a metric the curvature $2$-form is as follows:
\begin{eqnarray}
R^{01} & = & \frac{2 \dot{B}^2(\tau)- B(\tau) \,
\ddot{B}(\tau)}{B^6(\tau)}\, e^0 \, \wedge \,e^1 \nonumber\\
R^{02} & = & \frac{2 \dot{B}^2(\tau)- B(\tau) \,
\ddot{B}(\tau)}{B^6(\tau)}\, e^0 \, \wedge \,e^2 \nonumber\\
R^{12} & = & -\frac{ \dot{B}^2(\tau)}{B^6(\tau)} e^1 \, \wedge
\,e^2 \label{curvaformd3}
\end{eqnarray}
The Einstein equations, following from our lagrangian
(\ref{lag_sm}) are the following ones, in flat indices:
\begin{equation}
  2 \, G_{ab} \, = \, T_{ab} \quad ; \quad G_{ab} \equiv \mbox{Ric}_{ab}
  - \ft 12 \, R \,\eta_{ab} \quad ; \quad a,b=0,1,2
\label{3einsteino}
\end{equation}
With the above choice of the vielbein, the flat index Einstein
tensor is easily calculated and has the following form:
\begin{equation}
  G_{00}= \frac{\dot{B}^2(\tau)}{2 \,B^6(\tau)} \quad ; \quad
            G_{0i} = 0 \quad ; \quad G_{ij} = \frac{2\,
            \dot{B}^2(\tau)
            - B(\tau) \, \ddot{B}(\tau)}{2 \,B^6(\tau)} \,\delta_{ij}
            \quad ; \quad i,j = 1,2
\label{einsteinten}
\end{equation}
On the other hand, calculating the stress energy tensor of the
scalar matter in the background of the metric (\ref{confpiatto2})
we obtain (also in flat indices):
\begin{equation}
  T_{00} = \frac{1}{2 \, B^4(\tau)} \,\left(  \dot{\phi}^I  \dot{\phi}^J
h_{IJ}\right) \quad ;
  \quad T_{0i} =0 \quad ;
  \quad T_{ij} = \frac{1}{2 \, B^4(\tau)} \,\left(  \dot{\phi}^I  \dot{\phi}^J
h_{IJ}\right) \,
  \delta_{ij} \quad ; \quad i,j = 1,2
\label{stresstensore}
\end{equation}
where
\begin{equation}
            \dot{\phi}^I  \dot{\phi}^J h_{IJ} =
  \varpi^2
\label{omegasquare}
\end{equation}
is a constant independent from time as a consequence of the
geodesic equations (\ref{geodesiaque}). To prove this it suffices
to take a derivative in $\tau$ of ${\varpi^2}$ and verify that it
is zero upon use of eq.s (\ref{geodesiaque}).
\par
Hence in order to satisfy the coupled equations of gravity and
matter fields it is necessary that:
\begin{equation}
 2 \,  \frac{\dot{B}^2(\tau)}{2 \,B^6(\tau)} = 2\, \frac{2\,
            \dot{B}^2(\tau)
            - B(\tau) \, \ddot{B}(\tau)}{2 \,B^6(\tau)} = \frac{1}{2 \, B^4(\tau)}
            \,{\varpi^2}
\label{einsteinocon}
\end{equation}
The first of the above equalities implies:
\begin{equation}
  \dot{B} = k \, B \quad \Rightarrow \quad B(\tau) =
  \exp [ k \, \tau]
\label{kdefi}
\end{equation}
where $k$ is some constant. The second equality is satisfied if:
\begin{equation}
  k = \pm \ft 1{\sqrt{2}} \, |\varpi| =
  \pm \ft 1{\sqrt{2}} \,\sqrt{\dot{\phi}^I  \dot{\phi}^J h_{IJ}}
\label{kvalue}
\end{equation}
In this way we have completely fixed the metric of the
three--dimensional space as determined by the solution of the
geodesic equations for the scalar matter:
\begin{equation}
  ds^2_{3D} = \exp \left[ 4 k \,\tau\right]  d\tau^2 - \exp \left[ 2 k
  \,\tau\right]\, \left (dx_1^2 + dx_2^2 \right)
\label{orporpo}
\end{equation}
with the parameter $k$ given by eq.(\ref{kvalue}).
\subsection{Geodesic equations in target space and the Nomizu
operator} Having clarified how the three dimensional metric is
determined in terms of the solutions of the sigma model, we
concentrate on this latter. We focus on the geodesic equations
(\ref{geodesiaque}) and in order to study them, we rely on the
solvable Lie group description of the target manifold going to an
anholonomic basis for the tangent vectors to the geodesic.
Since gravity is decoupled from the scalars, we deal with a rigid
sigma-model where the fields depend only on time
 \begin{equation} \mathcal{L}^{\sigma-model}  \propto
 h_{IJ}(\phi)\dot{\phi}^I\dot{\phi}^J
 \label{lagra3d1}
 \end{equation}
 As was mentioned before, the equations of motion in this case
 reduce to the geodesic equations for the metric $h_{IJ}(\phi)$
 and time plays the role of a parameter along the geodesics
(see eq.(\ref{geodesiaque})).
 Since $h_{IJ}(\phi)$ is the metric of a scalar manifold
 which is a maximally non-compact coset  $\mathcal{M} = \mathrm{U/H}$, we
 can derive this metric from a coset representative
 $\mathbb{L}(\phi)\in U$
 \begin{equation}
 h_{IJ}(\phi) = \mathrm{Tr}(\mathbb{P}_K
 \IL^{-1}\partial_I\IL\IL^{-1}\partial_J\IL)
 \end{equation}
$\mathbb{P}_K$ being a projection operator on the coset directions
of the Lie algebra $\mathbb{U}$ to be discussed in a moment. To
this effect we introduce the following general notation. We make
the orthogonal split of the $\mathbb{U}$ Lie algebra:
\begin{equation}
  \mathbb{U} = \mathbb{H} \oplus \mathbb{{K}}
\label{orthosplitto}
\end{equation}
where $\mathbb{H} \subset \mathbb{U}$ is the maximal compact
subalgebra and $\mathbb{{K}}$ its orthogonal complement.
 We adopt the following normalizations for the generators in each
 subspace:
\begin{eqnarray}
\nonumber && \mathbb{U} = \mathrm{Span}\{H_i, E_{\alpha}, E_{-\alpha}\} \\
&& \mathbb{K}=\mathrm{Span}\{K_A\} = \mathrm{Span}\{H_i,
\ft 1{\sqrt{2}}(E_{\alpha} + E_{-\alpha})\} \\ && \nonumber
\mathbb{H} = \mathrm{Span}\{t_{\alpha}\} =
\mathrm{Span}\{(E_{\alpha} - E_{-\alpha})\}
\end{eqnarray}
The ${\mathbb{U}}$ Lie-algebra valued  left invariant one-form
\begin{equation}
\Omega = \mathbb{L}^{-1} d \mathbb{L} \,=\, V^AK_A +
\omega^{\alpha}t_{\alpha}  \label{solvodecompo}
\end{equation}
is in general expanded along all the generators of $\mathbb{U}$
(not only along $\mathbb{{K}}$) and $V = V^A\, K_A $ corresponds
to the coset manifold vielbein while
 $\omega=\omega^\alpha\, t_\alpha$ corresponds to the
coset manifold $\mathrm{H}$--connection.
\par
 As it is well known neither the
coset representative $\mathbb{L}(\phi)$, nor the one-form $\Omega$
are unique. Indeed $\mathbb{L}$ is defined up to multiplication on
the right by an element of the compact subgroup $h\in H$. This is
a gauge invariance which can be fixed in such a way that the coset
representative lies in the  solvable group ${Solv}(\mathrm{U/H})$ obtained by
exponentiating the solvable subalgebra $Solv(\mathbb{U}/\mathbb{H})$
\begin{equation}
\mathbb{L} \left( \phi \right) \,=\, \exp{\left( Solv(\mathbb{U}/\mathbb{H}) \cdot \phi
\right)} \label{solcoset}
\end{equation}
In the case of $\mathrm{U}/\mathrm{H}$ being maximally non compact
$Solv(\mathbb{U}/\mathbb{H})$ coincides with the Borel subalgebra and
therefore it is spanned by the collection of all Cartan generators
and  step-operators associated with positive roots, as we already
stated, namely:
\begin{equation}
Solv(\mathbb{U}/\mathbb{H}) = \mathrm{Span} \,\left\{ T_A\right\}  =
\mathrm{Span} \, \left \{H_i, E_{\alpha} \right \}
\label{solvospan}
\end{equation}
If the coset representative $ \mathbb{L}$ is chosen to be a
solvable group element, as in eq. (\ref{solcoset}), namely if we
are in the solvable parametrization of the coset, we can also
write:
\begin{equation}
            \Omega = \mathbb{L}^{-1} d\mathbb{L} = \widetilde{V}^i \, \mathcal{H}_i
\, + \,
  \widetilde{V}^\alpha \, E_\alpha = \widetilde{V}^A \, T_A
\label{solvomega}
\end{equation}
since $\Omega$ is contained in the solvable subalgebra
$Solv(\mathbb{U}/\mathbb{H}) \subset \mathbb{U}$. Eq.s (\ref{solvodecompo})
and (\ref{solvomega}) are compatible if and only if:
\begin{equation}
  V^\alpha = \sqrt{2} \, \omega^\alpha
\label{solvocondo}
\end{equation}
In this case we can identify $\widetilde{V}^i={V}^i$ and
$\widetilde{V}^\alpha = \sqrt{2} \, V^\alpha$; eq.(\ref{solvocondo}) is the
solvability condition for a coset
representative.
\par
Hence we can just rewrite the metric of our maximally non--compact
manifold $\mathrm{U} / \mathrm{H}$ as follows:
\begin{equation}
ds^{2}_{\mathrm{U}/\mathrm{H}} \,=\, \sum_{A=1}^{\mbox{dim}
\mathrm{U}}  V^A \,\otimes\, V^A = \tilde{V}^i\oplus \tilde{V}^i +
\ft 12 \tilde{V}^{\alpha}\oplus \tilde{V}^{\alpha}
\label{metricUH}
\end{equation}
It is interesting to discuss what are the residual $\mathrm{H}$--gauge
transformations that remain available after the solvable gauge
condition (\ref{solvocondo}) has been imposed. To this effect we
consider the multiplication
\begin{equation}
  \mathbb{L} \mapsto \mathbb{L} \, h = \overline{\mathbb{L}}
\label{hgauga}
\end{equation}
where
\begin{equation}
  h = \exp \, \left[  \theta^\alpha  \, t_\alpha  \right]
\label{helemo}
\end{equation}
is a finite element of the $\mathrm{H}$ subgroup singled out by generic
parameters theta.
 For any such element we can always write:
\begin{eqnarray}
h^{-1} \, t_\alpha  \, h & = & A(\theta)_\alpha^{\phantom{\alpha }\beta} \,t_\beta
\nonumber\\
h^{-1} \, K_A  \, h &  = & D(\theta)_A^{\phantom{A }B} \,K_B
\label{duematrici}
\end{eqnarray}
where the matrix $A(\theta)$ is the adjoint representation of $h$
and $D(\theta)$ is the $D$--representation of the same group
element. We obtain:
\begin{eqnarray}
            \overline{\Omega} \equiv \overline{\mathbb{L}}^{-1} \,
            d\overline{\mathbb{L}} & = & h^{-1}dh + h^{-1} \, \Omega \, h
            \nonumber\\
            & = & \underbrace{h^{-1}\, dh  + h^{-1} \,
            \omega \, h}_{=\,\,\overline{\omega}} \, + \underbrace{\, h^{-1} \, V \,
            h}_{=\,\,\overline{V}}
\label{omegabarra}
\end{eqnarray}
where
\begin{eqnarray}
\overline{\omega}^\alpha & = & \frac {1}{\mbox{tr}(t_\alpha ^2)}
\,\mbox{tr} \left( h^{-1}\, dh \, t_\alpha \right)  \, + \,
 \omega^\beta  \, A(\theta)_{\beta}^{\phantom{\beta}\alpha}\nonumber\\
\overline{V}^\alpha & = & V^\beta \, D(\theta)_{\beta}
^{\phantom{\beta}\alpha} \, + \, V^i \,
D(\theta)_i^{\phantom{i}\alpha} \, \label{bongo}
\end{eqnarray}
Suppose now that the coset representative $\mathbb{L}$ is
solvable, namely it satisfies eq.(\ref{solvocondo}). The coset
representative $\overline{\mathbb{L}}$ will still satisfy the same
condition if the $h$-compensator satisfies the following
condition:
\begin{equation}
 \frac{\sqrt{2}}{\mbox{tr}(t_\alpha ^2)} \, \,\mbox{tr} \left( h^{-1}(\theta)\,
dh(\theta) \, t_\alpha
  \right)= V^\beta \, \left(  A(\theta)_\beta^{\phantom{\beta}\alpha}
  \, -\, D(\theta)_\beta^{\phantom{\beta}\alpha} \right) \, + \,  V^i \,
  D(\theta)_i^{\phantom{i}\alpha}
\label{daequa}
\end{equation}
The above equations are a set of $n=\#roots = \mbox{dim}H$
differential equations on the parameters $\theta^\alpha$ of the
$h$--subgroup element (\textit{compensator}). In the following we
will use such set of equations as the basis of an algorithm to
produce solutions of the geodesic equations (\ref{geodesiaque}).

Given these preliminaries, we can establish a new notation. We
introduce tangent vectors to the geodesics in the anholonomic
basis:
\begin{equation}
\Phi^A \,=\, \widetilde{V}^A_I\left(\phi\right) \dot{\phi}^I
\label{anholotang}
\end{equation}
which are functions only of time: $\Phi^A \,=\,\Phi^A(t)$. In this
basis the field equations reduce to
\begin{equation}
\label{D=3feqn} \dot{\Phi}^A \,+\, \Gamma^A_{BC} \Phi^B \Phi^C
\,=\,0
\end{equation}
where now $\Gamma^A_{BC}$ are the components of the Levi-Civita
connection in the chosen anholonomic basis. Explicitly they are
related to the components of the Levi Civita connection in an
arbitrary holonomic basis by:
\begin{equation}
\Gamma^A_{BC}\,= \Gamma^I_{JK}V^A_I V^J_B V^K_C - \partial_K
(V^A_J)V^J_B V^K_C \label{capindgamma}
\end{equation}
where the inverse vielbein is defined in the usual way:
\begin{equation}
V^A_I \, V^I_B \,=\, \delta^A_B \label{invervielb}
\end{equation}
The most important point here is that,  the connection
$\Gamma^A_{BC}$ can be identified with the \textit{Nomizu
connection} defined on a solvable Lie algebra, if the coset
representative $\mathbb{L}$ from which we construct the vielbein
via eq.(\ref{solvodecompo}), is solvable, \textit{namely if and
only if} the solvability condition (\ref{solvocondo}) is
satisfied. In fact, as we can read in \cite{alekseevskii}, once we
have defined over $Solv$ a non degenerate, positive definite and
symmetric form:
\begin{eqnarray}
\langle \,,\, \rangle & \; : \; Solv \otimes Solv \longrightarrow \mathbb{R}
\nonumber \\
\langle X \,,\, Y \rangle & \; = \; \langle Y \,,\, X \rangle
\end{eqnarray}
whose lifting to the manifold produces the metric, the covariant
derivative is defined through the \textbf{Nomizu operator}:
\begin{equation}
\forall X \in Solv \,:\, \mathbb{L}_X: Solv \longrightarrow Solv
\end{equation}
so that
\begin{equation}
\forall X,Y,Z \in Solv \,:\, 2 \langle Z \,,\, \mathbb{L}_X Y
\rangle \,=\, \langle Z, \left[ X,Y \right] \rangle \,-\, \langle
X, \left[ Y,Z \right] \rangle \,-\, \langle Y, \left[ X,Z \right]
\rangle \label{Nomizuoper}
\end{equation}
while the Riemann curvature 2-form is given by the commutator of
two Nomizu operators:
\begin{equation}
R^W_{\phantom{W}Z} \left( X,Y \right) \,=\, \langle W \,,\,
\left\{ \left[ \mathbb{L}_X , \mathbb{L}_Y \right] \,-\,
\mathbb{L}_{\left[X,Y\right]} \right\} Z \rangle
\label{Nomizucurv}
\end{equation}
This implies that the covariant derivative explicitly reads:
\begin{equation}
\mathbb{L}_X \,Y \,=\, \Gamma_{XY}^Z \,Z \label{Gammonedefi}
\end{equation}
where
\begin{equation}
\Gamma_{XY}^Z \,=\,\ft 12\left( \langle Z, \left[ X,Y \right]
\rangle \,-\, \langle X, \left[ Y,Z \right] \rangle \,-\, \langle
Y, \left[ X,Z \right] \rangle\right) \, \frac{1}{<Z,Z>} \,\qquad\,
\forall X,Y,Z \in Solv \label{Nomizuconne}
\end{equation}
\par
In concrete, the non degenerate, positive definite, symmetric form
on the solvable Lie algebra which agrees with equation
(\ref{metricUH}) is defined by setting:
\begin{eqnarray}
\langle \mathcal{H}_i \,,\, \mathcal{H}_j \rangle & \,=\, & 2 \, \delta_{ij}
\nonumber \\
\langle \mathcal{H}_i \,,\, E_\alpha \rangle & \,=\, & 0 \nonumber \\
\langle E_\alpha \,,\, E_\beta \rangle & \,=\, &
\delta_{\alpha,\beta}
\end{eqnarray}
$\forall \mathcal{H}_i ,\, \mathcal{H}_j \,\in\,
\mathrm{CSA}_{\mathrm{E_{8(8)}}} $ and $\forall E_\alpha$, step
operator associated to a positive root $\alpha$ of
$\mathrm{E_{8(8)}}$. Then the Nomizu connection (which is
constant) is very easy to calculate. We have:
\begin{equation}
\begin{array}{lll}
\Gamma^i_{jk}& \,=&\,  0 \nonumber \\
\Gamma^i_{\alpha\beta} &\,=& \,  \ft 12\left(-\langle
E_\alpha,\,\left[E_\beta,\,H^i\right]\rangle - \langle
E_\beta,\,\left[E_\alpha,\,H^i\right]\rangle\right)\,=\,\ft 12 \,
\alpha^i \delta_{\alpha\beta}
\nonumber \\
\Gamma^{\alpha}_{ij}   &\,=&
\, \Gamma^{\alpha}_{i\beta} \,=\, \Gamma^i_{j\alpha} \,=\,0 \nonumber \\
\Gamma^{\alpha}_{\beta i} &\,=& \, \ft 12\left(\langle
E^\alpha,\,\left[E_\beta,\,H_i\right]\rangle -
\langle E_\beta,\,\left[H_i,\,E^\alpha\right]\rangle\right)\,=\,
-\alpha_i\,\delta^{\alpha}_{\beta}\nonumber\\
\Gamma^{\alpha+\beta}_{\alpha\beta}   &\,=&
\, -\Gamma^{\alpha+\beta}_{\beta\alpha}\,=\,\ft 12 N_{\alpha\beta}\nonumber\\
\Gamma^{\alpha}_{\alpha+\beta\,\beta}  &\,=&
 \, \Gamma^{\alpha}_{\beta\,\alpha+\beta} \,=\,\ft 12 N_{\alpha\beta}
\end{array}
 \label{Nomizuconne2}
\end{equation}
where $N_{\alpha\beta}$ is defined by the commutator:
\begin{equation}
\left[ E_\alpha \,,\, E_\beta \right] \,=\,
N_{\alpha\beta}\,E_{\alpha + \beta} \label{nalfabeta}
\end{equation}
which has to be worked out in the algebra.
\footnote{The values of the constants $N_{\alpha\beta}$, that enable to construct
explicitly the representation of $E_{8(8)}$, used in this paper,
 are given in the hidden appendix. To see it, download the source file, delete the
tag $end\{document\}$
 after the bibliography and $LaTeX$.}
Notice that
$\Gamma^Z_{XY}\neq \Gamma^Z_{YX}$ since its expression consists of
the first term which is antisymmetric in $(X,\,Y)$ and the sum of
the last two which is symmetric. The component
$\Gamma^{\alpha}_{\beta i}$ consists of the sum of two equal
contributions from the antisymmetric and symmetric part, the same
contributions cancel in  $\Gamma^{\alpha}_{i\beta}$ which indeed
vanishes.
 By substituting the explicit expression of the Nomizu connection
in (\ref{D=3feqn}) and introducing for the further convenience new
names for the tangent vectors along the Cartan generators
$\chi^i\equiv\Phi^i $ we have the equations:
\begin{eqnarray}
\label{D=3feqn_2} \dot{\chi}^i & \,+\, &  \, \ft 12
\,\sum_{\alpha\in \Delta_+}  \alpha^i \Phi_\alpha^2 \,=\, 0
\nonumber \\
\dot{\Phi}^\alpha &\,+\, &  \,\sum_{\beta\in
\Delta_+}\,N_{\alpha\beta} \Phi^{\beta}  \Phi^{\alpha+\beta} -
\alpha_i\,\chi^i\Phi^\alpha\,=\,0
\end{eqnarray}
Eq.s (\ref{D=3feqn_2} ) encode all the algebraic structure of the
$D=3$ sigma model and due to our oxidation algorithm of the
original supergravity in ten dimensions.
\par

All this means that, thanks to the solvability of the algebra (and
also to the fact that we know the explicit form of the connection
via the Nomizu operator), we have reduced the entire problem of
finding time dependent backgrounds for either type II A or type II
B superstrings or M-theory to the integration of a system of
differential equations firmly based on the algebraic structure of
$E_{8(8)}$. This is a system of non-linear differential equations,
and from this point of view it might seem hopeless to be solved.
Yet, due to its underlying algebraic structure, one can use its
isometries to generate the complete integral depending on as many
integration constants as the number of equations in the system.
This is the compensator algorithm we alluded to above, which we
shortly outline. To this effect we discuss the role of initial
conditions for the tangent vectors to the geodesics. There exist a
number of possibilities for such conditions that can truncate the
whole system to  smaller and simpler ones. The simplest choice is
to put all root-vectors to zero in the origin. This will ensure
that root-vectors will remain zero at all later times and the
system will reduce to
\begin{equation}
\dot{\chi}^i = 0 \label{dchi=0}
\end{equation}
The solution of such a reduced system is trivial and consists of a
constant vector $\tilde{V}^A = (\chi^i, 0)$.  If we apply an
$\mathrm{H}$--rotation to this tangent vector
\begin{equation}
  \overline{V}^A = V^B \, D(\theta)_B^{\phantom{B}A}
\label{rotatedV}
\end{equation}
we produce a new one, yet, for generic $\mathrm{H}$--rotations we will
break the solvable gauge, so that the result no longer produces a
solution of eq.s(\ref{D=3feqn_2}). However, if we restrict the
$\theta^\alpha$ parameters of the rotation to satisfy condition
(\ref{daequa}), then the solvable gauge is preserved and the
rotated tangent vector $\overline{V}^A$ is still a solution of
eq.s(\ref{D=3feqn_2}). Hence a general algorithm to solve the
differential system (\ref{D=3feqn_2}) has been outlined. One
starts from the trivial solution in eq.(\ref{dchi=0}) and then
tries to solve the differential equation for the  theta parameter
corresponding to one particular $\mathrm{H}$--generator $t_\alpha =
E_\alpha -E_{-\alpha}$. Applying this rotation to the trivial
solution we obtain a new non trivial one. Then starting from such
a new solution we can repeat the procedure and try to solve again
the differential equation for the theta parameter relative to a
new generator. If we succeed we obtain a further new solution of
the original system and we can repeat the procedure a third time
for a third generator, iteratively. Indeed, considering
eq.(\ref{daequa}) we see that if $h(t)$ is just a general element
of the subgroup $\mathrm{H}$, the system is
 rather difficult to solve, yet if we choose
 a rotation around a single axis $h_{\alpha_0} =
 e^{\theta^{\alpha_0}(t)t_{\alpha_0}}$,  then
$\frac{1}{\mathrm{Tr}(t_{\alpha_0}^2)}\mathrm{Tr}(h^{-1}dh
t_{\alpha_0}) = \dot{\theta^{\alpha_0}}$ and, if all the other
equations for $\alpha \ne \alpha_0$ are identically satisfied, as
it will turn out to be the case in the examples we consider, then
the system reduces to only one first order differential equation
on the angle $\theta^{\alpha_0}(t)$.
\par
 We name such an algorithm \textit{the compensator method} and we will illustrate it
in the next
section with specific examples.
\section{The $\mathrm{A_2}$ toy model as a paradigma}
\label{exampsolv} In this section we consider explicit examples of
solutions of the geodesic problem in  the case of an
$\mathrm{A_2}$ simple algebra. Later we will consider the possible
embeddings of such an algebra into the $\mathrm{E_8}$ algebra, so
that the  solutions we construct here will be promoted to
particular solutions of the full $\mathrm{E_{8(8)}/SO(16)}$ sigma
model. The diverse embeddings will correspond to diverse
oxidations of the same three dimensional configuration to $D=10$
configurations.  In other words there exist various   non abelian
solvable subalgebras $S_5 \subset Solv(\mathrm{E_8}/\mathrm{SO(16)})$ of dimension
$5$ which by  means of a linear transformation can be  identified
as the solvable Lie algebra of the simple Lie algebra $A_2$,
namely the solvable Lie algebra description of the coset manifold:
\begin{equation}
 \mathcal{M}_5 \equiv \exp \left [Solv(\mathrm{A_2}) \right ] \,
 \cong \, \frac{\mathrm{SL(3,\mathbb{R})}}{\mathrm{SO(3)}}
\label{solv5}
\end{equation}
The  detailed study of this model provides our paradigma for the
general solution of the complete theory based on the coset
manifold $\mathrm{E_{8(8)}/SO(16)}$. We emphasize that the
possibility of choosing a \textit{normal form} for the initial
tangent vector to the geodesic  allows to reduce the system of
first order equations to a much simpler set, as we started to
discuss in the previous section in general terms. Such a normal
form can be chosen in different ways. In particular it can always
be chosen so that it contains only Cartan generators. When this is
done the system   is always exactly solvable and in terms of pure
exponentials. The solution obtained in this way provides a
representative for the orbit of geodesics modulo isometries. We
can then generate new solutions of the differential system
(\ref{D=3feqn_2}) by the compensator method we described in the
previous section.
  In this section we
illustrate such an algorithm in the case of the toy $\mathrm{A_2}$
model. The resulting solutions have not only a tuitional interest,
rather they provide  examples of solutions of the full
$\mathrm{E_{8(8)}}$ system and hence of full supergravity. It
suffices to embed the $\mathrm{A_2}$ Lie algebra in the full
algebra $\mathrm{E_{8(8)}}$. We will discuss such embeddings and
the corresponding oxidations of our sigma model solutions in later
sections.
\subsection{Structure of the $\mathrm{A_2}$ system}
Our model consists of 5 scalar fields, which parametrize a coset
manifold $\mathcal{M}_5 = \mathrm{SL(3)/SO(3)}$. Our chosen
conventions are as follows. The two simple roots of
$\mathrm{SL(3)}$ are:
\begin{equation}
\beta_1  = \left\{ \sqrt{2} \, , \, 0 \right\}, \hspace{0.5cm}
\beta_2 = \left\{ - \ft 1{\sqrt{2}}\, ,
\,\sqrt{\ft 32}\right\} \label{simplea2}
\end{equation}
and the third positive root, which is the highest is:
\begin{equation}
  \beta_3 = \beta_1 + \beta_2 = \left\{  \ft 1{\sqrt{2}}\, ,
\,\sqrt{\ft 32}\right\} \label{highroota2}
\end{equation}
Furthermore the step operator $E_{\beta_3}$ is defined through
the commutator:
\begin{equation}
  E_{\beta_3} = \left[ E_{\beta_1} \, , \,
  E_{\beta_2}\right]
\label{steppotre}
\end{equation}
and this completely fixes all conventions for the Lie algebra
structure constants.
\par
The three generators of the maximally compact subgroup are defined
as:
\begin{equation}
  t_1 = E_{\beta _1} -E_{-\beta _1} \quad , \quad t_2 = E_{\beta _2} -E_{-\beta _2}
  \quad , \quad t_3 = E_{\beta _3} -E_{-\beta _3}
\label{generso3}
\end{equation}
and they satisfy the standard commutation relations:
\begin{equation}
  \left[ t_i \, , \, t_j \right] \, = \, \epsilon _{ijk} \, t_k
\label{standeso3}
\end{equation}
In the orthogonal decomposition of the Lie algebra:
\begin{equation}
  \mathrm{A_2} = \mathrm{SO(3)} \, \oplus \, \mathbb{K}_5
\label{orthodeco}
\end{equation}
the $5$-dimensional subspace $\mathbb{K}_5$  is identified with
the tangent space to $\mathcal{M}_5$ and corresponds to the $j=2$
representation of ${\mathrm{SO(3)}}$
 \begin{equation}
[t_{\beta},K_A] = Y_{\beta A}^BK_B
\end{equation}
 This subspace is spanned by the following generators:
\begin{equation}
 \mathbb{ K}_5 = \mathrm{Span}\left\{ H_1 \, , \, H_2 \, , \, \ft 1{\sqrt{2}} \left(
E_{\beta _1}
  + E_{-\beta _1}\right)
\, , \, \ft 1{\sqrt{2}} \left( E_{\beta _2} + E_{-\beta
_2}\right)\, , \, \ft 1{\sqrt{2}} \left( E_{\beta _3} +
E_{-\beta _3}\right)\right\} \label{K5generi}
\end{equation}
\par
Applying to this case the general formulae (\ref{D=3feqn_2}) based
on the Nomizu connection (\ref{Nomizuconne2}) we obtain the
differential system:
\begin{eqnarray}
 &&\dot{\chi}_1(t) + \ft 1{\sqrt{2}}\Phi^2_1(t) -
  \ft 1{2\sqrt{2}}\Phi_2^2(t) +
  \ft 1{2\sqrt{2}}\Phi_3^2(t)   =  0 \nonumber\\
 &&\dot{\chi}_2(t) + \ft {\sqrt{3}}{2\sqrt{2}}\Phi^2_2(t)
  + \ft {\sqrt{3}}{2\sqrt{2}}\Phi^2_3(t)  =  0
            \nonumber\\
            && \dot{\Phi}_1(t)  + {{\Phi }_2}(t)\,{{\Phi }_3}(t) -
  {\sqrt{2}}\,{{\Phi }_1}(t)\,
            {{\chi }_1}(t)
            =  0 \nonumber\\
&& \dot{\Phi}_2(t) - {{\Phi }_1}(t)\,
                     {{\Phi }_3}(t)   +
  \ft 1{\sqrt{2}}\Phi_2(t)
                     \chi_1(t) -
  \sqrt{\ft 32}
            \Phi_2(t)\chi_2(t)
            = 0\nonumber\\
&& \dot{\Phi}_3(t) -\ft 1{\sqrt{2}}\Phi_3(t)\chi_1(t) -
  \sqrt{\ft 32}\Phi_3(t)\chi_2(t)
            = 0 \label{A2system} \end{eqnarray}
In order to solve this differential system of equations we recall
their geometrical meaning. They are the geodesics equations for
the manifold (\ref{solv5}) written in flat indices, namely in an
anholonomic frame. Any geodesics is completely determined by two
data: the initial point $p_0 \in M_5$ and the initial tangent
vector $\overrightarrow{t}_0 \in T(M_5)$ at time $t=0$. Since our
manifold is homogeneous, all points are equivalent and we can just
choose the origin of the coset manifold. Since we are interested
in determining the orbits of geodesics modulo the action of the
isometry group, the relevant question is the following: {\it in
how many irreducible representations} of the tangent group
$\mathrm{SO(3)}$ does the tangent space decompose? The answer is
simple: the $5$ dimensional tangent space is irreducible and
corresponds to the $j=2$ representation of $\mathrm{SO(3)}$. The
next question is:\textit{ what is the normal form of such a representation}
and how many parameters does it contain. The answer is again
simple. A spin two representation is just a symmetric traceless
tensor $g_{ij}$ in three dimensions. By means of $\mathrm{SO(3)}$
rotations we can reduce it to a diagonal form and the essential
parameters are its eigenvalues, namely two parameters, since the third eigenvalue is
minus the sum of the other two, being the matrix traceless. So by
means of  $\mathrm{SO(3)}$ rotations a generic $5$-dimensional
tangent vector can be brought to contain only two parameters. This
argument is also evident from the consideration that $5-3=2$,
namely by means of the three $\mathrm{SO(3)}$ parameters we can
set three components of the $5$-dimensional vector to zero.
\par
We can also analyze the normal form of the $5$--dimensional
representation from the point of view of eigenstates of the
angular momentum third component $t_3$. This latter has skew
eigenvalues $\pm 2, \pm 1 $ and $0$. The transformation of the
matrix $g=\left\{ g_{ij} \right\} $ under any generator $t$ of the
$\mathrm{SO(3)}$ Lie algebra is
\begin{equation}
  \delta \, g \, = \, \left[ t \, , \, g \right]
\label{trasforg}
\end{equation}
so that the pair of skew eigenstates of the generator $t_3$, as given in eq.
(\ref{tgene3}), pertaining to the skew eigenvalues $\pm 2$ is provided by the
symmetric matrices of the form:
\begin{equation}
g_{(\pm 2)} =  \left(\begin{array}{ccc}
  a & 0 & b \\
  0 & 0 & 0 \\
  b & 0 & -a
\end{array} \right)
\label{eigepm2}
\end{equation}
which can be diagonalized through $\mathrm{SO(3)}$ rotations
(actually $\mathrm{SO(2)}$ in this case) and brought to the normal
form:
\begin{equation}
g_{2} =  \left(\begin{array}{ccc}
  \sqrt{a^2 + b^2} & 0 & 0 \\
 0 & 0 & 0 \\
  0 & 0 & -\sqrt{a^2 + b^2}
\end{array} \right)
\label{eige2}
\end{equation}
which is just one of the two in the pair of skew eigenstates. On
the other hand the symmetric traceless matrix that corresponds to
the null eigenstate of $t_3$ is:
\begin{equation}
  g_{(0)} = \left( \begin{array}{ccc}
             s & 0 & 0 \\
             0 & -2s & 0 \\
             0 & 0 & s \
  \end{array}\right)
\label{eige0}
\end{equation}
A superposition $g_{2} + g_{(0)}$ provides the most general
diagonal traceless symmetric matrix, namely the \textit{normal
form} to which any state in the $j=2$ irreducible representation
can be brought by means of $SO(3)$ rotations.
\par
Alternatively, since the $j=2$ representation is provided by the
tangent space to the $\mathcal{M}_5$ manifold, spanned by the
coset generators of $\mathrm{SL(3,\mathbb{R})}$ not lying in the
compact $\mathrm{SO(3)}$ subalgebra, we can identify the normal
form of a $5$--dimensional vector as one with non vanishing
components only in the directions of the Cartan generators.
Indeed, by means of $\mathrm{SO(3)}$ rotations any vector can be
brought to such a form and the counting of independent parameters
coincides, namely two. This is a completely general statement for
maximally non compact coset manifolds. The rank of the coset is
equal to the number of independent parameters in the normal form
of the $\mathbb{H}$ representation provided by the coset subspace
$\mathbb{K}$.
\par
Relying on these considerations, let us consider the explicit
representation of the group $\mathrm{SO(3)}$ on the tangent space
to our manifold $\mathcal{M}_5$ and how, by means of its
transformation we can bring the initial tangent vector to our
geodesic to our desired normal form. Indeed our  aim is to solve
 the geodesic equations (\ref{A2system}) fixing
initial conditions:
\begin{equation}
  \left\{ \chi_1(0)\, , \, \chi_2(0) \, , \,\Phi_1(0)\, , \, \Phi_2(0) \,
  , \, \Phi_3(0)\right \} \, = \,\tilde{V} \, =
  \, \left\{ \tilde{V}_1, \tilde{V}_2, \dots ,\tilde{V}_5 \right \}
\label{pincus}
\end{equation}
where $\tilde{V}$ is the normal form of the $5$ vector.
To this effect it is convenient to inspect the representative
matrices of $\mathrm{SO(3)}$ on the tangent space. The three
generators of the maximally compact subgroup were  defined in
(\ref{generso3}) and in the basis of $\mathbb{K}_5$ provided by
the generators (\ref{K5generi}) the $ 5 \times 5$ matrices
representing $\mathrm{SO(3)}$ are:
\par
\begin{eqnarray}
\label{j2generi}
t_1^{[5]}  &=& \left(\matrix{ 0 & 0 &
             -2 & 0 & 0 \cr 0 & 0 & 0 & 0 &
            0 \cr 2 & 0 & 0 & 0 & 0 \cr 0 &
            0 & 0 & 0 & 1 \cr 0 & 0 & 0 &
             -1 & 0 \cr  } \right),
 \hspace{0.2cm}
t_2^{[5]}  =  \left(\matrix{ 0 & 0 & 0 & 1 & 0 \cr 0 & 0 &
            0 & -{
                     \sqrt{3}} & 0 \cr 0 & 0 & 0 &
            0 & -1 \cr -1 & {
                     \sqrt{3}} & 0 & 0 & 0 \cr 0 &
            0 & 1 & 0 & 0 \cr  }\right) \\ \nonumber
            t_3^{[5]} & =&  \left(\matrix{ 0 & 0 & 0 & 0 &
             -1 \cr 0 & 0 & 0 & 0 & -{
                        \sqrt{3}} \cr 0 & 0 & 0 &
             -1 & 0 \cr 0 & 0 & 1 & 0 &
            0 \cr 1 & {
                     \sqrt{3}} & 0 & 0 & 0 \cr  } \right)
\end{eqnarray}
These matrices have  the expected skew eigenvalues:
\begin{equation}
  \left( \pm 2,\, \pm 1,\, 0 \right)
\end{equation}
For the generator $t^{[5]}_3$ the corresponding eigenvectors are:
\begin{equation}
  \begin{array}{lcl}
             \mbox{eigenvalue $0$} &
             \Rightarrow & \{ -{\sqrt{3}},1,0,0,0\} \\
             \mbox{eigenvalues $\pm 1$} & \Rightarrow &
             \left \{\begin{array}{l}
                        \{ 0,0,1 ,1,0\} \\
                                \{ 0,0,-1 ,1,0\} \\
             \end{array} \right.\\
             \mbox{eigenvalues $\pm 2$ } & \Rightarrow &
             \left \{ \begin{array}{l}
                                \{ \frac{1}{{\sqrt{2}}},
  {\sqrt{\frac{3}{2}}},0,0,
  {\sqrt{2}}\} \\
                                \{ - \frac{1}{{\sqrt{2}}}
                      ,-{\sqrt{\frac{3}{2}}},
  0,0,{\sqrt{2}}\} \\
\end{array} \right.\
  \end{array}
\label{eigevalori}
\end{equation}
So reduced to normal form the $5$-vector of initial condition is
 a linear combination of  the vectors $\overrightarrow{g}_{\pm 2}=\{ \pm
\frac{1}{{\sqrt{2}}},
 \pm {\sqrt{\frac{3}{2}}},0,0,
  {\sqrt{2}}\}$ with the vector $\overrightarrow{g}_{0}=\{ -{\sqrt{3}},1,0,0,0\}$.
In particular writing:
\begin{eqnarray}
  \tilde{V}_{\mbox{normal form}} &=& a \, \overrightarrow{g}_{0}
+ b \left( \overrightarrow{g}_{+2} - \overrightarrow{g}_{-2}\right) \nonumber\\
 &=& \left( -\sqrt{3} \, a \, + \sqrt{2} \, b ,a \, + \, \sqrt{6} \,
 b, 0,0,0 \right)
\label{normalforma}
\end{eqnarray}
we obtain an initial tangent vector that has non vanishing
components only in the directions of the Cartan generators. \footnote{Indeed,
starting from the Cartan subalgebra,
we can generate the whole $\mathbb{K}$ space by applying the adjoint action of the
$\mathbb{H}$
subalgebra $Ad_h H_i = h^{\alpha}[H_i,t_{\alpha}] = \sqrt{2}\alpha_i
h^{\alpha}K_{\alpha}$.} For
reasons of later convenience we parametrize the initial normal
tangent vector as follows:
\begin{equation}
  \tilde{V}_{\mbox{normal form}} =\left (\frac
  {\omega  - \kappa }{4\sqrt{2}}, \frac{3\omega +\kappa }{4\sqrt{6}}, 0, 0, 0\right )
\label{generatvecto}
\end{equation}
and we conclude that we can find a generating solution of the
geodesic equations if we solve the first order system for the
tangent vectors (eq.s (\ref{A2system})) with the initial
conditions given by eq.(\ref{generatvecto}).
With such  conditions the differential system (\ref{A2system}) is
immediately solved  by:
\begin{eqnarray}
  \Phi^{(gen)}_1(t) &=& 0 , \quad \Phi^{(gen)}_2(t)=0 , \quad \Phi^{(gen)}_3(t) =0
  \nonumber\\
            \chi^{(gen)}_1(t) &=& \frac
  {\omega  - \kappa }{4\sqrt{2}} , \quad \chi^{(gen)}_2(t) = \frac{3\,\omega +\kappa
}{4\sqrt{6}}
\label{A2genersolut}
\end{eqnarray}
From this generating solution we can obtain new ones by performing
$\mathrm{SO(3)}$ rotations such that they keep the solvable
parametrization of the coset stable. In particular by rotating
along the three possible rotation axes we can switch on the root
fields $\Phi_\beta(t)$, one by one. This procedure is discussed
in  section \ref{diffecompe}.
\subsection{Scalar fields of the $A_2$ model}
In order to find the solutions for the scalar fields $\phi^I$, we
have to construct explicitly the $\mathrm{SL(3,\mathbb{R})/SO(3)}$
coset representative $\mathbb{L}$. First, we fix the parametrization
of the coset representative as follows
\begin{equation}
  \mathbb{L}=\exp\left[ \varphi^3(t) \, E_3\right]\,
\exp[\varphi^1(t) \, E_1 \, + \,\varphi^2(t) \, E_2] \,
  \exp\left[ h^1(t) \, H_1\,+\, h^2(t) \,
  H_2\right] \,
\label{cosettus3}
\end{equation}
Note that here we have ordered the exponentials by height grading,
first the highest root of level two, then the simple roots of
level one, finally the Cartan generators of level zero. As we will appreciate
in eq.s (\ref{formeident}), this is
crucial in order to interpret the scalar fields $\varphi_i$ as the
components of the corresponding $p$-forms, in oxidation. Choosing
the following normalizations for the generators of the fundamental
defining representation of the group $\mathrm{SL(3,\mathbb{R})}$:
\begin{eqnarray}
H_1  =  \left(\matrix{ \frac{1}{{\sqrt{2}}} & 0 & 0 \cr 0 & -
\frac{1}{{\sqrt{2}}}   & 0 \cr 0 & 0 & 0 \cr
             } \right),\quad
H_2  = \left(\matrix{ \frac{1}{{\sqrt{6}}} & 0 & 0 \cr 0 &
\frac{1}
            {{\sqrt{6}}} & 0 \cr 0 & 0 & -{\sqrt{\frac{2}
                                {3}}} \cr  } \right)
\label{cartanini}
\end{eqnarray}
and
\begin{eqnarray}
E^1  =  \left( \matrix{ 0 & 1 & 0 \cr 0 & 0 & 0 \cr 0 & 0 & 0 \cr
} \right), \quad E^2  = \left(\matrix{ 0 & 0 & 0 \cr 0 & 0 & 1 \cr 0 & 0
& 0 \cr  } \right), \quad E^3  =  \left(\matrix{ 0 & 0 & 1 \cr 0 & 0 & 0
\cr 0 & 0 & 0 \cr  }\right) \label{Esteppini}
\end{eqnarray}
we construct a coset representative $\mathbb{L} \in
\mathrm{SL(3,\mathbb{R})/SO(3)}$ explicitly as the following upper
triangular matrix:
\begin{eqnarray}
  \mathbb{L} & = & \left( \matrix{ e^{\frac{h_1(t)}{\sqrt{2}} +
\frac{h_2(t)}{\sqrt{6}}} &
            e^{- \frac{{h_1}(t)}{\sqrt{2}} +
                        \frac{h_2(t)}{\sqrt{6}}}\varphi_1(t) &
e^{-\sqrt{\frac{2}{3}}h_2(t)}
                        (\ft 12\varphi_1(t)\varphi_2(t) +
             \varphi_3(t)) \cr 0 & e^
            {-\frac{{h_1}(t)}{{\sqrt{2}}} + \frac{{h_2}(t)}{{\sqrt{6}}}} &
            e^{-\sqrt{\frac{2}{3}}h_2(t)}\varphi_2(t) \cr 0 & 0 & e^
            {- \sqrt{\frac{2}{3}}h_2(t)}} \right)
\label{explicoset}
\end{eqnarray}
Then we calculate the vielbein components through the formula:
\begin{equation}
  V^I = \mbox{Tr}\left[\mathbb{L}^{-1} \frac{d}{dt}\mathbb{L} \,
  \mathbf{K}_5^I\right]
\label{Vivielbe}
\end{equation}
where $\mathbf{K}_5^I$ are the generators of the coset defined  in
eq.(\ref{K5generi}). The vielbein $V^A$ can be found explicitly as a
function of time, recalling that in the solvable gauge it is
connected with the solutions of the eq.s (\ref{A2system}) by the
formula $\tilde{V}^i={V}^i \, , \,  \tilde{V}^\beta  = \Phi^\beta  = \sqrt{2}V^\beta
$. We obtain the
following equations:
\begin{equation}
\begin{array}{lclcl}
V^1 & = &\dot{h}_1(t) & = & \chi_1(t) \\
V^2 & = &\dot{h}_2(t) & = & \chi_2(t) \\
V^3 & = & e^{-{\sqrt{2}}\,{h_1}(t)} \, \frac{1}{\sqrt{2}}\,{\dot{\varphi}_1}(t)&=&
\frac{1}{\sqrt{2}}\Phi_1(t)\\
V^4 & = &e^{\frac{{h_1}(t) - {\sqrt{3}}\,{h_2}(t)}{{\sqrt{2}}}}\,
\frac{1}{{\sqrt{2}}}\,{\dot{\varphi }_2}\,(t)&=& \frac{1}{\sqrt{2}}\Phi_2(t)\\
V^5 & = & e^{-\frac{{h_1}(t) +
{\sqrt{3}}\,{h_2}(t)}{{\sqrt{2}}}}\,\frac{ 1}{2\,
             {\sqrt{2}}\,}\,\left( {{\varphi }_2}(t)\,{\dot{\varphi }_1}\,(t) -
             {{\varphi }_1}(t)\,{\dot{\varphi }_2}\,(t) + 2\,{\dot{\varphi
}_3}\,(t)\right)  &=&
                     \frac{1}{\sqrt{2}}\Phi_3(t)\
                     \end{array}
\label{seconequaz}
\end{equation}
where in the last column we are supposed to write whatever
functions of the time $t$ we have found as solutions of the
differential equations (\ref{A2system}) for the tangent vectors. For future use in
the oxidation procedure it is convenient to give a name to the
following combination of derivatives:
\begin{equation}
  W(t) =  {{\varphi }_2}(t)\,{{\varphi }_1}\,'(t) -
             {{\varphi }_1}(t)\,{{\varphi }_2}\,'(t) + 2\,{{\varphi }_3}\,'(t)
\label{Wdefi}
\end{equation}
and rewrite the last of equations (\ref{seconequaz}) as follows:
\begin{equation}
\Phi_3(t) = \ft 12 e^{-\frac{{h_1}(t) +
{\sqrt{3}}\,{h_2}(t)}{{\sqrt{2}}}}\,W(t)  \
\label{fi3equaw}
\end{equation}
In particular, the generating solution for the tangent vectors
(inserting $\chi^1 = \frac{\omega - \kappa}{4\sqrt{2}}$, $\chi^2 =
\frac{3\omega + \kappa}{4\sqrt{6}}$, $\Phi^1 = 0$, $\Phi^2 = 0$,
$\Phi^3 = 0$) gives, up to irrelevant integration constants, the following scalar
fields:
\begin{eqnarray}
&& \nonumber h_1(t) = \frac{(\omega - \kappa)t}{4\sqrt{2}},\hspace{0.5cm}
h_2(t) = \frac{(3\omega + \kappa)t}{4\sqrt{6}}, \\
&& \varphi_1 (t) =0, \hspace{0.5cm} \varphi_2 (t) =0,
\hspace{0.5cm} \varphi_3 (t) = 0
\end{eqnarray}
\subsection{Differential equations for the $\mathrm{H}$-compensators and the generation of
new solutions}
\label{diffecompe}
Non trivial solutions of the system (\ref{A2system}) can now be  obtained
from the generating solution (\ref{A2genersolut}) by means of a
suitable $\mathrm{H}$-subgroup compensating transformation, applying to the present case
the general procedure of the compensator method outlined at the end of section
\ref{geodesinomi}.
In previous paragraphs we have already collected all the ingredients which are
necessary to
construct the explicit form of eq.s (\ref{daequa}). Indeed from
eq.s (\ref{cartanini}), (\ref{Esteppini}), by recalling the definition
(\ref{generso3}), we immediately obtain the three generators $t_i$ of
the compact subgroup $\mathrm{SO(3)}$ in the $3$--dimensional
representation which is also the adjoint:
\begin{equation}
  t_1^{[3]}=\left(\begin{array}{ccc}
             0 & 1 & 0 \\
             -1 & 0 & 0 \\
             0 & 0 & 0 \
  \end{array} \right), \quad t_2^{[3]}=\left(\begin{array}{ccc}
             0 & 0 & 0 \\
             0 & 0 & 1 \\
             0 & -1 & 0 \
  \end{array} \right),\quad t_3^{[3]}=\left(\begin{array}{ccc}
             0 & 0 & 1 \\
             0 & 0 & 0 \\
             -1 & 0 & 0 \
  \end{array} \right)
\label{tgene3}
\end{equation}
On the other hand in eq. (\ref{j2generi}), we constructed the
generators $t_i$ in the $5$--dimensional $j=2$ representation,
spanned by the vielbein. Hence introducing
a compensating group element $h \in \mathrm{SO(3)}$, parametrized by three
time dependent angles in the following way:
\begin{equation}
  h=\exp \left[ \theta _3(t) \, t_3 \right] \, \exp \left[ \theta _2(t) \, t_2 \right]
  \, \exp \left[ \theta _1(t) \, t_1 \right]
\label{compensah}
\end{equation}
we immediately obtain the explicit form of
the adjoint matrix $A(\theta)$ and of the matrix $D(\theta)$, by
setting:
\begin{eqnarray}
  A(\theta)&=&\exp \left[ \theta _3(t) \, t_3^{[3]} \right] \, \exp \left[ \theta
_2(t) \, t_2^{[3]} \right]
  \, \exp \left[ \theta _1(t) \, t_1^{[3]} \right]\nonumber\\
            D(\theta)&=&\exp \left[ \theta _3(t) \, t_3^{[5]} \right] \, \exp \left[
\theta
_2(t) \, t_2^{[5]} \right]
  \, \exp \left[ \theta _1(t) \, t_1^{[5]} \right]
\label{compensahrep}
\end{eqnarray}
Inserting the normal form vector (\ref{generatvecto}) and the above
defined matrices $A(\theta)$ and $D(\theta)$ into the differential system
(\ref{daequa}) we obtain the following explicit differential
equations for the three time dependent $\theta$-parameters:
\begin{eqnarray}
\nonumber &&{\dot{\theta} _3}(t)=\ft 14\omega \,
                                 \sin 2\,\theta _3(t) \\
\nonumber && {\dot{\theta} _2}(t) = \ft 18 \left[ \kappa  + \omega
\,\cos 2\,\theta _3(t) \right] \,
            \sin 2\,\theta _2(t)\\ \nonumber
&& {\dot{\theta} _1}(t) = - \ft 1{16} [ \kappa  + \kappa \,\cos 2\,{{\theta
}_2}(t) + \omega [\cos2\theta_2(t) - 3]\cos2\theta_3(t)]
\,\sin 2\,{{\theta }_1}(t) + \\ && + \ft 12\,\omega \,{\sin^2
{{\theta }_1}(t)}\,\sin {{\theta }_2}(t)\,
            \sin 2\,{{\theta }_3}(t)
            \label{A2thetas}
\end{eqnarray}
At the same time the rotated tangent vector reads as follows in terms
of the chosen angles:
\begin{eqnarray}
V_{\mbox{rot}}  & \equiv & \overrightarrow{v}_{\mbox{n.f.}}\,D(\theta)
  \nonumber\\
V_{\mbox{rot}}^1 &=& \ft 1{16\sqrt{2}}\left\{- \cos 2\theta_1
                                \left[2\kappa + 2\kappa \cos 2\theta_2  +
                                          \omega\cos 2(\theta_2
-\theta_3)\right.\right.\nonumber\\&&
\left.\left. -
6\omega \cos 2\theta_3  + \omega \cos 2(\theta_2 + \theta_3)  \right]
 + 8\omega \sin 2\theta_1 \sin \theta_2
                     \sin 2\theta_3  \right \} \nonumber\\
V_{\mbox{rot}}^2 &=& \ft 1{16\sqrt{6}}\left \{-2\,\kappa  + 6\,\kappa \,\cos
2\,{{\theta }_2}  +
             3\,\omega \,\cos 2\,\left( {{\theta }_2}  - {{\theta }_3}  \right)
                     \right.
\nonumber\\
&& \left.
             +
             6\,\omega \,\cos 2\,{{\theta }_3}  +
             3\,\omega \,\cos 2\,\left( {{\theta }_2}  + {{\theta }_3}  \right)
\right\}
                        \nonumber\\
V_{\mbox{rot}}^3 &=& \ft 1{16\sqrt{2}}\left\{-\left[ 2\,\kappa
 + 2\,\kappa \,\cos 2\,{{\theta }_2}  +
                                          \omega \,\cos 2\,\left( {{\theta }_2}  -
{{\theta }_3}  \right)  -
                                          6\,\omega \,\cos 2\,{{\theta }_3}
\right.\right.\nonumber\\
&&\left.\left.+
                                          \omega \,\cos 2\,\left( {{\theta }_2}  +
{{\theta }_3}  \right)
                                          \right] \,\sin 2\,{{\theta }_1}  -
             8\,\omega \,\cos 2\,{{\theta }_1} \,\sin {{\theta }_2} \,
                     \sin 2\,{{\theta }_3} \right\}  \nonumber\\
V_{\mbox{rot}}^4 &=&\ft 1{4\sqrt{2}} \left \{\cos {{\theta }_1} \,\left( {\kappa } +
                                \omega \,\cos 2\,{{\theta }_3} \right) \,\sin
2\,{{\theta }_2}  +
             2\,\omega \,\cos {{\theta }_2} \,\sin {{\theta }_1} \,
                     \sin 2\,{{\theta }_3} \right \}\nonumber\\
V_{\mbox{rot}}^5 &=& \ft 1{4\sqrt{2}}\left \{- \left( \kappa  + \omega \,\cos
(2\,{{\theta }_3} ) \right) \,
                                \sin {{\theta }_1}\,\sin 2\,{{\theta }_2}
                                \right.\nonumber\\
                                &&\left. +
             2\,\omega \,\cos {{\theta }_1} \,\cos {{\theta }_2} \,
                     \sin 2\,{{\theta }_3} \right \}
\label{ruotatone}
\end{eqnarray}
In this way finding solutions of the original differential system for
tangent vectors is reduced to the problem of finding solutions of the
differential system for the compensating angles (\ref{A2thetas}).
The main property of this latter system is that it can be solved
iteratively. By inspection we see that the first of eq.s
(\ref{A2thetas}) is a single differential equation in separable
variables for the angle $\theta_3$. Inserting the resulting solution
into the second of eq.s (\ref{A2thetas}) produces a new differential
equation in separable variables for $\theta_2$ which can also be
solved by direct integration. Inserting these results into the last
equation produces instead a non--linear differential equation for
$\theta_1$ which is not with separable variables and reads as
follows:
\begin{equation}
  p_1(t)\sin 2\,{{\theta }_1}(t) + p_2(t){\sin^2 {{\theta }_1}(t)} +
  {{\theta }_1}'(t)=0
\label{generth1}
\end{equation}
In eq.(\ref{generth1}) $p_i(t)$ are two functions of time determined by the previous
solutions for $\theta_{2,3}(t)$.
Explicitly they read:
\begin{eqnarray}
p_1(t) & = & \ft 1{32}\left\{ 2\,\kappa  + 2\,\kappa \,\cos 2\,{{\theta }_2}(t) +
            \omega \,\cos 2\,\left[ {{\theta }_2}(t) - {{\theta }_3}(t) \right]
\right.\nonumber\\
            &&\left. -
            6\,\omega \,\cos 2\,{{\theta }_3}(t) +
            \omega \,\cos 2\,\left[ {{\theta }_2}(t) + {{\theta }_3}(t) \right]
\right\}
\nonumber\\
p_2(t) & = & -\ft 12 \omega \sin \theta_2(t)\sin 2\theta_3(t)
\label{expp12}
\end{eqnarray}
and we can evaluate them using the general solutions of the first two
equations in (\ref{A2thetas}), namely:
\begin{eqnarray}
\theta_3(t) & = & -\arcsin \left[ \frac{e^{\frac{t \, \omega}{2}}}
                     {{\sqrt{e^{t\,\omega } + e^{\omega \,{{\lambda }_3}}}}}
\right]\nonumber\\
\theta_2(t) & = &  -\arcsin\frac{e^
                                {\frac{t(\kappa  + \omega)  + \lambda_2}{4}}}{\sqrt{e^
                                 {t\omega} + e^{\omega\lambda_3}+e^
                                {\frac{t(\kappa  + \omega)  + \lambda_2}{2}}}}
\label{th23}
\end{eqnarray}
where $\lambda_{2,3}$ are two integration constants.
Equation (\ref{generth1}) is actually an integrable differential
equation. Indeed
multiplying (\ref{generth1}) by $1/ \sin ^{2}\theta_1$
and introducing the new depending variable $y(t) = \cot \theta_1$,
(\ref{generth1}) becomes actually the following linear differential equation
for $y(t)$
\begin{equation}
 2y(t) p_1(t) + p_2(t) - y(t)^\prime =0
 \label{trasformata}
 \end{equation}
which can easily be solved.
Hence the general integral of (\ref{generth1}) reads as follows:
\begin{equation}
  {{{{\theta }_1}(t)}\rightarrow
          {-\mbox{arccot}\left [e^{2\,\int {p_1}(t)\,dt}\,
                      \left( -\int \frac{{p_2}(t)}{e^{2\,\int {p_1}(t)\,dt}}\,dt +
                                {{\lambda }_1} \right) \right]}}
\label{generint1}
\end{equation}
where $\lambda_1$ is a third integration constant.
\par
In this way the system of eq.s (\ref{A2thetas}) has obtained a fully
general solution containing three integration constants. By inserting
this general solution into equation (\ref{ruotatone}) one also
obtains a complete general solution of the original differential
system for the tangent vectors containing five integration constants
$\omega,\kappa, \lambda_1 , \lambda_2 , \lambda_3$, as many as the
first order equations in the system.
\par
Let us consider for instance the choice $\lambda_2=\lambda_3=0$. In
this case the solution (\ref{th23}) for the rotation angles $\theta_{3,2}$
reduces to:
\begin{eqnarray}
\theta_3(t) & \hookrightarrow & -\arcsin \frac{e^{\frac{t\omega}{2}}}
            {{\sqrt{
                                 1 + e^{t\,\omega } }}} \nonumber\\
\theta_2(t) & \hookrightarrow & -\arcsin \frac{e^{\frac{t\,\left( \kappa  + \omega
\right) }{4}}}
            {{\sqrt{1 + e^{t\,\omega } +
                                  e^{\frac{t\,\left( \kappa  + \omega  \right) }{2}}}}}
\label{thet32spec}
\end{eqnarray}
and by replacing this result into the integrals we get:
\begin{eqnarray}
  2\,\int {p_1}(t)\,dt & = & \frac{t\,\left( \kappa  - \omega  \right)  + 4\,\log (1
+ e^{t\,\omega }) -
            2\,\log (1 + e^{t\,\omega } +
                      e^{\frac{t\,\left( \kappa  + \omega  \right) }{2}})}{4}
\label{integral1}
\end{eqnarray}
and
\begin{equation}
  -\int \frac{{p_2}(t)}{e^{2\,\int {p_1}(t)\,dt}}\,dt = \frac{1}{1 + e^{t\,\omega }}
\label{integral2}
\end{equation}
Substituting the above explicit integrations into eq.(\ref{generint1}) we obtain:
\begin{eqnarray}
  \theta_1 & \hookrightarrow & {{
            {\mbox{arccot}\left[\frac{e^{\frac{t\,\left( \kappa  - \omega  \right)
}{4}}\,
                                \left( 1 + \left( 1 + e^{t\,\omega } \right)
\,{{\lambda }_1} \right) }
                                {{\sqrt{1 + e^{t\,\omega } +
                                          e^{\frac{t\,\left( \kappa  + \omega
\right) }{2}}}}}\right]}}}
\label{the3spec}
\end{eqnarray}
that together with eq.s (\ref{thet32spec}) provides an explicit
solution of equations (\ref{A2thetas}). We can replace such a result
in eq.(\ref{ruotatone}) and obtain the tangent vectors after three
rotations. Yet as it is evident form eq.(\ref{ruotatone}) the first two
rotations are already sufficient to obtain a solution where all the
entries of the $5$--dimensional tangent vector are non vanishing and
hence all the root fields are excited. In the sequel we will consider
the two solutions obtained by means of the first rotation and by
means of the first plus the second. They will constitute our
paradigma of how the full system can be eventually solved. These
solutions however, as we discuss in later sections, are not only
interesting as toy models and examples. Indeed through oxidation they
can be promoted to very interesting backgrounds of ten dimensional
supergravity that make contact with the physics of $S$--branes.
\subsubsection{Solution of the differential equations for the tangent vectors with
two Cartan  and
one nilpotent field}
Let us consider the system (\ref{A2thetas}) and put
\begin{equation}
  \theta_1=\theta_2= \mbox{const}=0
\label{th1th2=0}
\end{equation}
This identically solves the last two equations and we are left with
the first whose general integral was already given in
eq.(\ref{th23}). By choosing the integration constant $\lambda_3=0$
we can also write:
\begin{equation}
\theta_3(t) = \arccos \frac{1}{{\sqrt{1 + e^{t\omega  }}}}
\label{sol3th}
\end{equation}
By inserting (\ref{sol3th}) and
(\ref{th1th2=0}) into (\ref{ruotatone}) we obtain the desired
solution for the tangent vectors:
\begin{eqnarray}
\chi_1(t) & = & -\frac{ \kappa  + \omega \,\tanh \frac{t\,\omega }{2}  }
  {4\,{\sqrt{2}}}, \hspace{0.5cm}
\chi_2(t)  =  \frac{\kappa  - 3\,\omega \,\tanh \frac{t\,\omega
}{2}}{4\,{\sqrt{6}}},\nonumber\\
                                 \Phi_1(t) & = & 0, \hspace{0.5cm}
\Phi_2(t)  =  0, \hspace{0.5cm}  \Phi_3(t) =
\frac{\omega}{\sqrt{(1 + e^{- t\omega})(1 + e^{t\omega})}}
\label{generetsolu}
\end{eqnarray}
where one root field is excited.
\par
Next we address the problem of solving the equations for the
scalar fields, namely eq.s(\ref{seconequaz}), which are
immediately integrated, obtaining:
\begin{eqnarray}
h_1(t) & = &-\frac{t\kappa + 2
             \log ({\cosh\frac{t\,\omega }{2}})}{4\,{\sqrt{2}}}, \hspace{0.5cm}
             h_2(t)  =  \frac{t\kappa  - 6\,\log (\cosh \frac{t\,\omega
}{2})}{4\,{\sqrt{6}}},\nonumber\\
\varphi_1(t) &=& 0, \hspace{0.5cm} \varphi_2(t) = 0,
\hspace{0.5cm}
  \varphi_3(t) = \frac{\sqrt{2}}{1+e^{t\,\omega}}
\label{finsol}
\end{eqnarray}
We can now insert eq.s (\ref{finsol}) into the form of the coset
representative (\ref{explicoset}) and we obtain the geodesic as a
map of the time line into the solvable group manifold and hence
into the coset manifold depending on your taste for
interpretation:
\begin{equation}
  \mathbb{R}_t \, \hookrightarrow \, \exp \left [Solv(\mathrm{A_2}) \right
  ]\simeq \frac{\mathrm{SL(3,\mathbb{R})}}{\mathrm{SO(3)}}
\label{mojna}
\end{equation}
In section (\ref{oxide1a2}) the oxidation of this sigma--model
solution to a full fledged supergravity background in $D=10$ is studied.
\subsubsection{Solution of the differential equations for the tangent vectors with
two Cartan and
three nilpotent fields}
Then we continue the hierarchical solution of  the (\ref{A2thetas}) differential system
by considering the next rotation $\theta_2$.
We set $\theta_1= \mbox{const} = 0$ and we replace in eq.s (\ref{A2thetas}) the
solution (\ref{sol3th}) for $\theta_3$, with $\lambda_3=0$. The first and the last
differential equations are identically satisfied. The second
equation was already solved in eq.(\ref{th23}).
By choice of the irrelevant integration variable $\lambda_2=0$ a
convenient
solution of the above equation is provided by the following time
dependent angle:
             \begin{equation}
             \theta_2(t) = -\arcsin \frac{e^{\frac{t\,\left( \kappa  + \omega
\right) }{4}}}
            {{\sqrt{1 + e^{t\,\omega } +
                                  e^{\frac{t\,\left( \kappa  + \omega  \right) }{2}}}}} \label{sol2th}
             \end{equation}
Inserting (\ref{sol3th}) and (\ref{sol2th}) into (\ref{ruotatone}) we obtain :
\begin{eqnarray}
\chi_1(t) & = &  \frac{- {\left( 1 + e^{t\,\omega } \right) }^2\,\kappa    -
             \left( -1 + e^{t\,\omega } \right) \,
                     \left( 1 + e^{t\,\omega } +
                                2\,e^{\frac{t\,\left( \kappa  + \omega  \right)
}{2}} \right) \,\omega }
             {4\,{\sqrt{2}}\,\left( 1 + e^{t\,\omega } \right) \,
             \left( 1 + e^{t\,\omega } + e^
                                {\frac{t\,\left( \kappa  + \omega  \right) }{2}}
\right) } \nonumber\\
\chi_2(t) & = & \frac{\left( 1 + e^{t\,\omega } -
                                2\,e^{\frac{t\,\left( \kappa  + \omega  \right)
}{2}} \right) \,\kappa  -
             3\,\left( -1 + e^{t\,\omega } \right) \,\omega }{4\,{\sqrt{6}}\,
             \left( 1 + e^{t\,\omega } + e^
                                {\frac{t\,\left( \kappa  + \omega  \right) }{2}}
\right) }\nonumber\\
                                 \Phi_1(t) & = & - \frac{e^{\frac{t\,\left( \kappa
+ 3\,\omega  \right)
}{4}}\,\omega }
            {\left( 1 + e^{t\,\omega } \right) \,
                     {\sqrt{1 + e^{t\,\omega } +
                                 e^{\frac{t\,\left( \kappa  + \omega  \right)
}{2}}}}} \nonumber\\
\Phi_2(t) & = & \frac{e^{\frac{t\,\left( \kappa  + \omega  \right) }{4}}\,
            \left( \kappa  + e^{t\,\omega }\,\left( \kappa  - \omega  \right)  +
                     \omega  \right) }{2\,{\sqrt{1 + e^{t\,\omega }}}\,
            \left( 1 + e^{t\,\omega } + e^
                      {\frac{t\,\left( \kappa  + \omega  \right) }{2}} \right) }
\nonumber\\
\Phi_3(t) & = & \frac{e^{\frac{t\,\omega }{2}}\,\omega }
  {{\sqrt{\left( 1 + e^{t\,\omega } \right) \,
                      \left( 1 + e^{t\,\omega } +
                                e^{\frac{t\,\left( \kappa  + \omega  \right) }{2}}
\right)
                      }}}
\label{2rotasolu}
\end{eqnarray}
Integrating  eq.s(\ref{seconequaz}) with this new choice of the
left hand side we obtain:
\begin{eqnarray}
h_1(t) & = & \frac{t\,\left( -\kappa  + \omega  \right)  - 4\,\log (1 + e^{t\,\omega
}) +
             2\,\log (1 + e^{t\,\omega } +
                                e^{\frac{t\,\left( \kappa  + \omega  \right)
}{2}})}{4\,{\sqrt{2}}} \nonumber\\
h_2(t) & = & \frac{t\,\left( \kappa  + 3\,\omega  \right)  -
             6\,\log (1 + e^{t\,\omega } +
                                e^{\frac{t\,\left( \kappa  + \omega  \right)
}{2}})}{4\,{\sqrt{6}}}\nonumber\\
\varphi_1(t) &=&\frac{1}{1 + e^{t\,\omega }}\nonumber\\
\varphi_2(t) &=& - \frac{\left( 1 + e^{t\,\omega } \right) }
             {1 + e^{t\,\omega } + e^{\frac{t\,\left( \kappa  + \omega  \right) }{2}}}
             \nonumber\\
\varphi_3(t) &=&   -\frac{1}{2\left(1+e^{t\omega} + e^{\ft 12
t(\kappa+\omega)}\right)}          \nonumber\\
  W(t) &=& \frac{2\,e^{t\,\omega }\,\omega }
  {\left( 1 + e^{t\,\omega } \right) \,
             \left( 1 + e^{t\,\omega } + e^
                                {\frac{t\,\left( \kappa  + \omega  \right) }{2}}
\right) }
\label{2finsol}
\end{eqnarray}
In section \ref{oxide2a2} we will see how this $\sigma$-model
solution can be oxided, among other choices, to an interesting
$S3/S1$-brane solution of type II B supergravity.
\section{The $\mathrm{E_8}$ Lie algebra: Reduction, Oxidation and subalgebra
embeddings} \label{genoxide} We come now to a close examination of
the $\mathrm{E_8}$ Lie algebra and we show how the hierarchical
dimensional \textit{reduction/oxidation}
\cite{Keurentjes:2002xc}--\cite{Julia:1980gr} of supergravity
backgrounds is algebraically encoded in the hierarchical embedding
of subalgebras into the $\mathrm{E_8}$ algebra. Similarly the
structure of the bosonic lagrangians of type II A/B supergravities
in $D=10$ \cite{Campbell:zc,II B} is encoded in the decomposition
of the solvable Lie algebra
$Solv(\mathrm{E_{8(8)}}/\mathrm{SO(16)})$ according to irreducible
representations of two $\mathrm{GL(7,\mathbb{R})}$ subgroups which
we shall denote by $\mathrm{GL(7,\mathbb{R})_{A/B} }\subset
\mathrm{E_{8(8)}}$, respectively associated with the moduli space
of flat metrics on a torus $T^7$ in compactified type II A or type
II B theory
\cite{Andrianopoli:1996bq,Andrianopoli:1996zg,dualiza1}. Although
these two $\mathrm{GL(7,\mathbb{R})}$ subalgebras are equivalent,
namely it is possible to map one into the other by an
$\mathrm{E_{8(8)}}$ automorphism (a Weyl transformation), they can
be extended to two \emph{inequivalent}
$\mathrm{GL(8,\mathbb{R})_{A/B}}$ subalgebras by adding the seven
roots corresponding to dualized Kaluza--Klein vectors in the Type
II A and II B settings, their negative counterparts and a further
Cartan generator.\par In order to carry out our programme we begin
by spelling out the $\mathrm{E_{8}}$ Lie algebra in our chosen
conventions.
\par
Using the Cartan--Weyl basis the Lie algebra can be written in the
standard form:
\begin{eqnarray}
\left[ \mathcal{H}_i \, ,\, \mathcal{H}_j \right] &=& 0 \nonumber\\
\left[ \mathcal{H}_i \,,\, E_\alpha \right] &=&\alpha_i \,
E_\alpha \quad \quad \quad \quad\quad
\forall \alpha \in \Delta_+ \nonumber\\
\left[ \mathcal{H}_i \,,\, E_{-\alpha} \right] &=&- \alpha_i \,
E_{-\alpha} \nonumber\\
\left[ E_\alpha \,,\, E_\beta \right] &=&
N_{\alpha\beta}\,E_{\alpha + \beta}\quad \quad \quad \mbox{if
$\alpha + \beta \in
\Delta_+$}\nonumber\\
\left[ E_\alpha \,,\, E_\beta \right] &=& 0\quad \quad \quad \quad
\quad \quad \mbox{if $\alpha + \beta
\notin \Delta_+$}\nonumber\\
\left[ E_\alpha \,,\, E_{-\beta} \right]& = & \delta_{\alpha\beta}
\, \, \alpha^i \, {H}_i \label{cartaweyl}
\end{eqnarray}
where $\mathcal{H}_i$ are the $8$ Cartan generators, $E_{\alpha}$
are the $120$ step operators associated with the positive roots
$\alpha \in \Delta_+$.\footnote{The values of the constants $N_{\alpha\beta}$, that enable to construct
explicitly the representation of $E_{8(8)}$, used in this paper,
 are given in the hidden appendix. To see it, download the source file, delete the
tag $end\{document\}$
 after the bibliography and $LaTeX$.}
\par
 Our choice of the  simple roots as vectors in an Euclidean
$\mathbb{R}^8$ space is the following one
\begin{eqnarray}
\alpha_1 & = & \{ 0,
                                1, -1, 0, 0, 0, 0, 0\}  \nonumber\\
\alpha_2 & = & \{ 0,
                                0, 1, -1, 0, 0, 0, 0\}  \nonumber\\
\alpha_3 & = & \{ 0,
                                0, 0, 1, -1, 0, 0, 0\}  \nonumber\\
 \alpha_4 & = & \{ 0,
                                0, 0, 0, 1, -1, 0, 0\}  \nonumber\\
 \alpha_5 & = & \{ 0,
                                0, 0, 0, 0, 1, -1, 0\}  \nonumber\\
 \alpha_6 & = & \{ 0,
                                0, 0, 0, 0, 1, 1, 0\}  \nonumber\\
 \alpha_7 & = & \{ - \ft 12, - \ft 12, - \ft 12 , - \ft 12 , - \ft 12  , - \ft 12,
-\ft 12, -\ft 12\}  \nonumber\\
 \alpha_8 & = & \{ 1, -1, 0, 0, 0, 0, 0 , 0 \}  \nonumber
\label{simplerutte}
\end{eqnarray}
The Dynkin diagrams corresponding to
$\mathrm{GL(7,\mathbb{R})}_{A/B} $ are defined by the following
simple roots:
\begin{eqnarray}
\mathrm{GL(7,\mathbb{R})}_{A}&\leftrightarrow
&\{\alpha_1,\,\alpha_2,\,\alpha_3,\,\alpha_4,\,\alpha_6,\,\alpha_8\}\nonumber\\
\mathrm{GL(7,\mathbb{R})}_{B}&\leftrightarrow
&\{\alpha_1,\,\alpha_2,\,\alpha_3,\,\alpha_4,\,\alpha_5,\,\alpha_8\}\label{sl7ab}
\end{eqnarray}
\iffigs
\begin{figure}
\begin{center}
\epsfxsize =12cm \epsffile{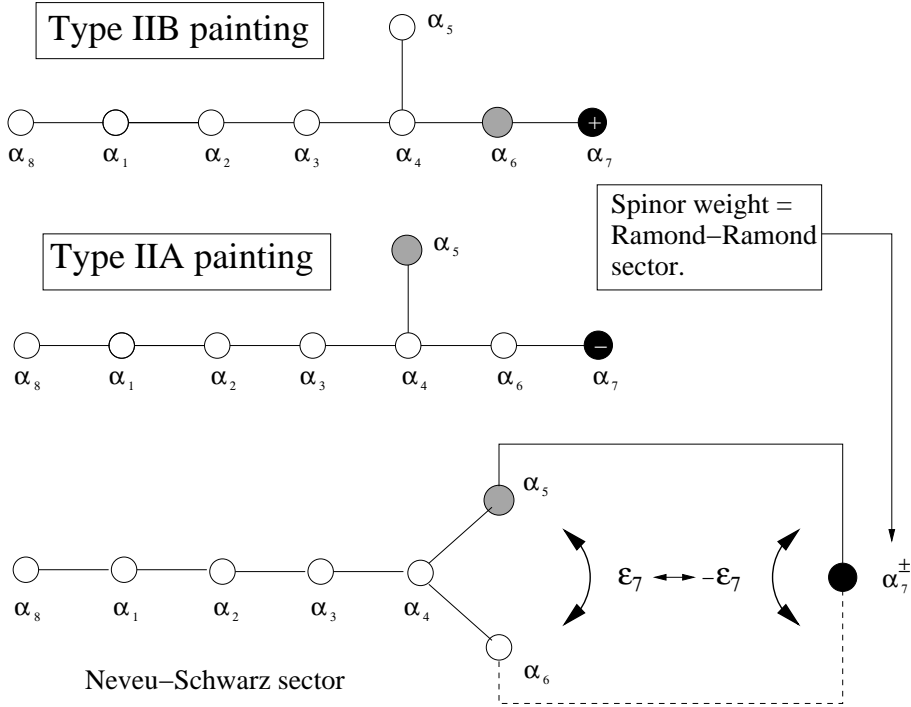} \caption{There are two
different ways of embedding the $\mathrm{SL(7,R)}$ Lie algebra in
$E_{8(8)}$ which correspond to the type II A and type II B
interpretation of the same sigma model. This can also be seen as a
different way of painting the same Dynkin diagram with blobs that
are either associated with the metric (white) or with the
$B$--field (gray) or with the Ramond--Ramond field (black).
Furthermore the T--duality transforming the A painting into the B
one is just the change in sign of the $\epsilon_7$ vector in
Euclidean space. Indeed this corresponds, physically to inverting
one of the torus radii $R_9 \rightarrow \alpha^\prime /R_9$.
 \label{pittodinko}} \hskip 1.5cm
\unitlength=1.1mm
\end{center}
\end{figure}
\fi
These two choices are illustrated in fig.\ref{pittodinko} where the
roots belonging to the $SL(7,\mathbb{R}) \subset GL(7,\mathbb{R})$
subgroup of the metric group are painted white. The roots eventually
corresponding to a B--field are instead painted black, while the root
eventually corresponding to a RR state are painted gray. As one sees
the difference between the A and B interpretation of the same Dynkin
diagram, named by us a \textit{painting} of the same, resides in the
fact that in the first case the RR root is linked to a metric, while
in the second it is linked to a B--field.
\par
In order to motivate the above identifications, let us start
recalling that the $T^7$ metric--moduli parametrize the coset $
{\mathcal M}^{A/B}_g\,=\,
\mathrm{GL}(7,\mathbb{R})_{A/B}/\mathrm{SO(7)}$ in the type II A
or B frameworks. If we describe ${\mathrm {\mathcal M}^{A/B}_g}$
as a solvable Lie group generated by the solvable Lie algebra
$Solv( {\mathcal M}^{A/B}_g)$ \cite{dualiza1,dualiza2} then its
coset representative $L^p{}_{\hat{q}}$ (in our notation the hatted
indices are rigid, i.e. are acted on by the compact isotropy
group) will be a solvable group element which, in virtue of the
Iwasawa decomposition can be expressed as the product of a matrix
${\mathcal N}^{-1T}$, which is the exponent of a nilpotent matrix,
times a diagonal one ${\mathcal H}^{-1}$: $L\,=\,{\mathcal N}^{-1
T}\,{\mathcal H}^{-1}$. Indeed the matrix ${\mathcal N}^{-1T}$ is
the exponential of the subalgebra $\mathcal{A}$ of $Solv(
{\mathcal M}^{A/B}_g)$ spanned by the shift operators
corresponding to the $\mathrm{GL}(7,\mathbb{R})_{A/B}$ positive
roots, while ${\mathcal H}^{-1}$ is the exponential of the
six--dimensional $\mathrm{GL}(7,\mathbb{R})_{A/B}$ Cartan
subalgebra. The vielbein $E_p{}^{\hat{q}}$ corresponding to the
$T^7$ metric $g_{pq}$ will have the following expression :
\begin{eqnarray}
E&=&L^{-1 T}\,=\,{\mathcal N}\,{\mathcal H}\,,\nonumber\\
g&=& E\,E^T\,=\,{\mathcal N}\,{\mathcal H}\,{\mathcal
H}^T\,{\mathcal N}^T\,. \label{NHHtNt}
\end{eqnarray}
The matrix ${\mathcal N}$ is non--trivial only if $T^7$ has
off--diagonal metric--moduli. In the case of a straight torus,
namely when $g_{pq}\,=\, e^{2\sigma_p}\,\delta_{pq}\,=\,
R_p^2\,\delta_{pq}$ the diagonal entries of ${\mathcal H}$ are
just the radii $R_p$: ${\mathcal
H}_p{}^{\hat{q}}\,=\,R_p\,\delta_p{}^{\hat{q}}$.\par
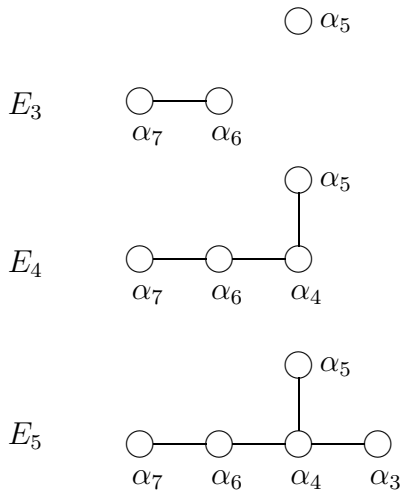
\begin{figure}
\centering
\begin{picture}(100,245)
\put (-70,160){$E_3$} \put (-20,165){\circle {10}} \put
(-23,150){$\alpha_7$} \put (-15,165){\line (1,0){20}} \put
(10,165){\circle {10}} \put (7,150){$\alpha_6$}
\put (40,195){\circle {10}} \put (48,193){$\alpha_5$}
\put (-70,100){$E_4$} \put (-20,105){\circle {10}} \put
(-23,90){$\alpha_7$} \put (-15,105){\line (1,0){20}} \put
(10,105){\circle {10}} \put (7,90){$\alpha_6$} \put (15,105){\line
(1,0){20}} \put (40,105){\circle {10}} \put (37,90){$\alpha_4$}
\put (40,135){\circle {10}} \put (48,132.8){$\alpha_5$} \put
(40,110){\line (0,1){20}}
\put (-70,35){$E_5$} \put (-20,35){\circle {10}} \put
(-23,20){$\alpha_7$} \put (-15,35){\line (1,0){20}} \put
(10,35){\circle {10}} \put (7,20){$\alpha_6$} \put (15,35){\line
(1,0){20}} \put (40,35){\circle {10}} \put (37,20){$\alpha_4$}
\put (40,65){\circle {10}} \put (48,62.8){$\alpha_5$} \put
(40,40){\line (0,1){20}} \put (45,35){\line (1,0){20}} \put
(70,35){\circle {10}} \put (67,20){$\alpha_{3}$}
\end{picture}
\vskip 1cm \caption{The Dynkin diagrams of $E_{3(3))}\subset
E_{4(4)}\subset E_{5(5)}$ and the labeling of simple roots }
\label{except1}
\end{figure}

\begin{figure}
\centering
\begin{picture}(100,245)
\put (-70,160){$E_6$} \put (-20,165){\circle {10}} \put
(-23,150){$\alpha_7$} \put (-15,165){\line (1,0){20}} \put
(10,165){\circle {10}} \put (7,150){$\alpha_6$} \put
(15,165){\line (1,0){20}} \put (40,165){\circle {10}} \put
(37,150){$\alpha_4$} \put (40,195){\circle {10}} \put
(48,193){$\alpha_5$} \put (40,170){\line (0,1){20}} \put
(45,165){\line (1,0){20}} \put (70,165){\circle {10}} \put
(67,150){$\alpha_{3}$} \put (75,165){\line (1,0){20}} \put
(100,165){\circle {10}} \put (97,150){$\alpha_{2}$}
\put (-70,100){$E_7$} \put (-20,105){\circle {10}} \put
(-23,90){$\alpha_7$} \put (-15,105){\line (1,0){20}} \put
(10,105){\circle {10}} \put (7,90){$\alpha_6$} \put (15,105){\line
(1,0){20}} \put (40,105){\circle {10}} \put (37,90){$\alpha_4$}
\put (40,135){\circle {10}} \put (48,132.8){$\alpha_5$} \put
(40,110){\line (0,1){20}} \put (45,105){\line (1,0){20}} \put
(70,105){\circle {10}} \put (67,90){$\alpha_{3}$} \put
(75,105){\line (1,0){20}} \put (100,105){\circle {10}} \put
(97,90){$\alpha_{2}$} \put (105,105){\line (1,0){20}} \put
(130,105){\circle {10}} \put (127,90){$\alpha_1$}
\put (-70,35){$E_8$} \put (-20,35){\circle {10}} \put
(-23,20){$\alpha_7$} \put (-15,35){\line (1,0){20}} \put
(10,35){\circle {10}} \put (7,20){$\alpha_6$} \put (15,35){\line
(1,0){20}} \put (40,35){\circle {10}} \put (37,20){$\alpha_4$}
\put (40,65){\circle {10}} \put (48,62.8){$\alpha_5$} \put
(40,40){\line (0,1){20}} \put (45,35){\line (1,0){20}} \put
(70,35){\circle {10}} \put (67,20){$\alpha_{3}$} \put
(75,35){\line (1,0){20}} \put (100,35){\circle {10}} \put
(97,20){$\alpha_{2}$} \put (105,35){\line (1,0){20}} \put
(130,35){\circle {10}} \put (127,20){$\alpha_1$} \put
(135,35){\line (1,0){20}} \put (160,35){\circle {10}} \put
(157,20){$\alpha_8$}
\end{picture}
\vskip 1cm \caption{The Dynkin diagrams of $E_{6(6))}\subset
E_{7(7)}\subset E_{8(8)}$ and the labeling of simple roots}
\label{except2}
\end{figure}
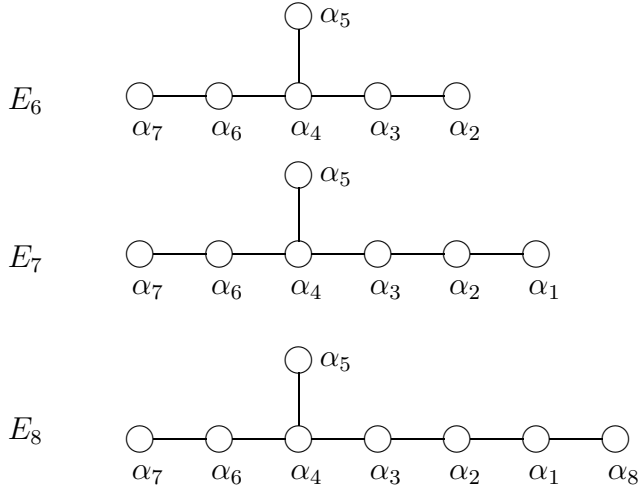

 The
decomposition of  $Solv(E_{8(8)}/SO(16))=Solv_8$ with respect to $Solv(GL(7,\mathbb{R})/SO(7))=Solv_7^{A/B}$ has the
following form:
\begin{eqnarray}
Solv_8 &=& Solv_7^{A/B}+ o(1,1)^{A/B}+ {\mathcal
A}^{[1]}+{\mathbf B}^{[1]}+{\mathbf B}^{[2]}+ \sum_{k} {\mathbf
C}_{A/B}^{[k]}
\end{eqnarray}
where $o(1,1)^{A/B}$ denote the Cartan generators
$H_{\alpha[7]+\epsilon_7},\,H_{\alpha[7]}$, which are parametrized
by  the ten dimensional dilaton $\phi $ in the Type II A and II B
settings respectively, ${\mathbf B}^{[2]}\,=\, B_{2+p \, ,
\,2+q}\, {\mathbf B}^{p\, \, q}$ is the subspace parametrized by
the internal components of the Kalb--Ramond field and ${\mathbf
C}_{A/B}^{[k]}\,=\, C_{2+p_1\,\dots\, 2+p_k}\,{\mathbf
C}^{p_1\,\dots\, p_k}$ the subspace spanned by the internal
components of the R--R k--form (in our conventions $C_{2+p_1,\dots
2+p_k}$ for $k>4$ are the {\em dualized} vectors $C_{2+q_1\,\dots
\, 2+q_{7-k},\mu}$, with $\epsilon^{p_1\dots p_k q_1\dots
q_{7-k}}\,\neq\,0$). Finally the nilpotent spaces
${\mathcal{A}}^{[1]}$ and ${\mathbf  B}^{[1]}$ are parametrized by
the dualized Kaluza--Klein and Kalb--Ramon vectors:
$g_{p\,\mu},\,B_{2+p,\,\mu}$. The nilpotent subspaces
$\{{\mathcal{A}}^{[1]},\,{\mathbf  B}^{[1]},\,{\mathbf
C}_{A/B}^{[k]}\}$ define order--$k$ antisymmetric tensorial
representations ${\mathbf
 T}^{[k]}\,=\,\{{\mathbf T}^{{p_1\dots p_k}}\} $ with respect to
 the adjoint action of $\mathrm{GL}(7,\mathbb{R})_{A/B}$:
 \begin{eqnarray}
{\mathbf E}&\in & \mathrm{GL}(7,\mathbb{R})_{A/B}\,:\,\,\,{\mathbf
E}\cdot {\mathbf T}^{{p_1\dots p_k}}\cdot {\mathbf
E}^{-1}\,=\,E^{p_1}{}_{q_1}\dots E^{p_k}{}_{q_k}{\mathbf
T}^{{q_1\dots q_k}}
 \end{eqnarray}
From the definitions (\ref{sl7ab}) we see that the shift
generators in $Solv_8 - Solv_7^{A}$ branch with respect to
$\mathrm{GL}(7,\mathbb{R})_{A}$ into the subspaces
${\mathcal{A}}^{[1]},\,{\mathbf  B}^{[1]},\,{\mathbf B}^{[2]}$ and
${\mathbf C}_{A}^{[k]}$, $k\,=\,1,\,3,\,5,\,7$ and those in
$Solv_8 - Solv_7^{B}$ branch with respect to
$\mathrm{GL}(7,\mathbb{R})_{B}$ into
${\mathcal{A}}^{[1]},\,{\mathbf  B}^{[1]},\,{\mathbf B}^{[2]}$ and
${\mathbf C}_{B}^{[k]}$, $k\,=\,0,\,2,\,4,\,6$. As far as the R--R
scalars are concerned, these representations correspond indeed to
the tensorial structure of the type II A spectrum
$C_{2+p},\,C_{2+p \, ,\, 2+q \, ,\, 2+r},\,C_{\mu},\,C_{2+p\, , \,
2+q\, , \, \mu}$ and type II B spectrum $C,\,C_{2+p,
2+q},\,C_{2+p, 2+q, 2+r, 2+s},\,C_{2+p,\mu}$. We can now define a
one-to-one correspondence between axions and ${\mathrm E}_{8(8)}$
positive roots. The $T^7$ moduli space is ${\rm SO}(7,7)_T/[{\rm
SO}(7)\times {\rm SO}(7)]\,=\, {\rm Exp}(Solv_T)$ parametrized by
the scalars $g_{2+p \, ,\,  2+q}$ and $B_{2+p\, ,\, 2+q}$, where:
\begin{eqnarray}
Solv_T &=&Solv_7^{A/B}+ {\mathbf  B}^{[2]}
\end{eqnarray}
In three dimensions the scalar fields deriving from the
dualization of $g_{p\mu}$ and $B_{p \,\mu}$ together with the dilaton
$\phi$ enlarge the manifold ${\rm SO}(7,7)_T/[{\rm SO}(7)\times
{\rm SO}(7)]$ to ${\rm SO}(8,8)/[{\rm SO}(8)\times {\rm
SO}(8)]\,=\, {\rm Exp}(Solv_{NS})$ where now:
\begin{eqnarray}
Solv_{NS} &=&Solv_7^{A/B}+o(1,1)+{\mathcal A  }^{[1]}+{\mathbf
B}^{[1]} +{\mathbf  B}^{[2]}\,,
\end{eqnarray}
This manifold is parametrized by the 64 NS scalar fields. If we
decompose $Solv_8$ with respect to $Solv_{NS}$ we may achieve an
intrinsic group--theoretical characterization of the NS and R--R
scalars. From this point of view the R--R scalar fields span the
64--dimensional subalgebra $Solv_8/Solv_{NS}$ which coincides with
a spinorial representation of ${\rm SO}(7,7)_T$ with a definite
chirality. Therefore the corresponding positive roots have grading
one with respect to the ${\rm SO}(7,7)_T$ spinorial root
$\alpha[7]$. Finally the higher--dimensional origin of the three
dimensional scalar fields can be determined by decomposing
$Solv_8$ with respect to the solvable algebra $Solv_{11-D}$
generating the scalar manifold ${\rm E}_{11-D(11-D)}/{\rm H}$ of
the D--dimensional maximal supergravity. This decomposition is
defined by the embedding of the higher--dimensional duality groups
${\rm E}_{11-D(11-D)}$ inside the three dimensional one. The
Dynkin diagrams of the $E_{11-D(11-D)}$ nested Lie algebras are
arranged according to the the pictures displayed in Fig.
\ref{except1} and Fig. \ref{except2}.\par
Let us now comment on the geometrical relation between the Type II
A and II B representations. The two ${\rm SL}(7,\mathbb{R})_{A/B}$
Dynkin diagrams are mapped into each other by the ${\rm SO}(8,8)$
outer authomorphism $\epsilon_7\rightarrow -\epsilon_7$ which
corresponds,  in the light of our parametrization of the ${\rm
E}_{8(8)}$ Cartan generators, to a T--duality along the direction
$x^9$. This transformation defines the correspondence between the
two inequivalent ${\rm SL}(8,\mathbb{R})_{A/B}$ inside which ${\rm
SL}(7,\mathbb{R})_{A/B}$ are embedded. To show that this operation
is indeed a T--duality (see \cite{Lu:1996ge} and also
\cite{Bertolini:1999uz} for a geometrical definition of T--duality
in the solvable Lie algebra formalism) let us recall the
parametrization of the Cartan subalgebra in our setup:
\begin{eqnarray}
\label{typeII AB}
\mbox{Type II B:}&&\nonumber\\
\vec{h}\cdot
\vec{H}&=&\sum_{p=1}^7\,\sigma_{p}\,(\epsilon_p-\epsilon_8)-\frac{\phi}{2}\,\alpha_7\,=\,
\sum_{p=1}^7\,\tilde{\sigma}_{p}\,(\epsilon_p-\epsilon_8)+2\phi\,\epsilon_8\nonumber\\
\mbox{Type II A:}&&\nonumber\\
\vec{h}\cdot
\vec{H}&=&\sum_{p=1}^7\,\sigma_{p}\,
(\epsilon^\prime_p-\epsilon^\prime_8)-\frac{\phi}{2}\,
(\alpha_7+\epsilon_7)\,=\,
\sum_{p=1}^7\,\tilde{\sigma}_{p}\,
(\epsilon^\prime_p-\epsilon^\prime_8)+2\phi\,\epsilon^\prime_8
\\ \epsilon^\prime_w
&=& \epsilon_w
\,,\,\,w\,\neq\,7\,,\,\,\,\epsilon^\prime_7\,=\,-\epsilon_7 \nonumber
\end{eqnarray}
where, in the case of a compactification on a straight torus,
$\sigma_p\,=\,\log{(R_p)}$ and
$\tilde{\sigma}_p\,=\,\log{(\tilde{R}_p)}$, $R_p$ and
$\tilde{R}_p$ being the $T^7$ radii in the ten--dimensional
Einstein-- or string--frame respectively.  Let us consider a
T--duality along directions $x^{i_1},\dots,\,x^{i_k} $:
$\tilde{R}_{i_r}\,\rightarrow\, 1/\tilde{R}_{i_r}$
($r=1,\dots,k,\,\,\alpha^\prime\,=\,1$). The transformation
$\epsilon_{i_r}\,\rightarrow\, -\epsilon_{i_r}$ in the expression
of $\vec{h}\cdot \vec{H}$ can be absorbed by the transformation
$\tilde{\sigma}_{i_r}\,\rightarrow\, -\tilde{\sigma}_{i_r}$ and
$\phi\,\rightarrow\, \phi-\sum_{r=1}^k\, \tilde{\sigma}_{i_r}$
which is indeed the effect of the T--duality. \par
 As a result of this analysis the precise one--to--one
correspondence between axions and positive roots can now be given
in the following form:
\begin{eqnarray}
\mbox{Type II B:}&&\nonumber\\
&&C_{2+p_1 \, \dots \,2+p_k}\,\leftrightarrow\, \alpha_7+\epsilon_{p_1}+\dots
\epsilon_{p_k} \,,\,\,\,(k\,=\,2,\,4)\,,\nonumber\\
&&C_{2+p\, , \,\mu}\,\leftrightarrow\, \alpha_7+\epsilon_{q_1}+\dots
\epsilon_{q_{6}} \,,\,\,\,(\epsilon^{p q_1\dots q_{6}}\neq 0)\,,\nonumber\\
&&B_{2+p\, , \, 2+q}\,\leftrightarrow\, \epsilon_{p}+ \epsilon_{q}\,,\nonumber\\
&&B_{2+p\, , \,\mu}\,\leftrightarrow\, -\epsilon_{p}- \epsilon_{8}\,,\nonumber\\
&&\gamma_{2+p}{}^{2+q}\,\leftrightarrow\,\epsilon_{p}- \epsilon_{q}\nonumber\\
&&\gamma_\mu{}^{2+q}\,\leftrightarrow\,\epsilon_{q}- \epsilon_{8}\nonumber\\
\mbox{Type II A:}&&\nonumber\\
&&C_{2+p_1\dots 2+p_k}\,\leftrightarrow\,
\alpha_7+\epsilon_7+\epsilon^\prime_{p_1}+\dots \epsilon^\prime_{p_k}
\,,\,\,\,(k\,=\,1,\,3)\,,\nonumber\\
&&C_{2+p_1 \, , \, 2+p_2\, , \,\mu}\,\leftrightarrow\,
\alpha_7+\epsilon_7+\epsilon^\prime_{q_1}+\dots \epsilon^\prime_{q_{5}}
\,,\,\,\,(\epsilon^{p_1 p_2 q_1\dots q_{5}}\neq 0)\,,\nonumber\\
&&C_{\mu}\,\leftrightarrow\, \alpha_7+\epsilon_7+\epsilon^\prime_{1}+\dots
\epsilon^\prime_{7}\,,\nonumber\\
&&B_{2+p\, , \, 2+q}\,\leftrightarrow\, \epsilon^\prime_{p}+
\epsilon^\prime_{q}\,,\nonumber\\
&&B_{2+p\, , \,\mu}\,\leftrightarrow\, -\epsilon^\prime_{p}-
\epsilon^\prime_{8}\,,\nonumber\\
&&\gamma_{2+p}{}^{2+q}\,\leftrightarrow\,\epsilon^\prime_{p}-
\epsilon^\prime_{q}\nonumber\\&&\gamma_\mu{}^{2+q}\,\leftrightarrow\,\epsilon^\prime_{q}-
\epsilon^\prime_{8}\nonumber\\
\end{eqnarray}
where $\gamma_{2+p}{}^{2+q}$ are the parameters entering the matrix
${\mathcal N}$ and which determine the off--diagonal entries of
the $T^7$ vielbein $E_p{}^{\hat{q}}$:
\begin{eqnarray}
\mathcal{N}&\equiv& \exp{(\gamma_{2+p}{}^{2+q}\,\mathcal{A}_q{}^p)}
\end{eqnarray}
for a precise definition of the above  exponential representation
see \cite{dualiza1}. The fields $\gamma_\mu{}^{q}$ denote the
scalars dual to the Kaluza--Klein vectors.

\section{Oxidation of the $\mathrm{A_2}$ solutions}
\label{occidoa2} In this section, as a working illustration of the
oxidation process we derive two full fledged $D=10$ supergravity
backgrounds corresponding to  the two $A_2$ sigma model solutions
derived in previous sections. As we already emphasized in our
introduction the correspondence is not one-to-one, rather it is
one-to-many. This has two reasons. First of all we can either
oxide to a type II A or to a type II B configuration. Secondly,
even within the same supergravity choice (A or B), there are
several different oxidations of the same abstract sigma model
solution, just as many as the different ways of embedding the
solvable $Solv(\mathrm{A_2})$ algebra into the solvable
$Solv(\mathrm{E_8}/SO(16))$ algebra. This embeddings lead to quite
different physical interpretations of the same abstract sigma
model solution.
\par
Our first task is the classification of these inequivalent embeddings.
\subsection{Possible embeddings of the $A_2$ algebra}
In order to study the possible embeddings it is convenient to rely on
a compact notation and on the following graded structure of the Solvable Lie
algebra $Solv(\mathrm{E_8}/SO(16))$ characterized by the following non vanishing commutators:
\begin{eqnarray}
\left[ \mathcal{A} \, ,\, \mathcal{A}\right]  & = & \mathcal{A} \nonumber\\
\left[\mathcal{A} \, ,\, \mathcal{A}^{[1]}\right]
&=&\mathcal{A}^{[1]}\nonumber\\
\left[ \mathcal{A} \, ,\, \mathbf{B}^{[1]}\right] & = & \mathbf{B}^{[1] } \nonumber\\
\left[ \mathcal{A} \, ,\, \mathbf{B}^{[2]}\right] & = & \mathbf{B}^{[2] } \nonumber\\
\left[ \mathcal{A} \, ,\, \mathbf{C}^{[k]}\right] & = & \mathbf{C}^{[k]} \nonumber\\
\left[ \mathbf{B}^{[2]} \, , \,\mathbf{B} ^{[1]}\right]
&=&\mathcal{A}^{[1]}\nonumber\\
\left[ \mathbf{B}^{[2]} \, ,\, \mathbf{C}^{[k]}\right] & = & \mathbf{C}^{[k+2]}
\nonumber\\
\left[ \mathbf{C}^{[k]} \, ,\, \mathbf{C}^{[6-k]}\right] & = & {\mathbf{B}}^{[1]}
\nonumber\\
\left[ \mathbf{C}^{[k]} \, ,\, \mathbf{C}^{[8-k]}\right] & = &
{\mathcal{A}}^{[1]} \label{gradastructa}
\end{eqnarray}
In eq. (\ref{gradastructa})  $\mathcal{A}^{[1]}$,
$\mathbf{B}^{[2]}$, $\mathbf{B}^{[1]}$ and $\mathbf{C}^{[k]}$ are
the spaces of nilpotent generators defined in the previous
section, while $\mathcal{A}$ is the $Solv_7^{A/B}$ Lie algebra. In
view of the above graded structure there are essentially $10$
physically different ways of embedding the $A_2$ algebra into
$\mathrm{E_{8(8)}}$.
\begin{description}
  \item[1] Every root $\beta_{1,2,3}$ is a metric generator ${\mathcal{A}}$.
  In this case the $A_2$ Lie algebra is embedded into the
$\mathrm{SL(7,\mathbb{R})}$ subalgebra of
  $\mathrm{E_{8(8)}}$ and the corresponding oxidation leads to a
  purely gravitational background of supergravity which is identical
  in the type II A or type II B theory.
  \item[2] The root $\beta_1$ corresponds to an off--diagonal element of
the internal metric, namely belongs to $\mathcal{A}$, while
$\beta_{2,3}$ correspond to scalars dual to the Kaluza--Klein
vectors $g_{2+p\, ,\,\mu}$ namely belong to $\mathcal{A}^{[1]}$.
This commutation relation follows from the fact that the
$\mathcal{A}^{[1]}$ generators transform in the ${\bf 7}$ of
$\mathrm{SL(7,\mathbb{R})}$.
\item [3] The two simple roots $\beta_{1,2}$ are respectively associated with  a
metric generator
  ${\mathcal{A}}$ and a $B$-field generator $\mathbf{B}^{[1]}$ corresponding to the dualized vector field
  deriving from the Kalb--Ramond two form. The
  composite root $\beta_3$ is associated with a second $B$-field generator $\mathbf{B}^{[1]}$.  This is so because the $\mathbf{B}^{[1]}$ transform in the ${\overline{7}}$
  of $\mathrm{SL(7,\mathbb{R})}$. In this case
  oxidation leads to a purely NS configuration, shared by type II A
  and type II B theories, involving the metric, the dilaton and the
  B-field alone.
  \item [4] The two simple roots $\beta_{1,2}$ are respectively associated with  a
metric generator
  ${\mathcal{A}}$ and a $B$-field generator $\mathbf{B}^{[2]}$ parametrized by a $B_{pq}$ scalar field.  The
  composite root $\beta_3$ is associated with a second $B$-field generator $\mathbf{B}^{[2]}$.
  This follows from the fact that $\mathbf{B}^{[2]}$ defines a representation ${\bf 21}$ of $\mathrm{SL(7,\mathbb{R})}$.
   In this case
  oxidation leads to a purely NS configuration, shared by type II A
  and type II B theories.
  \item[5] The two simple roots $\beta_{1,2}$ are respectively associated with  a
metric generator
  ${\mathcal{A}}$ and with a RR $k$--form generator $\mathbf{C}^{[k]}$.
  The composite root $\beta_3$ is associated with a second RR generator
  $\mathbf{C}^{[k]}$ in the same $[k]$--representation. This follows again from
  the fact that the $\mathbf{C}^{[k]}$ generators span an order
  --$k$ tensor representation of
  $\mathrm{SL(7,\mathbb{R})}$. In this case oxidation
  leads to different results in type II A and type II B theories,
  although the metric is the same for the two cases and it has non
  trivial off-diagonal parts.
\item[6] The root $\beta_1 $ is contained in $ \mathbf{B}^{[2]}$, namely it
describes an internal component of the B--field. The root $\beta_2
\, \in \, \mathbf{B}^{[1]}$ namely it corresponds to a B--field
with mixed indices. The root $\beta_3 \, \in \, \mathcal{A}^{[1]}$
is associated with a Kaluza--Klein vector.
  \item[8] The simple roots $\beta_{1,2}$ are respectively associated
  with a RR $k$-form generator $\mathbf{C}^{[k]}$ and with a B-field generator
$\mathbf{B}$. The composite root $\beta_3$ is associated with a
$k+2$ form generator $\mathbf{C}^{{k+2}}$. In this case the metric
is purely diagonal and we have non trivial B-fields  and RR forms.
Type II A and type II B oxidations are differ just in this latter
sector. The NS sector is the same for both.
\item[9] The two simple roots $\beta_{1,2}$ are respectively
associated with a RR generator $\mathbf{C}^{[k]}$ and a RR
generator $\mathbf{C}^{[6-k]}$. The composite root $\beta_3$ is
associated with a $\mathbf{B}^{[1]}$ generator. The oxidation
properties of this case are just similar to those of the previous
case. Also here the metric is diagonal.
\item[10] In type II B theory the roots $\beta_{1,2}$ belong to
$C^{[k]}$ and $C^{[8-k]}$ namely are associated with the internal
components of two different R--R forms, while $\beta_3 \, \in
\,\mathcal{A}^{[1]}$ describes a Kaluza--Klein vector.
\end{description}
\subsection{Choice of one embedding example}
As an illustration, out of the above list
we choose one example of embedding that has an immediate and nice
physical interpretation in terms of a brane system. We consider  the
case $4$, with a RR generator $\mathbf{C}^{[2]}$
and a B-field generator respectively associated with $\beta_{1,2}$ and
a $\mathbf{C}^{[4]}$ generator associated with the composite root
$\beta_{3}$.
In particular we set:
\begin{eqnarray}
  \beta_1 & \rightarrow & \mathbf{B}^{34}\nonumber\\
\beta_2 & \rightarrow & \mathbf{C}^{89}\nonumber\\
\beta_3 & \rightarrow & \mathbf{C}^{3489} \, \sim \, \mathbf{C}^{\mu  567}
\label{associati}
\end{eqnarray}
More precisely this corresponds to identifying $\beta _{1,2,3}$ with
the following roots of $E_{8(8)}$ according to their classification
given in the appendix:
\begin{equation}
  \begin{array}{cccclcc}
             \beta_1& \hookrightarrow &  \alpha [69] & = &\epsilon _1 + \epsilon _2 &
\leftrightarrow & B_{34}\\
             \beta_2 & \hookrightarrow & \alpha [15] & = & \alpha[7] +\epsilon _6 +
\epsilon
_7  & \leftrightarrow & C_{89}  \\
             \beta_3& \hookrightarrow & \alpha [80] & = & \alpha [7] +\epsilon _1 +
\epsilon _1 +
             \epsilon _6 + \epsilon _7&  \leftrightarrow & C_{3489} \sim C_{\mu 567} \
  \end{array}
\label{identificazie}
\end{equation}
where $\alpha[7]=\left\{ - \ft 12 ,  - \ft 12 , - \ft 12, - \ft 12, - \ft
12, - \ft 12, - \ft 12, - \ft 12 \right \}$ is the spinorial simple root
of $\mathrm{E_{8(8)}}$.
\par
Next given the explicit form of the two roots $ \beta_1  \hookrightarrow  \alpha
[69]$ and $\beta_2  \hookrightarrow  \alpha [15]$ we construct the $2$--dimensional
subspace of the Cartan
subalgebra which is orthogonal to the orthogonal complement of
$\alpha[69]$ and $\alpha[15]$ in $\mathbb{R}^8$. We immediately see
that this subspace is spanned by all $8$ vectors of the form:
\begin{equation}
  \overrightarrow{h} = \left\{ x,x,y,y,y,-y,-y,y \right\}
\label{hsubspace}
\end{equation}
so that we find:
\begin{equation}
  \overrightarrow{h} \cdot \overrightarrow{\alpha}[69] = 2x \quad ;
  \quad \overrightarrow{h} \cdot \overrightarrow{\alpha}[15] = - (x +
  3y)
\label{halp}
\end{equation}
Then we relate the fields $x$ and $y$ to the diagonal part of the ten
dimensional metric.
\par
To this effect we start from the
general relations between the ten-dimensional metric in the \textit{Einstein
frame} and the fields in three-dimensions evaluated in the $D=3$
Einstein frame, then we specialize such relations to our particular
case.
\paragraph{General relations in dimensional reduction}
The Einstein frame metric in $D=10$ can be written as:
\begin{equation}
  G_{MN}^{(Einstein)} = \left(\begin{array}{c|c}
             \exp [4 \phi_3 -\ft 12 \phi] \, g^{(E,3)}_{\mu\nu} + G_{ij} \,
             \mathcal{A}^i_\mu \, \mathcal{A}^i_\nu & G_{ik} \,
             \mathcal{A}^k_\mu  \\
             \hline
             G_{jk} \,
             \mathcal{A}^k_\nu  &G_{ij} \
  \end{array} \right)
\label{metrica10}
\end{equation}
where $g^{(E,3)}_{\mu\nu}$ is the three dimensional Einstein frame
metric (\ref{confpiatto2}) determined by the solution of the
$D=3$-sigma model via equations (\ref{einsteinocon}) and
(\ref{kdefi},\ref{kvalue}). On the other hand
$G_{ij}$ is  the Einstein frame metric in the internal seven
directions.
It parametrizes the coset:
\begin{equation}
 \frac{\mathrm{GL(7,\mathbb{R})}}{\mathrm{SO(7)}} = O(1,1) \, \times
 \, \frac{\mathrm{SL(7,\mathbb{R})}}{\mathrm{SO(7)}}
\label{7torusmoduli}
\end{equation}
In full generality, recalling eq.s(\ref{NHHtNt}) we can  set
(\cite{dualiza1,dualiza2}):
\begin{equation}
 G = E \,  E^T \quad ; \quad E = \mathcal{N} \, \mathcal{H}
\label{diag+ndiag}
\end{equation}
where, in this case:
\begin{equation}
  \mathcal{N}_{ij} = \delta_{ij}
\label{nilpotentN}
\end{equation}
since there are no roots associated with metric generators, while  the diagonal matrix:
\begin{equation}
  H_{ij} = \exp[\sigma_i] \, \delta_{ij}
\label{sigmadefi}
\end{equation}
parametrizes the degrees of freedom associated with the Cartan subalgebra
of $\mathrm{O(1,1)} \times \mathrm{SL(7,\mathbb{R})}$. The  relation
of the fields $\sigma_i$ with
the dilaton field and the Cartan fields of $E_{8(8)}$ is
obtained through the following general formulae:
\begin{eqnarray}
  \overrightarrow{h} &=&\sum_{p=1}^7 \, \left( \sigma_p + \ft 14 \phi\right)
  \,\overrightarrow{\epsilon}_p + 2 \phi_3  \, \overrightarrow{\epsilon} _8\nonumber\\
  \phi_3 & = & \ft 18 \, \phi - \ft 12 \sum_{p=1}^7 \, \sigma_p
\label{identicarta}
\end{eqnarray}
$\phi$ being the dilaton  in $D=10$ and $\phi_3$ its counterpart
in $D=3$. The above formula follows immediately from
eq.(\ref{typeII AB})
\par
We also stress the following general property of the parametrization
(\ref{metrica10}) for the $D=10$ metric:
\begin{equation}
  \sqrt{-\mbox{det}G} \, G^{00} \, = \, \sqrt{-\mbox{det}g} \,\, g^{00}
\label{property}
\end{equation}
having denoted $G$ the full Einstein metric in ten dimension and $g$
the Einstein metric in three dimension.
\par
\paragraph{Specializing to our example}
Hence, in our example  the ansatz for $G_{ij}$ is diagonal
\begin{equation}
  G_{ij} =   \exp \left[ 2
  \sigma_i\right] \, \delta_{ij} \quad ; \quad i=1,\dots,7
\label{sigmafildi}
\end{equation}
and we  obtain the following relation between the fields $x$
and $y$ and the diagonal entries of the metric and the dilaton:
\begin{eqnarray}
\phi = - \overrightarrow{h}\cdot \overrightarrow{\alpha}[7]& =  & x+y \nonumber\\
\sigma_{1,2} & = & \frac {3x-y}{4} \nonumber\\
\sigma_{3,4,5} & = & \frac{3y-x}{4} \nonumber\\
\sigma_{6,7}&=&-\frac{5y+x}{4}
\label{xyident}
\end{eqnarray}
Calling $\widetilde{h}_{1,2}$ the Cartan fields in the abstract $\mathrm{A_2}$
model discussed in section \ref{exampsolv}, we have:
\begin{equation}
  {\widetilde{h}}\cdot \beta_1 = \sqrt{2} \,
  \widetilde{h}_1 \quad ; \quad {\widetilde{h}}\cdot
  \beta_2 = - \ft 1{\sqrt{2}} \, \widetilde{h}_1 \, + \,
  \sqrt{\ft 32} \, \widetilde{h}_2
\label{htildeident}
\end{equation}
so that we can conclude:
\begin{equation}
  x =  \ft 1{\sqrt{2}} \, \widetilde{h}_1 \, \quad ; \quad y = -
  \ft 1{\sqrt{6}} \widetilde{h}_2
\label{xyversush}
\end{equation}
We can also immediately conclude that:
\begin{equation}
 Q^2 =  \frac{d}{dt}{h}\cdot \frac{d}{dt}{h}
  =\frac{d}{dt}{\widetilde{h}}\cdot \frac{d}{dt}{\widetilde{h}}
  = |\chi_1|^2 + |\chi_2|^2
\label{hnorme}
\end{equation}
On the other hand the interpretation of $Q^2$ is the following.
Consider the
parameter $\varpi^2$  appearing in the three-dimensional
metric determined from the sigma model by Einstein equations. It is
defined as:
\begin{equation}
  \varpi^2 = h_{IJ} \dot{\phi}^I \, \dot{\phi}^J = \sum_{i=1}^8
  |\chi_i|^2 + \sum_{\alpha=1}^{120} \,|\Phi|^2
\label{varpi}
\end{equation}
If we calculate $ \varpi^2$  using the generating solution or any other
solution obtained from it by compensating $\mathrm{H}$-transformations, its
value, which is a constant, does not change. So we have:
\begin{equation}
  \varpi^2 = \sum_{i=1}^8
  |\chi_i^{(\mbox{gen.sol.})}|^2
\label{varpigensol}
\end{equation}
and in the lifting of our $\mathrm{A_2}$ solutions we can conclude
that $Q^2=\varpi^2$. Let us calculate this crucial parameter for the
case of the non trivial $A_2$ solutions discussed above. By means
of straightforward algebra we get:
\begin{equation}
\varpi^2 =  |\chi_1^{(\mbox{gen.sol.})}|^2 + |\chi_2^{(\mbox{gen.sol.})}|^2  = \ft
1{24} \, \left(
\kappa^2 + 3 \, \omega^2 \right)
\label{finalvarpi}
\end{equation}
Next we turn to the identification of the $p$-forms. As we will
explicitly verify by checking type II B supergravity field
equations, the appropriately normalized identifications are the
following ones:
\begin{eqnarray}
B_{[2]} & = &  \varphi_1(t) \, dx_3 \, \wedge \, dx_4 \nonumber\\
C_{[2]} & = &  \varphi_2(t) \, dx_8 \, \wedge \,
dx_9 \nonumber\\
C_{[4]} &=& \varphi_3(t) \,dx_3 \, \wedge \,
dx_4\, \wedge \, dx_8 \, \wedge \,
dx_9 \, + \, U
\label{formeident}
\end{eqnarray}
where $U$ is the appropriate $4$--form needed to make the
corresponding field strength self dual.
\par
In this way recalling the normalizations of type II B field
strengths as given in appendix we get:
\begin{eqnarray}
F^{NS}_{[3]034} & = & \ft 16 \varphi_1 \, '(t)\nonumber\\
F^{RR}_{[3]089} & = & \ft 16 \varphi_2 \, '(t)\nonumber\\
F^{RR}_{[5]03489}&=&\ft 1{240} \, W(t) \, \nonumber\\
F^{RR}_{[5]12567}&=&\ft 1{240} \, W(t) \,
\sqrt{-\mbox{det}g} \, \frac{1}{g_{00} \, g_{33} \, g_{44} \, g_{88} \, g_{99}}
\label{Fformidenti}
\end{eqnarray}
and we recognize that the combination $W(t)$ defined in
eq.(\ref{Wdefi}) is just the self-dual $5$-form field strength
including Chern-Simons factors.
\subsubsection{Full oxidation of the $A_2$ solution with only one
root switched on} \label{oxide1a2} Let us now focus on the $A_2$
solution involving only the highest root (similar solutions were
obtained in \cite{pope1}-\cite{quevedo},\cite{otherSbranes}),
namely on eq.s (\ref{generetsolu}) and (\ref{finsol}). Inserting
the explicit form of the Cartan fields in eq.s(\ref{xyversush})
and then using (\ref{xyident}) we obtain the complete form of the
metric
\begin{eqnarray}
  ds^2 &=&-r^2_{[0]}(t) \,dt^2 + r^2_{[1|2]}(t) \, \left( dx_1^2 +
  dx_2^2 \right) + r^2_{[3|4]}(t) \, \left( dx_3^2 +
  dx_4^2 \right) \nonumber\\
  && + r^2_{[5|6|7]}(t) \, \left( dx_5^2 + dx_6^2 +
  dx_7^2\right) +
  r^2_{[8|9]}(t) \, \left( dx_8^2 +
  dx_9^2 \right)
  \label{a2firstmetric}
\end{eqnarray}
which is diagonal and it is parametrized
by five time dependent \textit{scale factors}
\begin{eqnarray}
r^2_{[0]}(t)&=& e^{t\,{\sqrt{\frac{{\kappa }^2}{3} + {\omega }^2}}}\,
  {\sqrt{\cosh \frac{t\,\omega }{2}}} \nonumber\\
r^2_{[1|2]}(t)&=&e^{\frac{t\,{\sqrt{\frac{{\kappa }^2}{3} + {\omega }^2}}}{2}}\,
  {\sqrt{\cosh \frac{t\,\omega }{2}}}\nonumber\\
  r^2_{[3|4]}(t)&=&\frac{1}{e^{\frac{t\,\kappa }{6}}\,{\sqrt{\cosh \frac{t\,\omega
  }{2}}}}\nonumber\\
r^2_{[5|6|7]}(t)&=&{\sqrt{\cosh \frac{t\,\omega }{2}}}\nonumber\\
r^2_{[8|9]}(t)&=&\frac{e^{\frac{t\,\kappa }{6}}}{{\sqrt{\cosh \frac{t\,\omega }{2}}}}
\label{a2firstscafacti}
\end{eqnarray}
We also obtain the explicit form of the dilaton, which turns out to be linear in
time:
\begin{equation}
 \phi= -\ft {1}{6} \, \kappa \, t
\label{a2firstdilaton}
\end{equation}
Calculating the Ricci tensor of the metric (\ref{a2firstmetric}) we
find that it is also diagonal and it has five independent eigenvalues
respectively given by:
\begin{eqnarray}
Ric_{00} & = & \frac{\left( {\kappa }^2 + 9\,{\omega }^2 + {\kappa }^2\,\cosh t\,\omega
                        \right) \,{\mbox{sech}^2(\frac{t\,\omega }{2})}}{288}
\nonumber\\
Ric_{11}=Ric_{22} & = & \frac{{\omega }^2\,{\mbox{sech}^2(\frac{t\,\omega }{2})}}
  {32\,e^{\frac{t\,{\sqrt{\frac{{\kappa }^2}{3} + {\omega }^2}}}{2}}} \nonumber\\
Ric_{33}=Ric_{44} & = & \frac{\omega^2\mbox{sech}^2(\frac{t\omega}{2})}
  {32\,e^{\frac{t\,{\sqrt{\frac{{\kappa }^2}{3} + {\omega }^2}}}{2}}} \nonumber\\
Ric_{55}=Ric_{66}=Ric_{77} & = & \frac{-\omega^2\mbox{sech}^3(\frac{t\omega}{2})}
  {32\,e^{\frac{t\,\left( \kappa  +
                                             6\,{\sqrt{\frac{{\kappa }^2}{3} +
{\omega }^2}} \right) }{6}}}
\nonumber\\
Ric_{88}=Ric_{99} & = & \frac{- \omega^2\mbox{sech}^3(\frac{t\omega}{2}) }
  {32\,e^{\frac{t\,\left( \kappa  +
                                             6\,{\sqrt{\frac{{\kappa }^2}{3} +
{\omega }^2}} \right) }{6}}}
\label{a2firstRicci}
\end{eqnarray}
On the other hand inserting the explicit values of scalar fields (\ref{finsol})
into equations (\ref{formeident}) we obtain:
\begin{eqnarray}
F^{NS}_{[3]} & = & 0 \nonumber\\
F^{RR}_{[3]} & = & 0 \nonumber\\
F^{RR}_{[5]} & = & \frac{\omega \,{dt}\wedge {{{dx}}_3}\wedge
                        {{{dx}}_4}\wedge {{{dx}}_8}\wedge {{{dx}}_9}
                     }{1 + \cosh t\,\omega } +
  \frac{\omega \,{{{dx}}_1}\wedge {{{dx}}_2}\wedge
                        {{{dx}}_5}\wedge {{{dx}}_6}\wedge {{{dx}}_7}
                     }{2}
\label{a2firstpform}
\end{eqnarray}
Considering eq.s(\ref{a2firstpform}) and (\ref{a2firstdilaton})
together the physical interpretation of the parameters $\omega$ and
$\kappa$ labeling the generating solution, becomes clear. They are
respectively associated to the charges of the $D3$ and $D5$ branes
which originate this classical supergravity solution. Indeed, as it
is obvious from the last of eq.s (\ref{a2firstpform}), there is a
dyonic $D3$-brane whose magnetic charge is uniformly distributed on
the Euclidean hyperplane 12567 while the electric charge is attached to the
 Minkowskian hyperplane 03489. The magnetic charge per unit volume
is $\omega/2$. With our choice of the $\mathrm{A_2}$ subalgebra,
there should also be a $D5$-brane magnetically dual to an
Euclidean $D$-string extending in the directions 89. In this
particular solution, where $\varphi_{1,2}=0$ the $F^{RR}_{[3]}$
vanishes, yet the presence of the $D5$ brane is revealed by the
dilaton. Indeed in a pure $D3$ brane solution the dilaton would be
constant. The linear behaviour (\ref{a2firstdilaton}) of $\phi$,
with coefficient $-\kappa/6$  is due to the $D5$ brane which
couples non trivially to the dilaton field. Such an interpretation
will become completely evident when we consider the oxidation of
the solution obtained from this by a further $\mathrm{SO(3)}$
rotation which switches on all the roots. This we do in the next
subsection. Then we will discuss how both oxidations do indeed
satisfy the field equations of type II B supergravity and we will
illustrate their physical properties as cosmic backgrounds.
\subsubsection{Full oxidation of the $\mathrm{A_2}$ solution with all three
roots switched on} \label{oxide2a2} Let us then turn to the
$\mathrm{A_2}$ solution involving all the three nilpotent fields,
namely to eq.s (\ref{2rotasolu}) and (\ref{2finsol}). Just as
before, by inserting the explicit form of the Cartan fields in
eq.s(\ref{xyversush}) and then using (\ref{xyident}) we obtain the
complete form of the new metric, which has the same diagonal
structure as in the previous example, namely
\begin{eqnarray}
  ds^2 &=&-\overline{r}^2_{[0]}(t) \,dt^2 + \overline{r}^2_{[1|2]}(t) \, \left(
dx_1^2 +
  dx_2^2 \right) + \overline{r}^2_{[3|4]}(t) \, \left( dx_3^2 +
  dx_4^2 \right) \nonumber\\
  && + \overline{r}^2_{[5|6|7]}(t) \, \left( dx_5^2 + dx_6^2 +
  dx_7^2\right) +
  \overline{r}^2_{[8|9]}(t) \, \left( dx_8^2 +
  dx_9^2 \right)
  \label{a2secondtmetric}
\end{eqnarray}
now, however, the  \textit{scale factors} are given by:
\begin{eqnarray}
\overline{r}^2_{[0]}(t)&=&e^{t\,\left( \frac{-\omega }{4} +
                                {\sqrt{\frac{{\kappa }^2}{3} + {\omega }^2}} \right)
}\,
  {\left( 1 + e^{t\,\omega } \right) }^{\frac{1}{4}}\,
  {\left( 1 + e^{t\,\omega } + e^
                                {\frac{t\,\left( \kappa  + \omega  \right) }{2}}
\right) }^{\frac{1}{4}} \nonumber\\
\overline{r}^2_{[1|2]}(t)&=&e^{\frac{t\,\left( -3\,\omega  +
                                          2\,{\sqrt{3}}\,{\sqrt{{\kappa }^2 +
3\,{\omega }^2}} \right) }{12}}\,
  {\left( 1 + e^{t\,\omega } \right) }^{\frac{1}{4}}\,
  {\left( 1 + e^{t\,\omega } + e^
                                {\frac{t\,\left( \kappa  + \omega  \right) }{2}}
\right) }^{\frac{1}{4}}\nonumber\\
\overline{r}^2_{[3|4]}(t)&=&\frac{e^{\frac{-\left( t\,\kappa  \right) }{6} +
\frac{t\,\omega }{4}}\,
             {\left( 1 + e^{t\,\omega } +
                                 e^{\frac{t\,\left( \kappa  + \omega  \right) }{2}}
\right) }^{\frac{1}{4}}
             }{{\left( 1 + e^{t\,\omega } \right) }^{\frac{3}{4}}}\nonumber\\
\overline{r}^2_{[5|6|7]}(t)&=&\frac{{\left( 1 + e^{t\,\omega } \right)
}^{\frac{1}{4}}\,
             {\left( 1 + e^{t\,\omega } +
                                 e^{\frac{t\,\left( \kappa  + \omega  \right) }{2}}
\right) }^{\frac{1}{4}}
             }{e^{\frac{t\,\omega }{4}}}\nonumber\\
\overline{r}^2_{[8|9]}(t)&=&\frac{e^{\frac{t\,\left( 2\,\kappa  + 3\,\omega  \right)
}{12}}\,
             {\left( 1 + e^{t\,\omega } \right) }^{\frac{1}{4}}}{{\left( 1 +
                                e^{t\,\omega } + e^{\frac{t\,\left( \kappa  + \omega
 \right) }{2}} \right)
                                }^{\frac{3}{4}}}
\label{a2secondscafacti}
\end{eqnarray}
and the dilaton is no longer linear in time, rather it is given by:
\begin{equation}
 \phi = \frac{-\left( t\,\kappa  \right)  - 3\,\log (1 + e^{t\,\omega }) +
             3\,\log (1 + e^{t\,\omega } +
                                e^{\frac{t\,\left( \kappa  + \omega  \right) }{2}})}{6}
\label{a2secondilaton}
\end{equation}
\iffigs
\begin{figure}
\begin{center}
\epsfxsize =7cm
{\epsffile{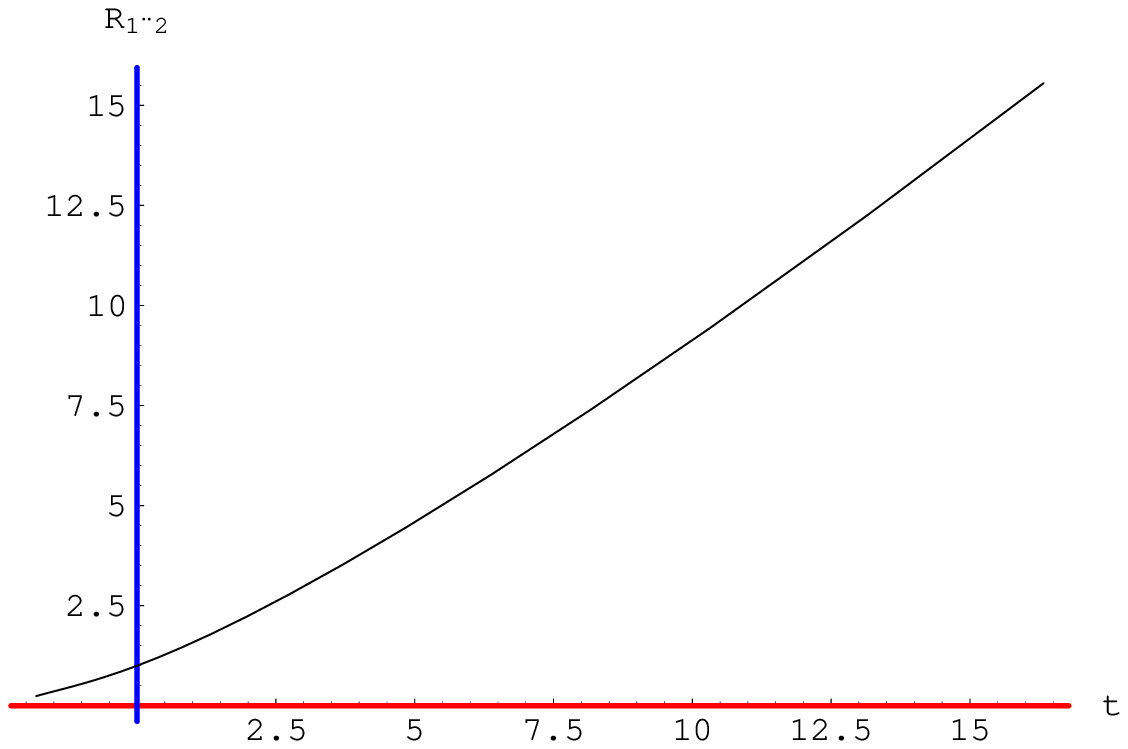}}
\epsfxsize =7cm
{\epsffile{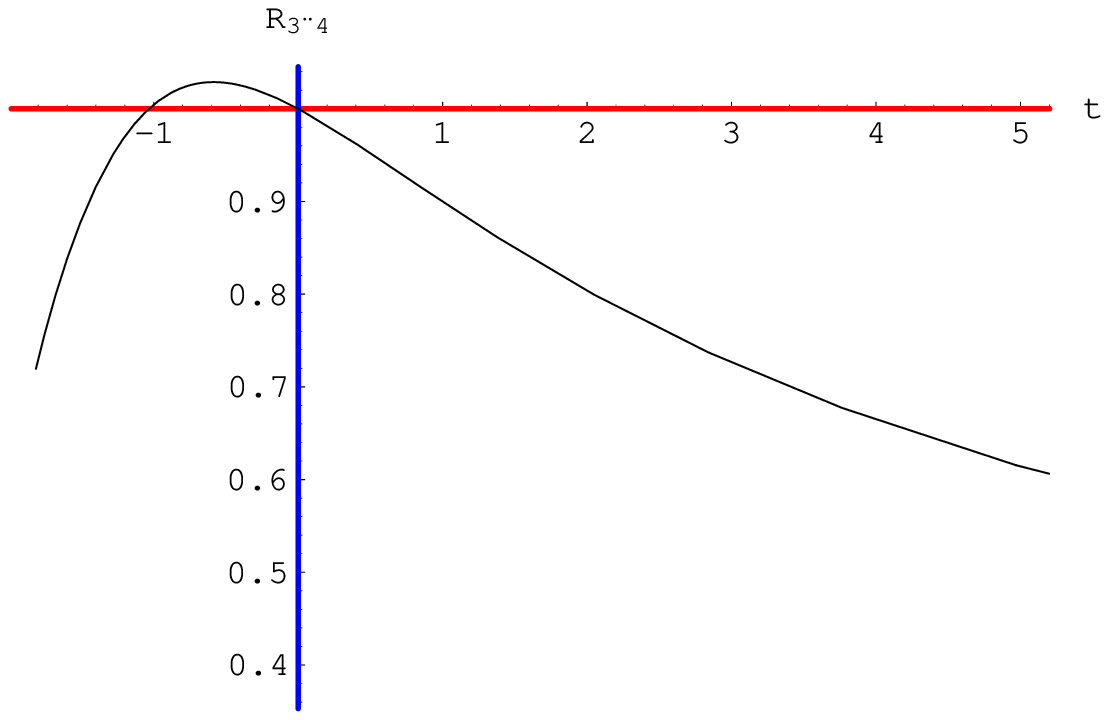}}
\epsfxsize =7cm
{\epsffile{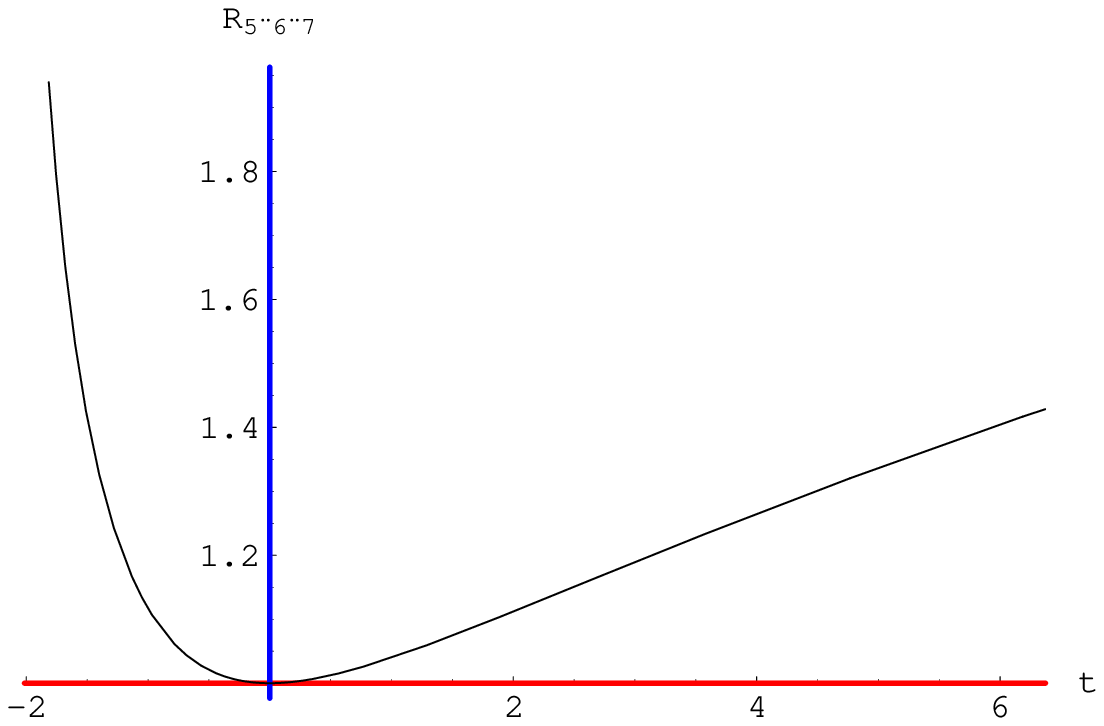}}
\epsfxsize =7cm
{\epsffile{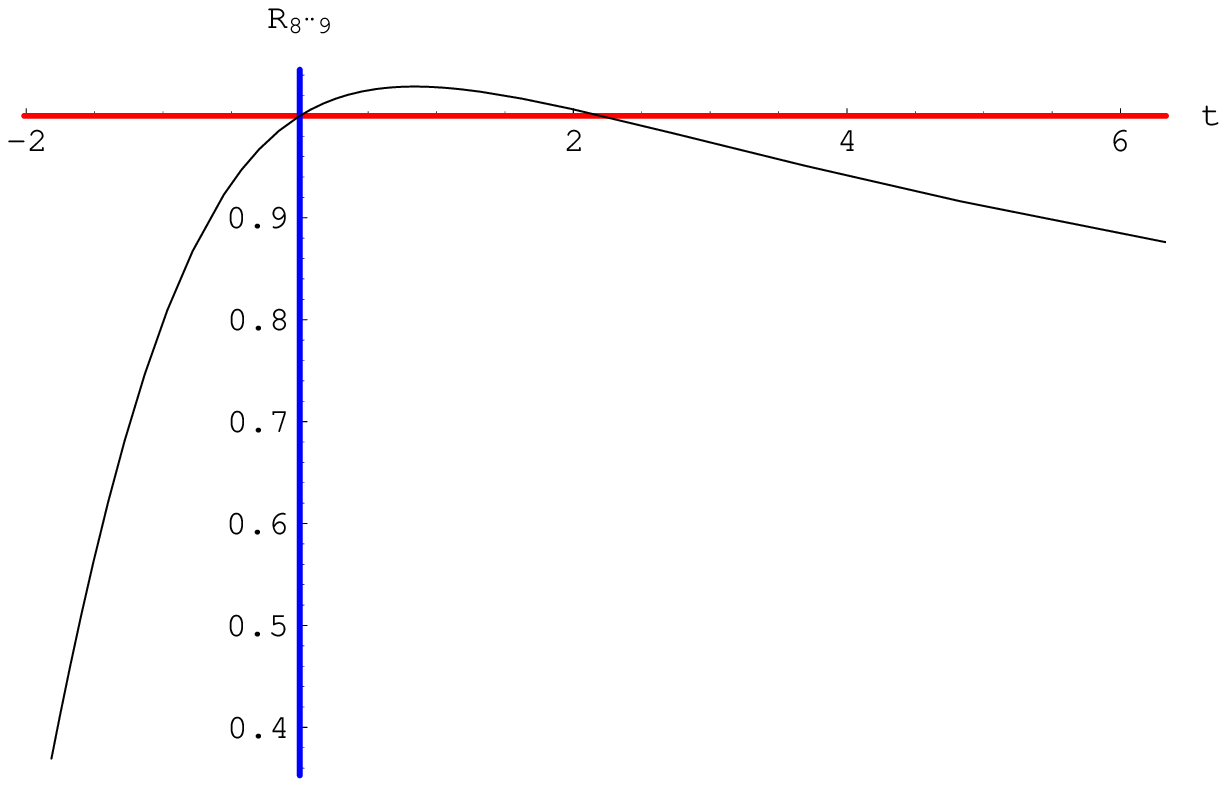}}
\caption{Plots of the scale factors $r^2_{[\alpha]}$, $\alpha= 1|2 \, ,\, 3|4 \, ,
\, 5|6|7 \, , \,
8|9$ as functions of the cosmic time $t=\tau(T)$ in the case of the choice of
parameters $\omega =1$, $\kappa=0.5$
and for the $A_2$ solution with only the highest root switched on.
\label{radiio1k05}}
\hskip 1.5cm \unitlength=1.1mm
\end{center}
\end{figure}
\fi
\iffigs
\begin{figure}
\begin{center}
\epsfxsize =7cm
{\epsffile{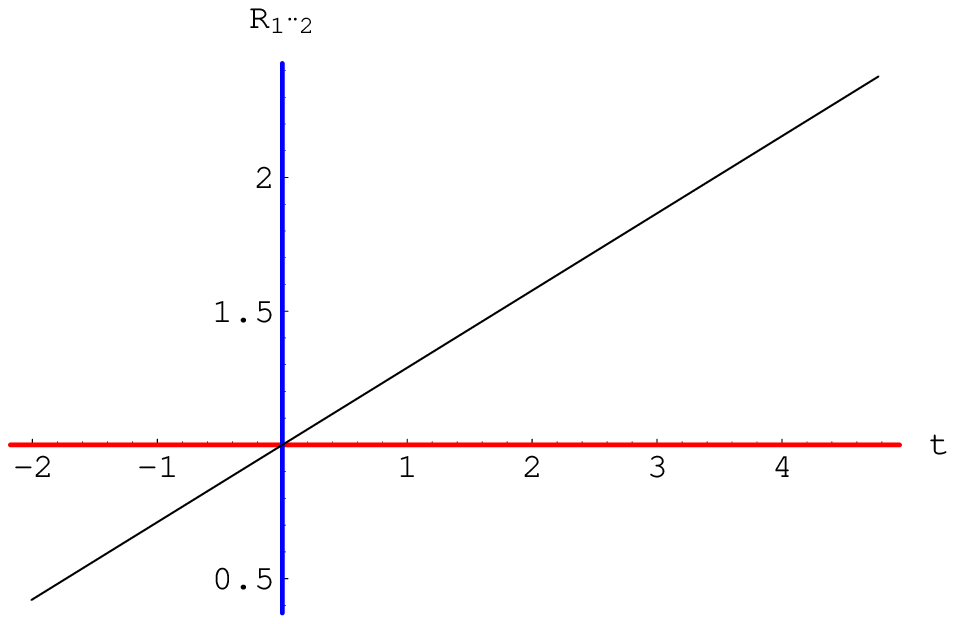}}
\epsfxsize =7cm
{\epsffile{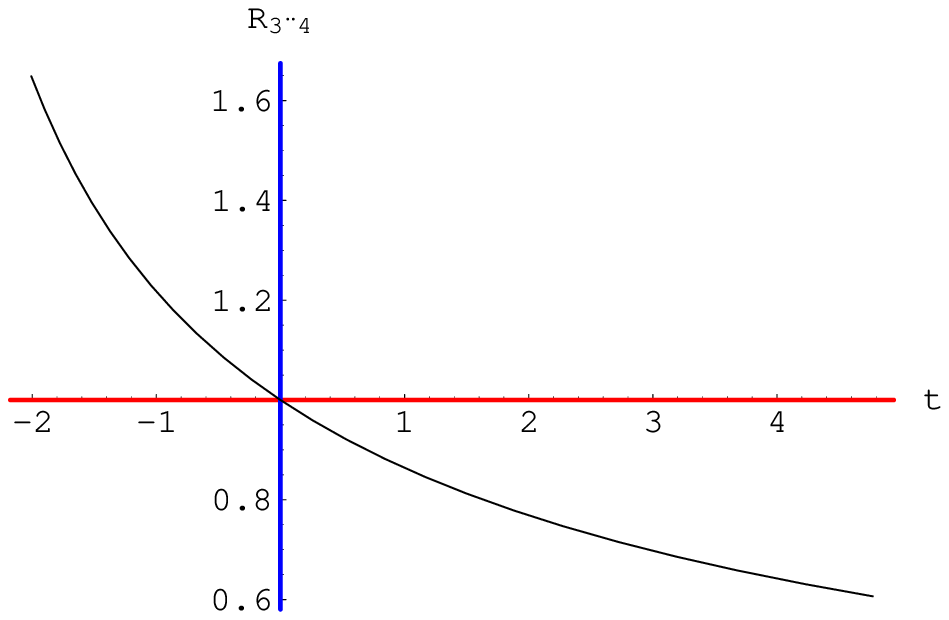}}
\epsfxsize =7cm
{\epsffile{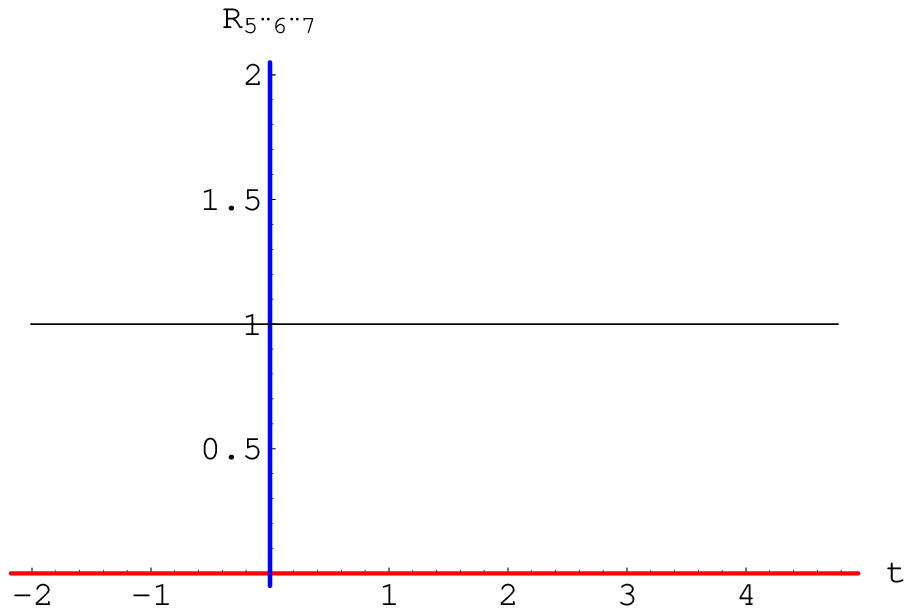}}
\epsfxsize =7cm
{\epsffile{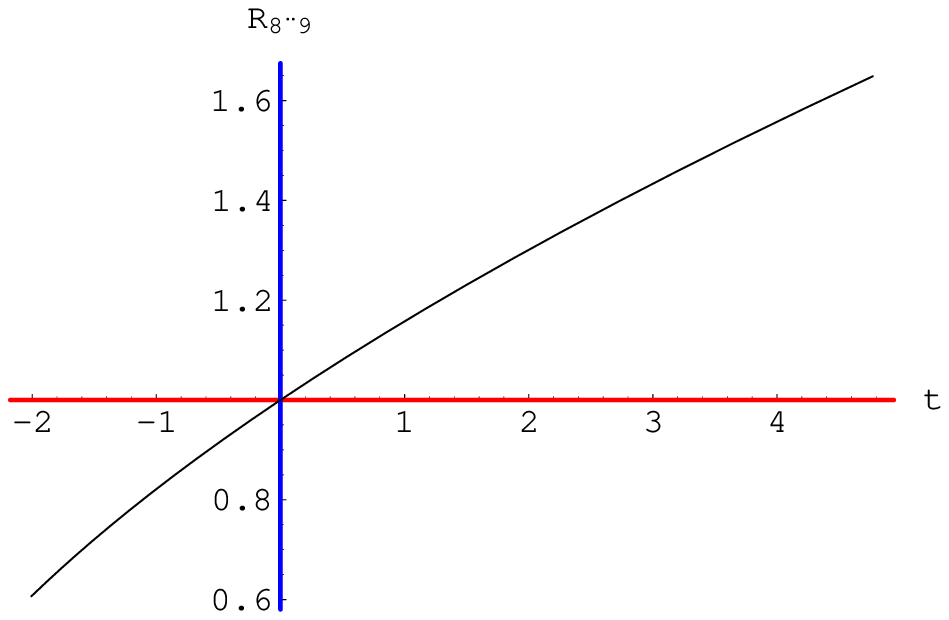}}
\caption{Plots of the scale factors $r^2_{[\alpha]}$, $\alpha= 1|2 \, ,\, 3|4 \, ,
\, 5|6|7 \, , \,
8|9$ as functions of the cosmic time $t=\tau(T)$ in the case of the choice of
parameters $\omega =0$, $\kappa=1$
and for the $A_2$ solution with only the highest root switched on.
\label{radiio0k1}}
\hskip 1.5cm \unitlength=1.1mm
\end{center}
\end{figure}
\fi
\iffigs
\begin{figure}
\begin{center}
\epsfxsize =7cm
{\epsffile{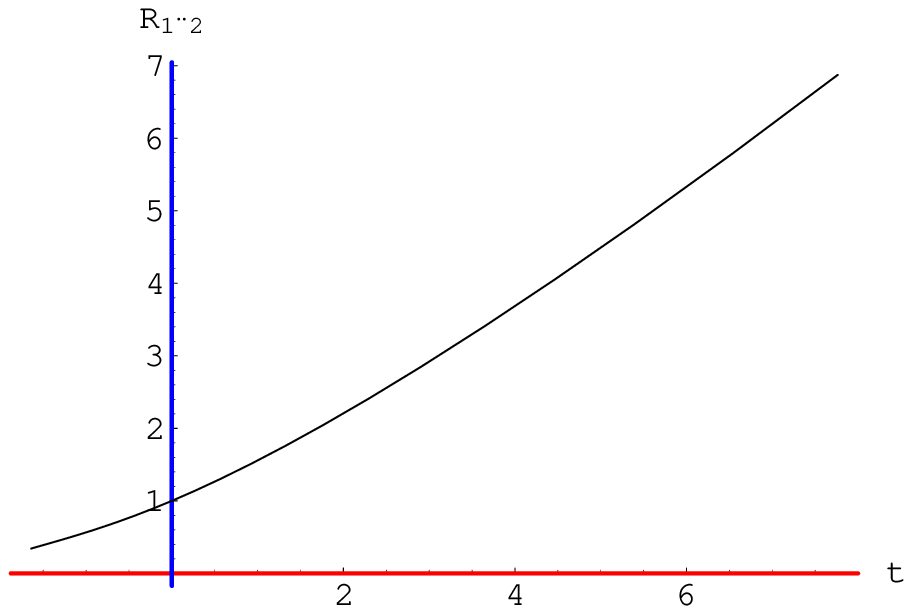}}
\epsfxsize =7cm
{\epsffile{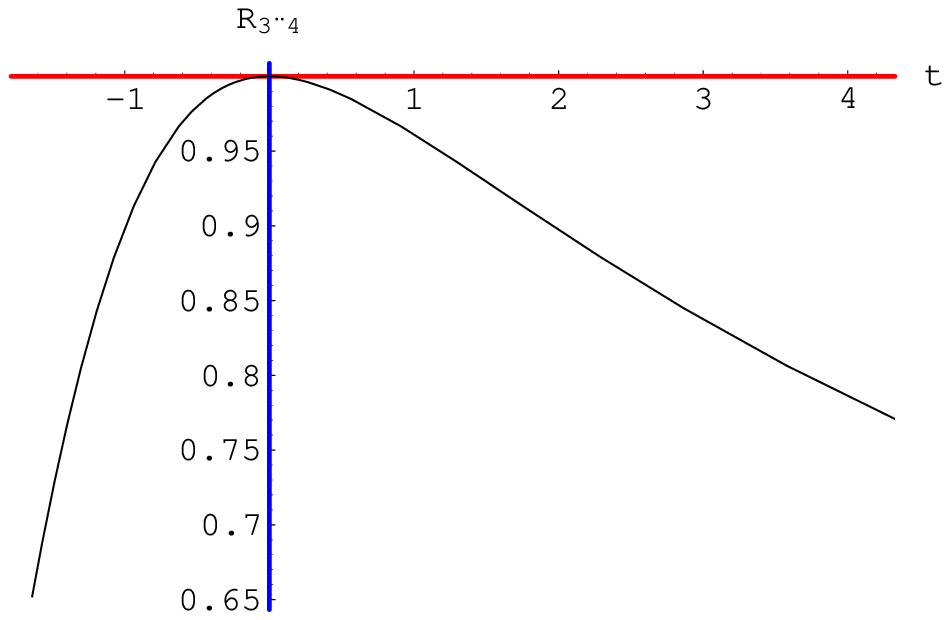}}
\epsfxsize =7cm
{\epsffile{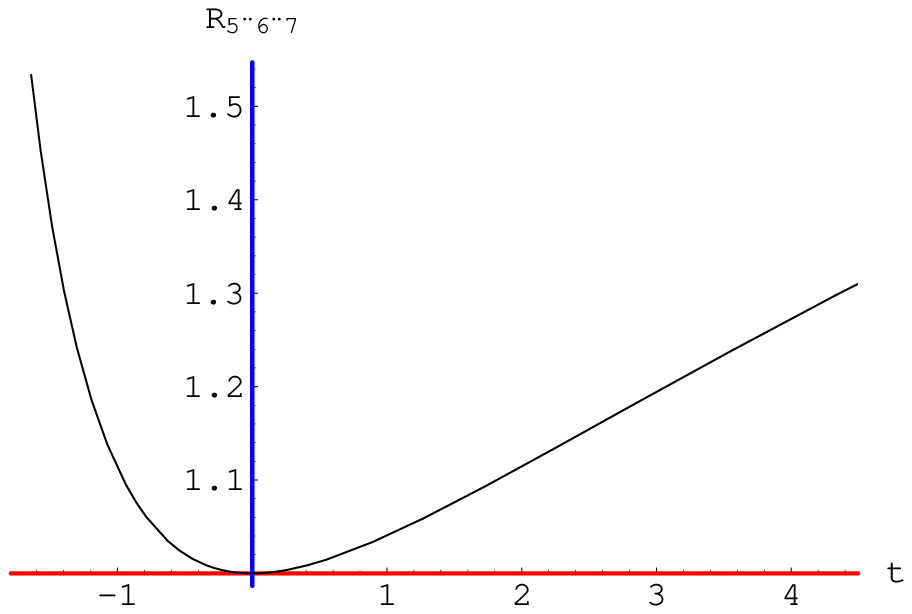}}
\epsfxsize =7cm
{\epsffile{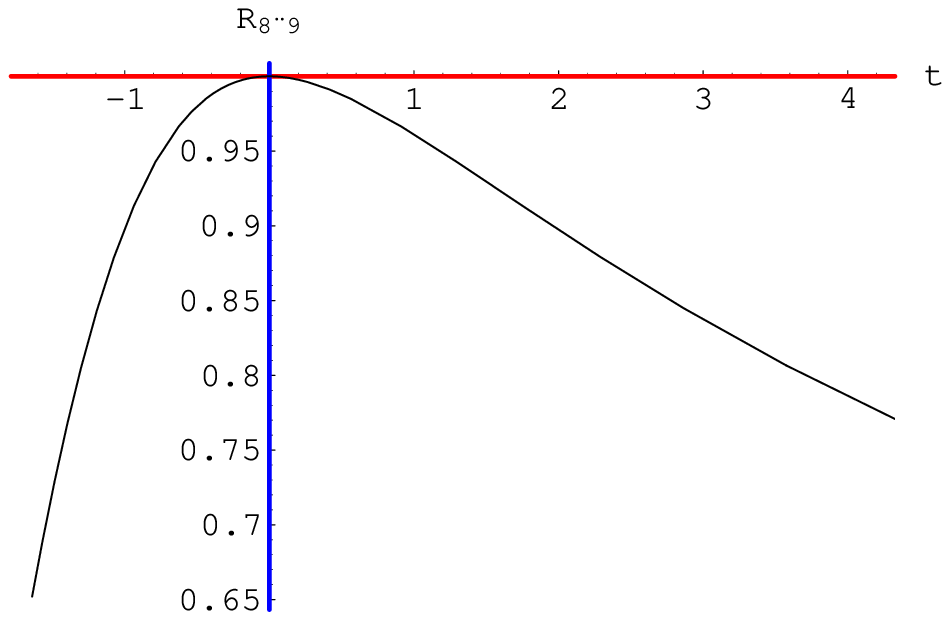}}
\caption{Plots of the scale factors $r^2_{[\alpha]}$, $\alpha= 1|2 \, ,\, 3|4 \, ,
\, 5|6|7 \, , \,
8|9$ as functions of the cosmic time $t=\tau(T)$ in the case of the choice of
parameters $\omega =1$, $\kappa=0$
and for the $A_2$ solution with only the highest root switched on.
\label{radiio1k0}}
\hskip 1.5cm \unitlength=1.1mm
\end{center}
\end{figure}
\fi
Calculating the Ricci tensor of the metric
(\ref{a2secondtmetric},\ref{a2secondscafacti}) we find it diagonal
with five different eigenvalues, just as in the previous case, but
with a modified time dependence, namely:
\begin{eqnarray}
Ric_{00} & = & \frac{1}{576\,{\left( 1 + e^{t\,\omega } \right) }^2\,
             {\left( 1 + e^{t\,\omega } +
                                 e^{\frac{t\,\left( \kappa  + \omega  \right) }{2}}
\right) }^2} \, \left (
                                 {\left( 1 + e^{t\,\omega } \right) }^2\,
            \left( 4 + 8\,e^{t\,\omega } + 4\,e^{2\,t\,\omega } +
                     23\,e^{\frac{t\,\left( \kappa  + \omega  \right) }{2}}
                     \right.\right.\nonumber\\
            && +\left.\left. e^{t\,\left( \kappa  + \omega  \right) } +
                     23\,e^{\frac{t\,\left( \kappa  + 3\,\omega  \right) }{2}}
\right) \,
            {\kappa }^2 - 6\,e^{\frac{t\,\left( \kappa  + \omega  \right) }{2}}\,
            \left( -1 + e^{2\,t\,\omega } \right) \,
            \left( 7 + 7\,e^{t\,\omega }  +
                     e^{\frac{t\,\left( \kappa  + \omega  \right) }{2}} \right)
\,\kappa \,\omega
                     \right.\nonumber\\
 &&
  \left. + 9\,\left( 8\,e^{t\,\omega } + 16\,e^{2\,t\,\omega } + 8\,e^{3\,t\,\omega } +
                     3\,e^{\frac{t\,\left( \kappa  + \omega  \right) }{2}} +
                     e^{t\,\left( \kappa  + \omega  \right) } +
                     10\,e^{t\,\left( \kappa  + 2\,\omega  \right)
                     }\right.\right.\nonumber\\
                     &&\left.\left. +
                     17\,e^{\frac{t\,\left( \kappa  + 3\,\omega  \right) }{2}} +
                     e^{t\,\left( \kappa  + 3\,\omega  \right) } +
                     17\,e^{\frac{t\,\left( \kappa  + 5\,\omega  \right) }{2}} +
                     3\,e^{\frac{t\,\left( \kappa  + 7\,\omega  \right) }{2}}
\right) \,
            {\omega }^2
                                 \right ) \nonumber\\
Ric_{11}&=&Ric_{22}\nonumber\\
 & = & \frac{1}{64\,{\left( 1 + e^{t\,\omega } \right) }^2\,
                     {\left( 1 + e^{t\,\omega } +
                                          e^{\frac{t\,\left( \kappa  + \omega
\right) }{2}} \right) }^2} \,\left(
e^{\frac{t\,\left( \omega  - {\sqrt{\frac{{\kappa }^2}{3} + {\omega }^2}}
\right)
                                          }{2}}\,\left( e^{\frac{t\,\left( \kappa  +
6\,\omega  \right) }{2}}\,
                     {\left( \kappa  - \omega  \right) }^2 \right.\right.\nonumber\\
                     && \left.\left. +
             8\,e^{\frac{t\,\omega }{2}}\,{\omega }^2 +
             16\,e^{\frac{3\,t\,\omega }{2}}\,{\omega }^2 +
             8\,e^{\frac{5\,t\,\omega }{2}}\,{\omega }^2 +
             4\,e^{t\,\left( \kappa  + \frac{3\,\omega }{2} \right) }\,{\omega }^2
\right.\right.\nonumber\\
                     && \left.\left. +
             e^{\frac{t\,\kappa }{2}}\,{\left( \kappa  + \omega  \right) }^2 +
             e^{\frac{t\,\left( \kappa  + 4\,\omega  \right) }{2}}\,
                     \left( 3\,{\kappa }^2 - 2\,\kappa \,\omega  + 11\,{\omega }^2
\right)
\right.\right.\nonumber\\
                     && \left.\left.  +
             e^{\frac{t\,\kappa }{2} + t\,\omega }\,
                     \left( 3\,{\kappa }^2 + 2\,\kappa \,\omega  + 11\,{\omega }^2
\right)
             \right) \right)\nonumber\\
Ric_{33}&=&Ric_{44}\nonumber\\
 & = &  \frac{1}{64\,e^
                        {\frac{t\,\left( \kappa  - 6\,\omega  +
                                                      6\,{\sqrt{\frac{{\kappa
}^2}{3} + {\omega }^2}}
\right) }{6}}\,
                     {\left( 1 + e^{t\,\omega } \right) }^3\,
                     {\left( 1 + e^{t\,\omega } +
                                          e^{\frac{t\,\left( \kappa  + \omega
\right) }{2}} \right) }^2} \left(
                                          e^{\frac{t\,\left( \kappa  + 6\,\omega
\right) }{2}}\,
            {\left( \kappa  - \omega  \right) }^2  \right. \nonumber\\
            && \left. -
  8\,e^{\frac{t\,\omega }{2}}\,{\omega }^2 -
  16\,e^{\frac{3\,t\,\omega }{2}}\,{\omega }^2 -
  8\,e^{\frac{5\,t\,\omega }{2}}\,{\omega }^2 -
  12\,e^{t\,\left( \kappa  + \frac{3\,\omega }{2} \right) }\,{\omega }^2
\right.\nonumber\\
  &&\left.  +
  e^{\frac{t\,\kappa }{2}}\,{\left( \kappa  + \omega  \right) }^2 +
  e^{\frac{t\,\left( \kappa  + 4\,\omega  \right) }{2}}\,
            \left( 3\,{\kappa }^2 - 2\,\kappa \,\omega  - 21\,{\omega }^2 \right)
\right.\nonumber\\
            &&\left. +
  e^{\frac{t\,\kappa }{2} + t\,\omega }\,
            \left( 3\,{\kappa }^2 + 2\,\kappa \,\omega  - 21\,{\omega }^2 \right)
                                          \right)\nonumber\\
                                          Ric_{55}&=&Ric_{66}=Ric_{77}\nonumber\\
                                            & = & \frac{1}{64\,{\left( 1 +
e^{t\,\omega } \right) }^2\,
                     {\left( 1 + e^{t\,\omega } +
                                          e^{\frac{t\,\left( \kappa  + \omega
\right) }{2}} \right) }^2} \, \left(
e^{\frac{t\,\left( \omega  - 2\,{\sqrt{\frac{{\kappa }^2}{3} + {\omega }^2}}
                                          \right) }{2}}\,\left( e^
                        {\frac{t\,\left( \kappa  + 6\,\omega  \right) }{2}}\,
                     {\left( \kappa  - \omega  \right) }^2 +
             8\,e^{\frac{t\,\omega }{2}}\,{\omega }^2 +
             \right.\right.\nonumber\\
            && \left.\left.
             16\,e^{\frac{3\,t\,\omega }{2}}\,{\omega }^2 +
             8\,e^{\frac{5\,t\,\omega }{2}}\,{\omega }^2 +
             4\,e^{t\,\left( \kappa  + \frac{3\,\omega }{2} \right) }\,{\omega }^2 +
             e^{\frac{t\,\kappa }{2}}\,{\left( \kappa  + \omega  \right) }^2
             \right.\right.\nonumber\\
             &&\left.\left.+
             e^{\frac{t\,\left( \kappa  + 4\,\omega  \right) }{2}}\,
                     \left( 3\,{\kappa }^2 - 2\,\kappa \,\omega  + 11\,{\omega }^2
\right)  +
             e^{\frac{t\,\kappa }{2} + t\,\omega }\,
                     \left( 3\,{\kappa }^2 + 2\,\kappa \,\omega  + 11\,{\omega }^2
\right)
             \right)
                                          \right) \nonumber\\
Ric_{88}&=&Ric_{99}\nonumber\\
 & = & \frac{-1}{64\,{\left( 1 + e^{t\,\omega } \right) }^2\,
                     {\left( 1 + e^{t\,\omega } +
                                          e^{\frac{t\,\left( \kappa  + \omega
\right) }{2}} \right)
                                          }^3}\, \left(e^{\frac{t\,\left( \kappa  +
6\,\omega  -
                                          6\,{\sqrt{\frac{{\kappa }^2}{3} + {\omega
}^2}} \right) }{6}}\,
  \left( 3\,e^{\frac{t\,\left( \kappa  + 6\,\omega  \right) }{2}}\,
                     {\left( \kappa  - \omega  \right) }^2 \right.\right.\nonumber\\
                     &&\left.\left. +
             8\,e^{\frac{t\,\omega }{2}}\,{\omega }^2 +
             16\,e^{\frac{3\,t\,\omega }{2}}\,{\omega }^2 +
             8\,e^{\frac{5\,t\,\omega }{2}}\,{\omega }^2 -
             4\,e^{t\,\left( \kappa  + \frac{3\,\omega }{2} \right) }\,{\omega }^2 +
             e^{\frac{t\,\left( \kappa  + 4\,\omega  \right) }{2}}\,
                     {\left( -3\,\kappa  + \omega  \right) }^2 \right.\right.
\nonumber\\
                     &&\left.\left. +
             3\,e^{\frac{t\,\kappa }{2}}\,{\left( \kappa  + \omega  \right) }^2 +
             e^{\frac{t\,\kappa }{2} + t\,\omega }\,
                     {\left( 3\,\kappa  + \omega  \right) }^2 \right)
                                          \right)
\label{a2secondRicci}
\end{eqnarray}
On the other hand inserting the explicit values of scalar fields (\ref{2finsol})
into equations (\ref{formeident}) we obtain:
\begin{eqnarray}
F^{NS}_{[3]} & = & -\ft 14 \omega {\mbox{sech}^2(\frac{t\,\omega }{2})}
 dt \, \wedge \, dx_3 \, \wedge \,
dx_4\nonumber\\
F^{RR}_{[3]} & = & \frac{e^{\frac{t\,\left( \kappa  + \omega  \right) }{2}}\,
             \left( \kappa  + e^{t\,\omega }\,
                                \left( \kappa  - \omega  \right)  + \omega  \right)
}{2\,
             {\left( 1 + e^{t\,\omega } +
                                 e^{\frac{t\,\left( \kappa  + \omega  \right) }{2}}
\right) }^2
             }\, dt \, \wedge \, dx_8 \, \wedge \, dx_9 \nonumber\\
F^{RR}_{[5]} & = & \frac{e^{t\,\omega }\,\omega \,
                        {dt}\wedge {{ {dx}}_3}\wedge
                        {{ {dx}}_4}\wedge {{ {dx}}_8}\wedge
                        {{ {dx}}_9}}{\left( 1 + e^{t\,\omega } \right) \,
                     \left( 1 + e^{t\,\omega } +
                                e^{\frac{t\,\left( \kappa  + \omega  \right) }{2}}
\right) } -
  \omega \,{{ {dx}}_1}\wedge {{ {dx}}_2}\wedge
             {{ {dx}}_5}\wedge {{ {dx}}_6}\wedge
             {{ {dx}}_7}
\label{a2secondpform}
\end{eqnarray}
This formula completes the oxidation also of the second sigma
model solution to a full fledged $D=10$ type II B configuration.
As expected in both cases the ten dimensional fields obtained by
oxidation satisfy the field equations of supergravity as
formulated in the appendix (see
eq.s(\ref{NSscalapr})-(\ref{einsteinequa})). We discuss this in
the next section.
\subsubsection{How the supergravity field equations are satisfied and
their cosmological interpretation}
Taking into account that the Ramond scalar $C_0$ vanishes the
effective bosonic field equations of supergravity reduce to:
\begin{eqnarray}
d\star d \phi & = & \ft 1 2 \left( e^{-\phi} \,F_3^{NS} \wedge \star F_3^{NS} -
e^{\phi} \,F_3^{RR} \wedge \star F_3^{RR} \right) \label{dilatequa}\\
0 & = & F_3^{NS} \wedge \star F_3^{NS} \label{ramscalequa}\\
d\left( e^{-\phi} F_3^{NS} \right) &=& -F_3^{RR} \wedge \star F_5^{RR}
\label{Bfielequa}\\
d\left( e^{\phi} F_3^{RR} \right) &=& F_3^{NS} \wedge \star F_5^{RR}
\label{Cfielequa}\\
d\left( \star F_3^{RR} \right) &=& - F_3^{NS} \wedge F_3^{RR}
\label{Cfielequabis}\\
-2 Ric_{MN}& =& \hat{T}_{MN} \label{einsteinequazia}
\end{eqnarray}
where the reduced stress energy tensor $\hat{T}_{MN}$ is the
superposition of two contributions that we respectively attribute to
the $D3$ brane and to the $D5$-brane, namely:
\begin{eqnarray}
  \hat{T}_{MN} &=& \hat{T}_{MN}^{[D3]} + \hat{T}_{MN}^{[D5]}\label{totstress}\\
\hat{T}_{MN}^{[D3]} & \equiv & 150
 {F}_{[5]{M}\cdot\cdot\cdot\cdot}
{F}_{[5]{N}}^{\phantom{{M}}\cdot\cdot\cdot\cdot}\label{d3stress}\\
\hat{T}_{MN}^{[D5]} & \equiv & \frac{1}{2}\partial_{{M}}\varphi\partial_{{N}}\varphi
+ 9 \left( e^{-\varphi}F_{[3]{M}\cdot\cdot}^{NS}\,
F_{[3]{N}}^{{NS}\phantom{{M}}\cdot\cdot} +e^{\varphi}{
F}_{[3]{M}\cdot\cdot}^{RR}
{ F}_{[3]{N}}^{RR\phantom{{M}}\cdot\cdot}\right)\nonumber\\
&& -\frac{3}{4}\,
g_{{MN}}\,\left(e^{-\varphi}F_{[3]\cdot\cdot\cdot}^{NS}
F_{[3]}^{NS\cdot\cdot\cdot}+e^{\varphi}{{F}}_{[3]\cdot\cdot\cdot}^{RR}{
F}^{RR\cdot\cdot\cdot}_{[3]}\right)\label{d5stress}
\end{eqnarray}
By means of laborious algebraic manipulations that can be easily performed
on a computer with the help of MATHEMATICA, we have explicitly
verified that in both cases, that of section \ref{oxide1a2} and that
of section \ref{oxide2a2} the field
eq.s(\ref{dilatequa})-(\ref{einsteinequazia}) are indeed satisfied,
so that the oxidation procedure we have described turns out to be well tuned and fully
correct.
\iffigs
\begin{figure}
\begin{center}
\epsfxsize =7cm
{\epsffile{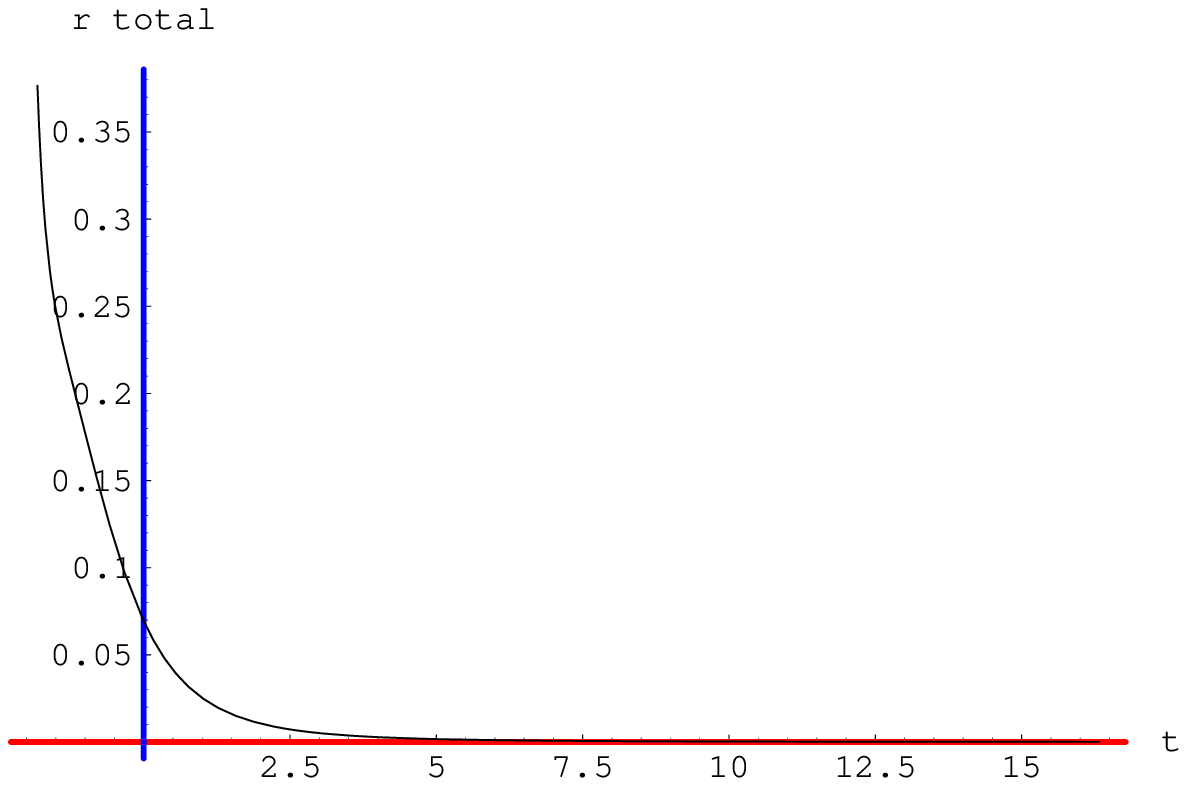}}
\epsfxsize =7cm
{\epsffile{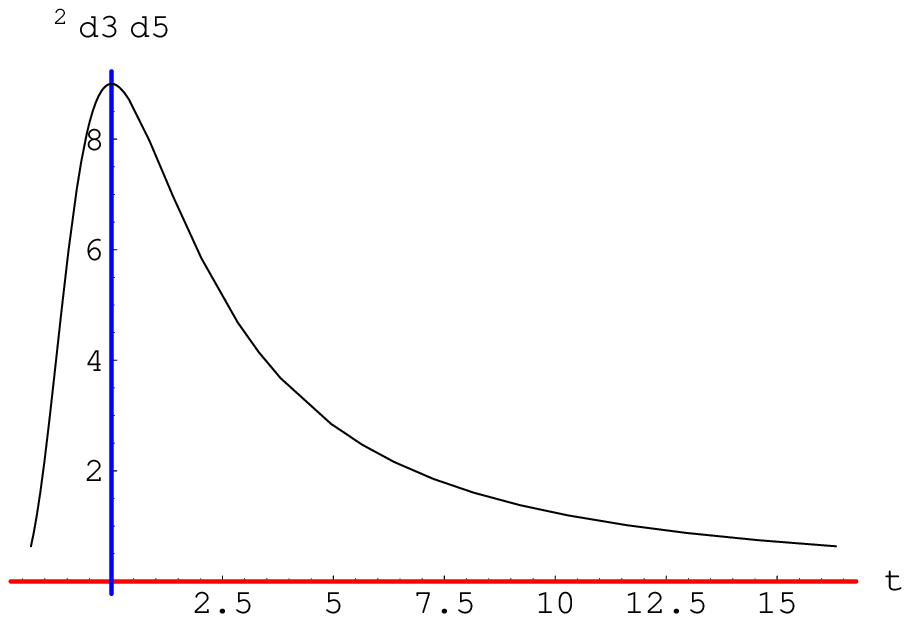}}
\epsfxsize =7cm
{\epsffile{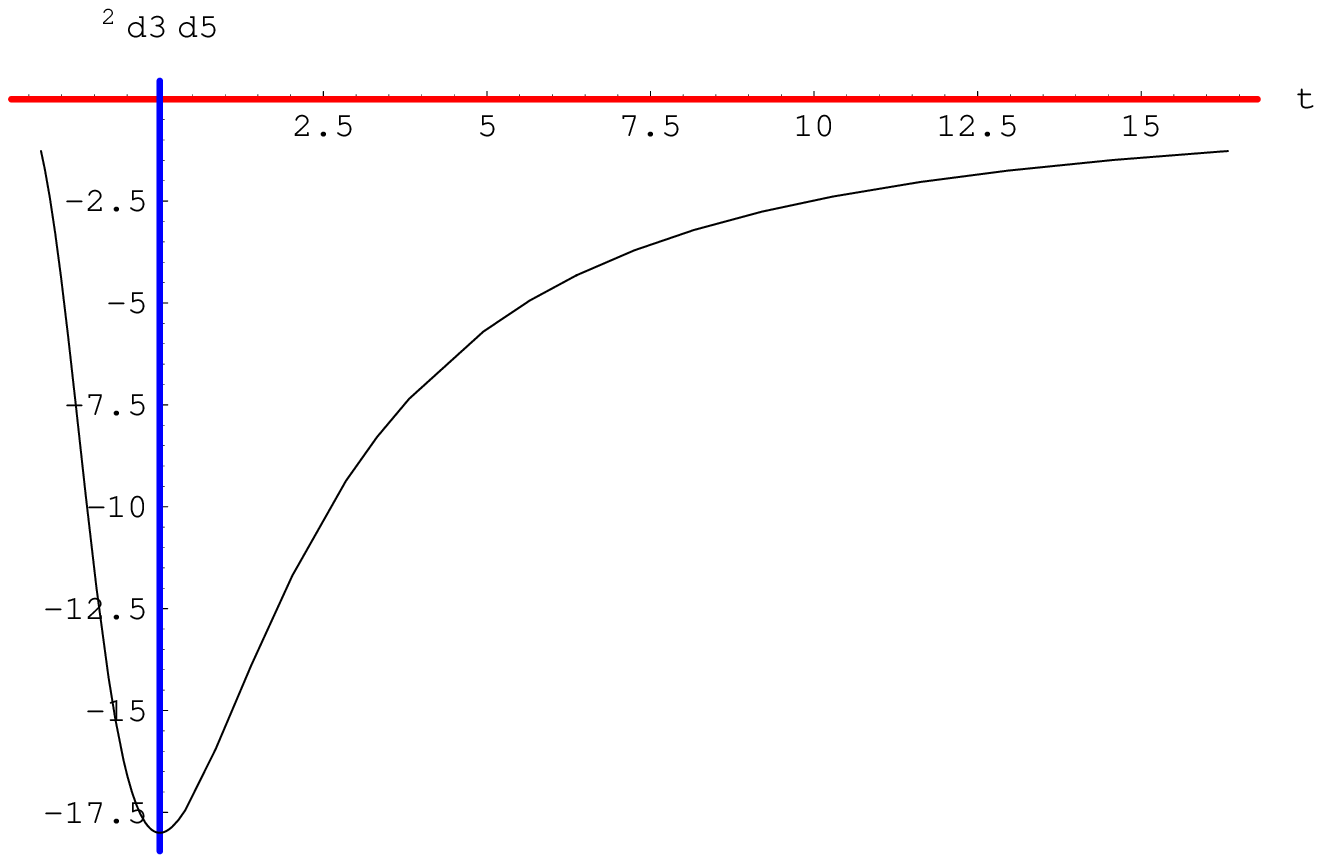}}
\caption{Plots of the energy densities  in the case of the choice of parameters
$\omega =1$, $\kappa=0.5$
and for the $A_2$ solution with only the highest root switched on.
The first picture plots the total density $\rho^{tot}(\tau)$. The
second picture plots the ratio $\rho^{d3}(\tau)/\rho^{tot}(\tau)$ and
the third plots the ratio $\rho^{d3}(\tau)/\rho^{d5}(\tau)$
\label{rhoo1k05}}
\hskip 1.5cm \unitlength=1.1mm
\end{center}
\end{figure}
\fi
\iffigs
\begin{figure}
\begin{center}
\epsfxsize =7cm
{\epsffile{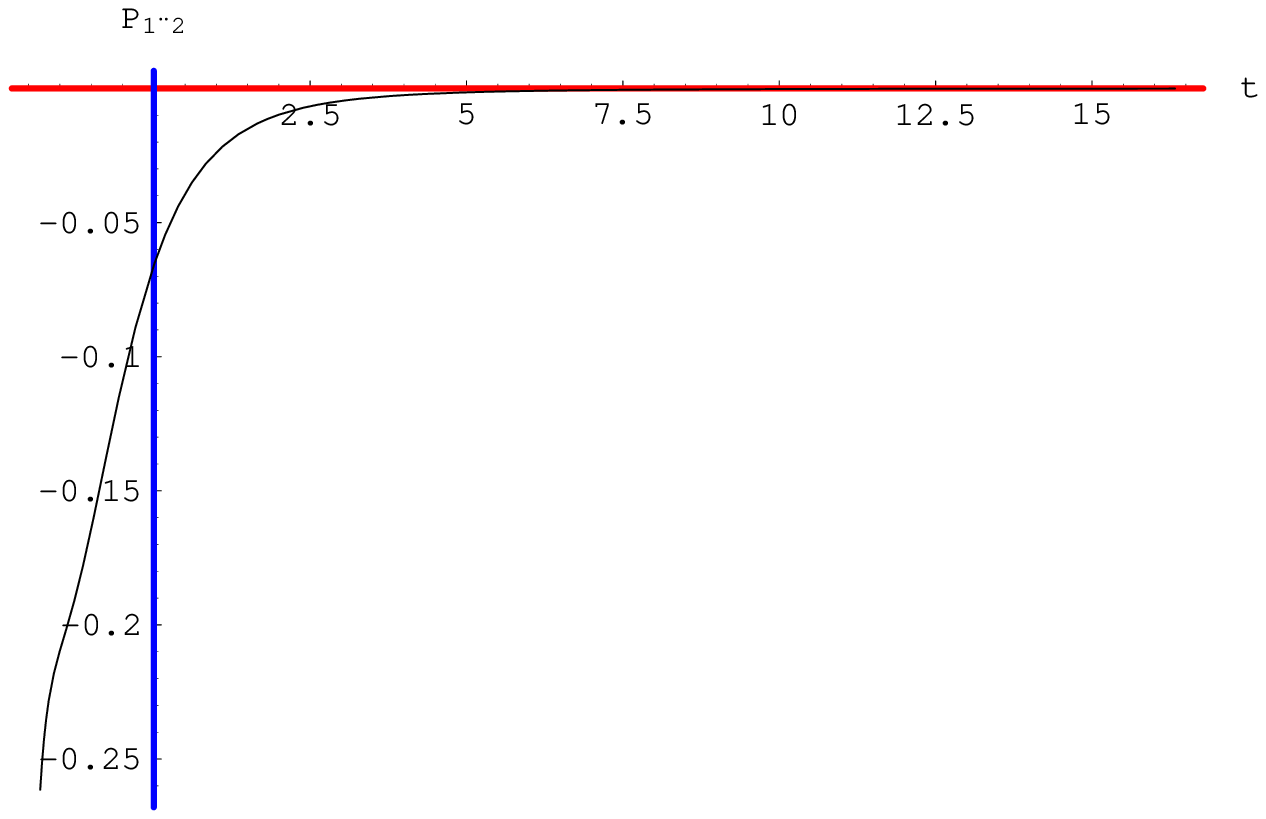}}
\epsfxsize =7cm
{\epsffile{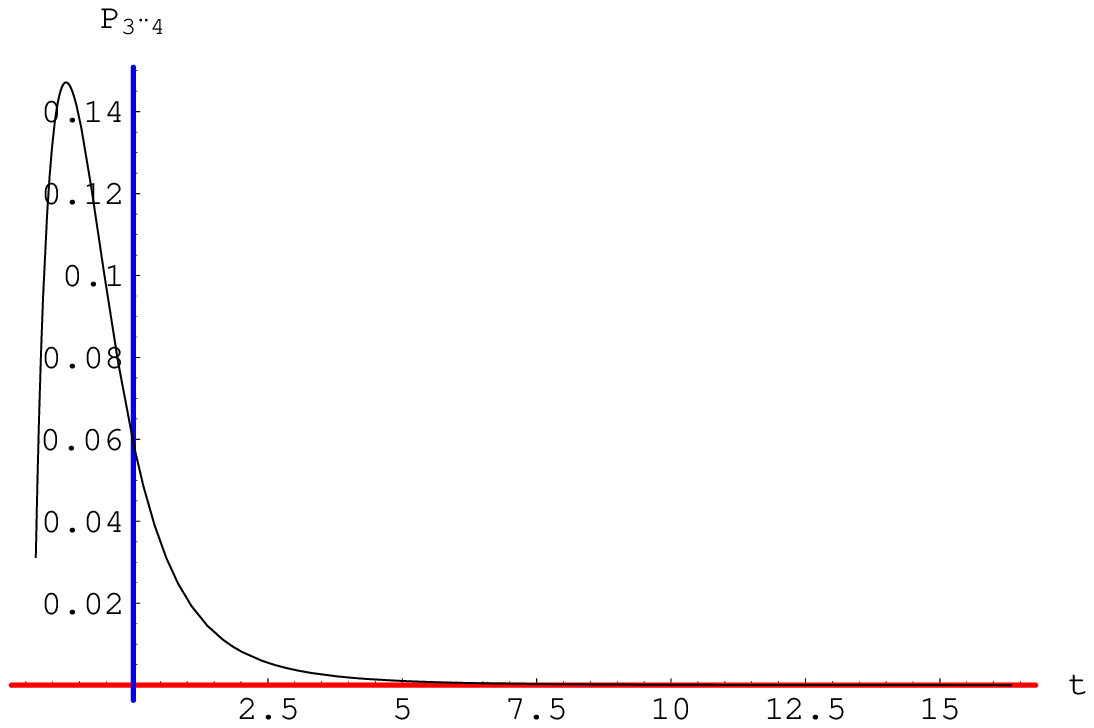}}
\epsfxsize =7cm
{\epsffile{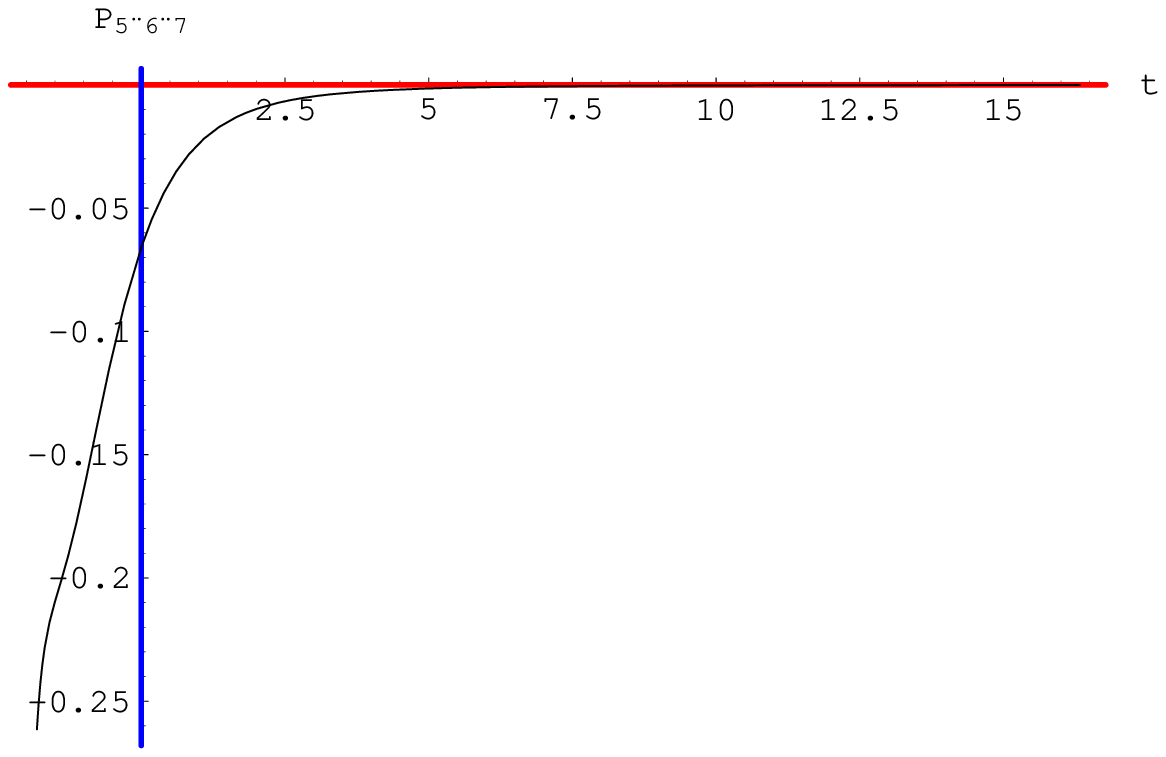}}
\epsfxsize =7cm
{\epsffile{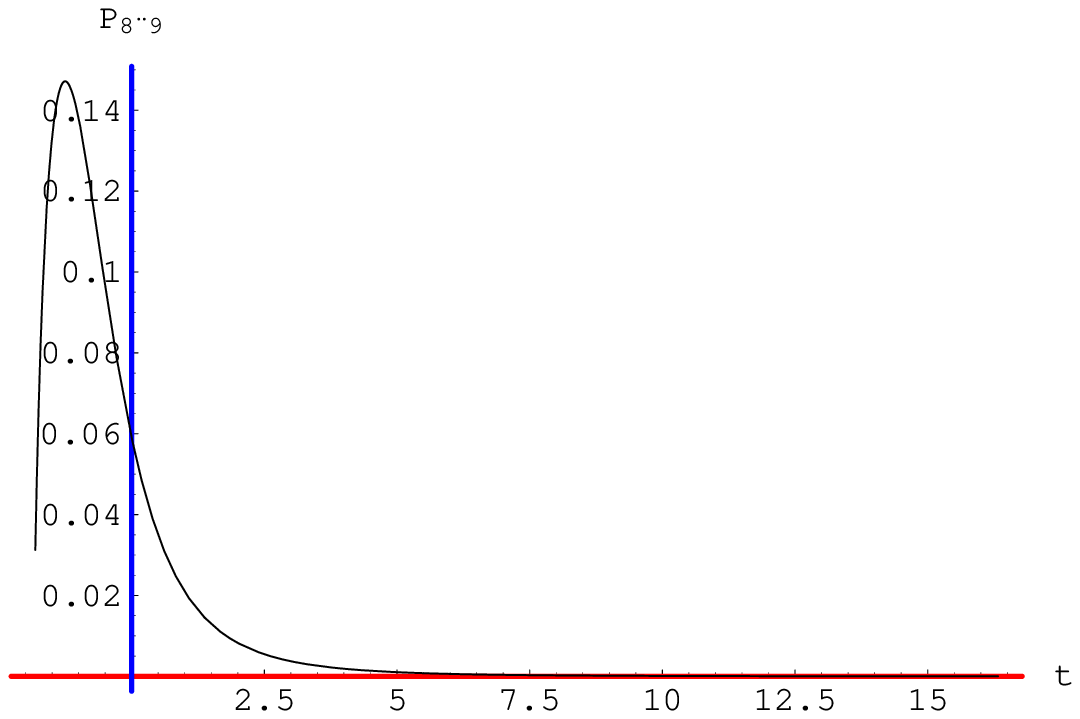}}
\caption{Plots of the pressure eigenvalues $P_{[\alpha]}$, $\alpha= 1|2 \, ,\, 3|4
\, , \, 5|6|7 \, , \,
8|9$ as functions of the cosmic time $t=\tau(T)$ in the case of the choice of
parameters $\omega =1$, $\kappa=0.5$
and for the $A_2$ solution with only the highest root switched on.
\label{pressuro1k05}}
\hskip 1.5cm \unitlength=1.1mm
\end{center}
\end{figure}
\fi In order to enlighten the physical meaning of the type II B
superstring backgrounds we have eventually constructed it is worth
to analyze the structure of the stress energy tensor. First,
reintroducing the missing traces we define:
\begin{eqnarray}
T_{MN}^{[D3]} &=& \hat{T}_{MN}^{[D3]}-\ft 12 \, g_{MN} \,
\hat{T}_{RS}^{[D3]}\, g^{RS} \nonumber\\
T_{MN}^{[D5]} &=& \hat{T}_{MN}^{[D5]}-\ft 12 \, g_{MN} \,
\hat{T}_{RS}^{[D5]}\, g^{RS} \nonumber\\
T_{MN}^{tot} &=&T_{MN}^{[D3]} + T_{MN}^{[D5]}
\label{stresscontra}
\end{eqnarray}
It turns out that the stress energy tensors are diagonal, just as the
metric, and have the form of a perfect
fluid, but with different pressure eigenvalues in the various
subspaces. Indeed we can write:
\begin{eqnarray}
T_{00}^{tot,\, D3,\, D5} & = & g_{00} \, \rho^{tot,\, D3,\, D5} \nonumber\\
T_{i_\alpha j_\alpha}^{tot,\, D3,\, D5} & = & -g_{i_\alpha j_\alpha} \,
P_{\alpha}^{tot,\, D3,\, D5}
\label{definizionipress}
\end{eqnarray}
where $\alpha$ denotes the four different submanifolds extending in
directions:
\begin{equation}
  \alpha= 1|2 \, , \, 3|4 \, , \, 5|6|7 \, ,\, 8|9
\label{directions}
\end{equation}
We can now analyze the specific properties of the two example of
solutions.
\subsubsection{Properties of the solution with just one
root switched on}
In the case of the time dependent background described in section (\ref{oxide1a2})
and obtained by oxiding the solution (\ref{finsol}) we obtain for the \textit{energy
densities}:
\begin{eqnarray}
\rho^{tot} & = & \frac{{\left( 1 + e^{t\,\omega } \right) }^2\,{\kappa }^2 +
             9\,e^{t\,\omega }\,{\omega }^2}{36\,
             e^{t\,{\sqrt{\frac{{\kappa }^2}{3} + {\omega }^2}}}\,
             {\left( 1 + e^{t\,\omega } \right) }^2\,
             {\sqrt{\cosh \frac{t\,\omega }{2}}}} \nonumber\\
\rho^{d3} & = & \frac{{\omega }^2}
  {16\,e^{t\,{\sqrt{\frac{{\kappa }^2}{3} + {\omega }^2}}}\,
             {\cosh^{\frac{5}{2}} \frac{t\,\omega }{2}}}\nonumber\\
\rho^{d5}    & =&\frac{{\kappa }^2}
  {36\,e^{t\,{\sqrt{\frac{{\kappa }^2}{3} + {\omega }^2}}}\,
             {\sqrt{\cosh \frac{t\,\omega }{2}}}}
\label{energydensitya2first}
\end{eqnarray}
for the \textit{total pressures}:
\begin{eqnarray}
P_{1|2}^{tot} & = &-\frac{\kappa^2 + 9\omega^2 + \kappa^2\cosh t\omega}{144
             e^{t\sqrt{\frac{\kappa^2}{3} + \omega^2}}
             \cosh^{\frac{5}{2}}\frac{t\omega}{2}} \nonumber\\
P_{3|4}^{tot} & = & -\frac{ e^{t\left( \omega  -
 \sqrt{\frac{\kappa^2}{3} + \omega^2}\right) } \left( \kappa^2 - 9\omega^2 +
 \kappa^2\cosh t\omega\right)}{36
             {\left( 1 + e^{t\omega} \right)}^2
             \sqrt{\cosh \frac{t\omega }{2}}}\nonumber\\
P_{5|6|7}^{tot} & = & -\frac{ \kappa^2 + 9\omega^2 + \kappa^2\cosh t\omega}
{144e^{t\sqrt{\frac{\kappa^2}{3} + \omega^2}}
             \cosh^{\frac{5}{2}}\frac{t\omega}{2}}\nonumber\\
P_{8|9}^{tot} & = &-\frac{e^{t\left(\omega -\sqrt{\frac{\kappa^2}{3} +
\omega^2} \right) }
\left(\kappa^2 - 9\omega^2 +\kappa^2\cosh t\omega\right)}{36
\left( 1 + e^{t\omega} \right)^2\sqrt{\cosh \frac{t\omega}{2}}}
\label{totpressio}
\end{eqnarray}
for the \textit{pressures associated with the $D3$ brane}:
\begin{eqnarray}
P_{1|2}^{D3} & = & \frac{-{\omega }^2}
  {16\,e^{t\,{\sqrt{\frac{{\kappa }^2}{3} + {\omega }^2}}}\,
             {\cosh^{\frac{5}{2}}\frac{t\omega }{2}}} \nonumber\\
P_{3|4}^{D3} & = & \frac{{\omega }^2}
  {16\,e^{t\,{\sqrt{\frac{{\kappa }^2}{3} + {\omega }^2}}}\,
             {\cosh^{\frac{5}{2}}\frac{t\omega }{2}}}\nonumber\\
P_{5|6|7}^{D3} & = & \frac{-{\omega }^2}
  {16\,e^{t\,{\sqrt{\frac{{\kappa }^2}{3} + {\omega }^2}}}\,
             {\cosh^{\frac{5}{2}}\frac{t\,\omega }{2}}}\nonumber\\
P_{8|9}^{D3} & = & \frac{{\omega }^2}
  {16\,e^{t\,{\sqrt{\frac{{\kappa }^2}{3} + {\omega }^2}}}\,
             {\cosh^{\frac{5}{2}}\frac{t\,\omega }{2}}}
\label{d3pressio}
\end{eqnarray}
and for the pressures associated with the $D5$ brane:
\begin{eqnarray}
P_{1|2}^{D5} & = & \frac{-{\kappa }^2}
  {72\,e^{t\,{\sqrt{\frac{{\kappa }^2}{3} + {\omega }^2}}}\,
             {\sqrt{\cosh \frac{t\,\omega }{2}}}} \nonumber\\
P_{3|4}^{D5} & = & \frac{-{\kappa }^2}
  {72\,e^{t\,{\sqrt{\frac{{\kappa }^2}{3} + {\omega }^2}}}\,
             {\sqrt{\cosh \frac{t\,\omega }{2}}}}\nonumber\\
P_{5|6|7}^{D5} & = & \frac{-{\kappa }^2}
  {72\,e^{t\,{\sqrt{\frac{{\kappa }^2}{3} + {\omega }^2}}}\,
             {\sqrt{\cosh \frac{t\,\omega }{2}}}}\nonumber\\
P_{8|9}^{D5} & = & \frac{-{\kappa }^2}
  {72\,e^{t\,{\sqrt{\frac{{\kappa }^2}{3} + {\omega }^2}}}\,
             {\sqrt{\cosh \frac{t\,\omega }{2}}}}
\label{d5pressio}
\end{eqnarray}
As we see from its analytic expression the total energy density is an
exponentially decreasing function of time which tends to zero at
asymptotically late times ($t \mapsto \infty$). What happens instead at
asymptotically early
times ($t \mapsto -\infty$) depends  on the value of $\kappa$.
For $\kappa=0$ we have $\lim_{t\mapsto-\infty} \rho^{tot}(t)=0$,
while for $\kappa \ne 0$ we always have $
\lim_{t\mapsto-\infty} \rho^{tot}(t)=\infty$. This is illustrated,
for instance, in figs.(8) and (10). This phenomenon is related to the
presence or absence of a $D5$ brane as it is evident from eq.s
(\ref{energydensitya2first}) which shows that the \textit{dilaton}-($D5$) brane
contribution to the energy density  is proportional to
$\kappa^2$ and it is always divergent at asymptotically early times,
while the $D3$ brane contribution tends to zero in the same regime.
\par
We also note, comparing eq.s(\ref{d5pressio}) with
eq.s(\ref{d3pressio}) that the pressure contributed by the
\textit{dilaton}--$D5$--brane
system is the same in all directions 1--9, while the pressure contributed by the
$D3$--brane system is just opposite in the direction 3489 and in the
transverse directions 12567. This is the origin of the \textit{cosmic billiard
phenomenon} that we observe in the behaviour of the metric scale
factors.
Indeed the presence of the $D3$-brane
causes, at a certain instant of time, a switch in the cosmic
expansion. Dimensions that were previously shrinking begin to expand
and dimensions that were expanding begin to shrink. It is like a ball
that hits a wall and inverts its speed. In the exact solution that we have
constructed through reduction to three dimensions this occurs in a
smooth way. There is a maximum and respectively a minimum in the
behaviour of certain scale factors, which is in relation with a
predominance of the $D3$--brane energy density with respect to the
total energy density. The cosmic $D3$--brane behaves just as an
instanton. Its contribution to the total energy is originally almost
zero, then it raises and dominates for some time, then it
exponentially decays again. This is the smooth exact realization of
the potential walls envisaged by Damour et al.
\par
To appreciate such a behaviour it is convenient to consider some
plots of the scale factors, the energy densities and the pressures.
In order to present such plots we first reduce the metric
(\ref{a2firstmetric}) to a standard cosmological form, by introducing
a new time variable $\tau$ such that:
\begin{equation}
  r_{[0]}(t) \, dt = d\tau
\label{taudefini1}
\end{equation}
Explicitly we set:
\begin{equation}
  \tau = \int_{0}^{T} \, r_{[0]}(t) \, dt
\label{cosmotime1}
\end{equation}
and inserting the explicit form of the scale factor $r_{[0]}(t)$ as
given in eq.(\ref{a2firstscafacti}) we obtain:
\begin{eqnarray}
\tau(T) &=&\frac{4\,2^{\frac{3}{4}}}{\omega  -
             4\,{\sqrt{\frac{{\kappa }^2}{3} + {\omega }^2}}}\,\left [ {_2 F_1}\left
(-  \ft {1}
                                             {8}   + \ft {{\sqrt{\frac{{\kappa
}^2}{3} + {\omega }^2}}}
                                          {2\,\omega },-  \ft {1}{4}   ,
                                \ft {7}{8} + \ft {{\sqrt{\frac{{\kappa }^2}{3} +
{\omega }^2}}}
                                          {2\,\omega },-1 \right) \right. \nonumber\\
                                          && \left. - e^
                                 {\frac{T\,\left( -\omega  +
                                                                4\,{\sqrt{\frac{{\kappa
}^2}{3} + {\omega
}^2}} \right)
}{8}}\,
                                {_2 F_1}\left (-  \ft{1}{8}    +
                                          \ft {{\sqrt{\frac{{\kappa }^2}{3} +
{\omega }^2}}}{2\,\omega },
                                 -  \ft {1}{4}   ,
                                 \ft {7}{8} + \ft {{\sqrt{\frac{{\kappa }^2}{3} +
{\omega }^2}}}
                                            {2\,\omega },-e^{T\,\omega } \right)
\right]
\label{tau1}
\end{eqnarray}
which expresses $\tau$ in terms of hypergeometric functions and
exponentials.
\iffigs
\begin{figure}
\begin{center}
\epsfxsize =7cm
{\epsffile{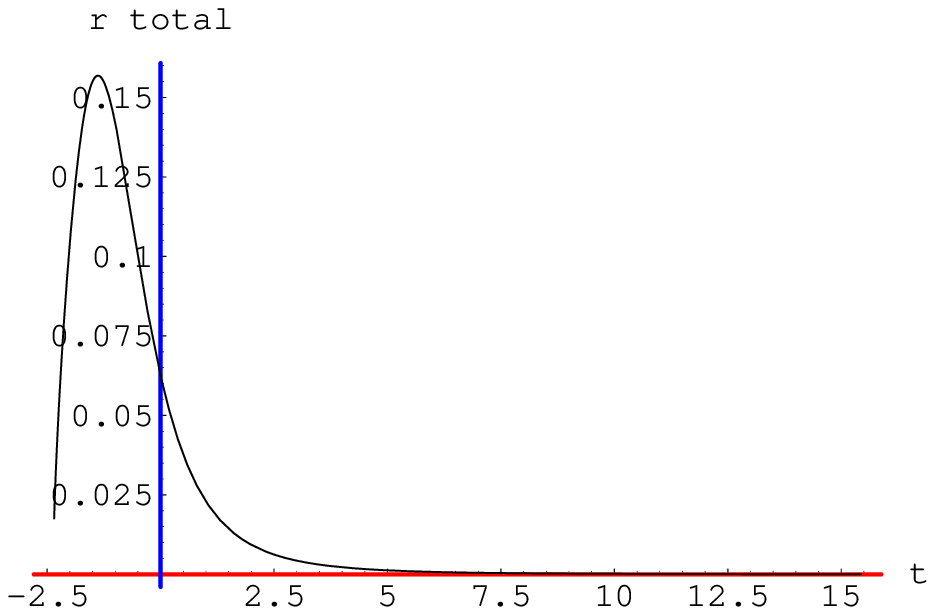}}
\caption{Plot of the  energy density as function of the cosmic time $t=\tau(T)$
in the case of a pure $D3$ brane system, namely for the the choice of parameters
$\omega =1$, $\kappa=0$
and for the $A_2$ solution with only the highest root switched on.
\label{rhototo1k0}}
\hskip 1.5cm \unitlength=1.1mm
\end{center}
\end{figure}
\fi
In fig.(\ref{radiio1k05}) we observe the billiard phenomenon in a
generic case where both parameters $\omega$ and $\kappa$ are non
vanishing. Since the value of $\omega$ can always be rescaled by a
rescaling of the original time coordinate $t$, we can just set it to $1$
and what matters is to distinguish the case $\omega \ne 0$ where the
$D3$ brane is present from the case $\omega = 0$  corresponding to
its absence. Hence fig.(\ref{radiio1k05}) corresponds to the presence
of both a $D3$--brane and a \textit{dilaton}--$D5$--brane system.
A very different behavior occurs in fig.(\ref{radiio0k1}) where
$\omega=0$. In this case there is no billiard and the dimensions
either shrink or expand uniformly.
On the other hand in fig.(\ref{radiio1k0}) we observe the pure
billiard phenomenon induced by the $D3$-brane in the case where no
\textit{dilaton}-$D5$--brane is present, namely when we set
$\kappa=0$. In this case, as we see, the parallel directions to the
Euclidean $D3$--brane, namely 3489 have exactly the same behaviour:
they first inflate and then they deflate, namely there is a maximum
in the scale factor. The transverse directions to the $D3$ brane 567
have the opposite behaviour. They display a minimum at the same point
where the parallel directions display a maximum.
In all cases the directions 12 corresponding to the spatial
directions of the three dimensional sigma model world suffer a
uniform expansion.
\par
Let us now consider the behavior of the energy densities.
In fig.(\ref{rhoo1k05}) we focus on the mixed case $\omega=1$,
$\kappa=0.5$ characterized by the presence of both a $D3$ brane and
\textit{dilaton}-$D5$--brane system. As we see the total energy
density exponentially decreases at late times and has a singularity
at asymptotically early times. This is like in a standard Big Bang
cosmological model with an indefinite expansion starting from an initial
singularity. Yet the ratio of the $D3$ energy with respect to the total
energy has a maximum at some instant of time and this is the cause of
the billiard phenomenon in the behaviour of the scale factors
respectively parallel and transverse to the $D3$ brane itself.
The two contributions to the energy density from the
$D3$--brane and from the dilaton have the same sign and the plot
of their ratio displays a maximum in correspondence with the billiard time.
With the same choice of parameters $\omega=1$,
$\kappa=0.5$ the physical behavior of the system can be appreciated
by looking at the plots of the pressure eigenvalues. They are
displayed in fig.(\ref{pressuro1k05}). We observe that the pressure
is negative in the directions transverse to the $D3$ brane 12 and
567. Slowly, but uniformly it increases to zero in these directions.
In the directions parallel to the brane the pressure is instead always positive
and it displays a sharp maximum at the instant of time where the
billiard phenomenon occurs.
\iffigs
\begin{figure}
\begin{center}
\epsfxsize =7cm
{\epsffile{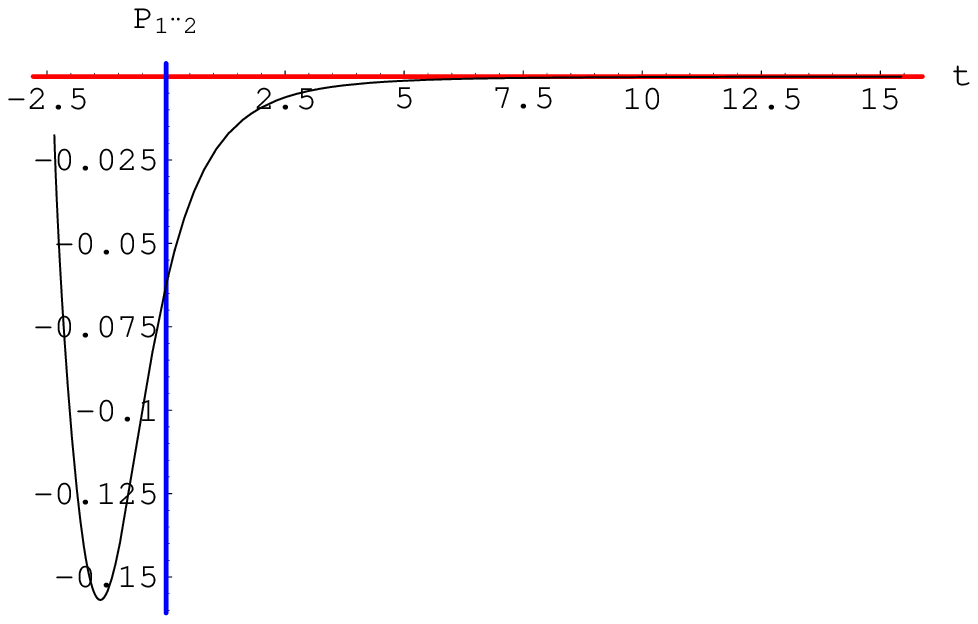}}
\epsfxsize =7cm
{\epsffile{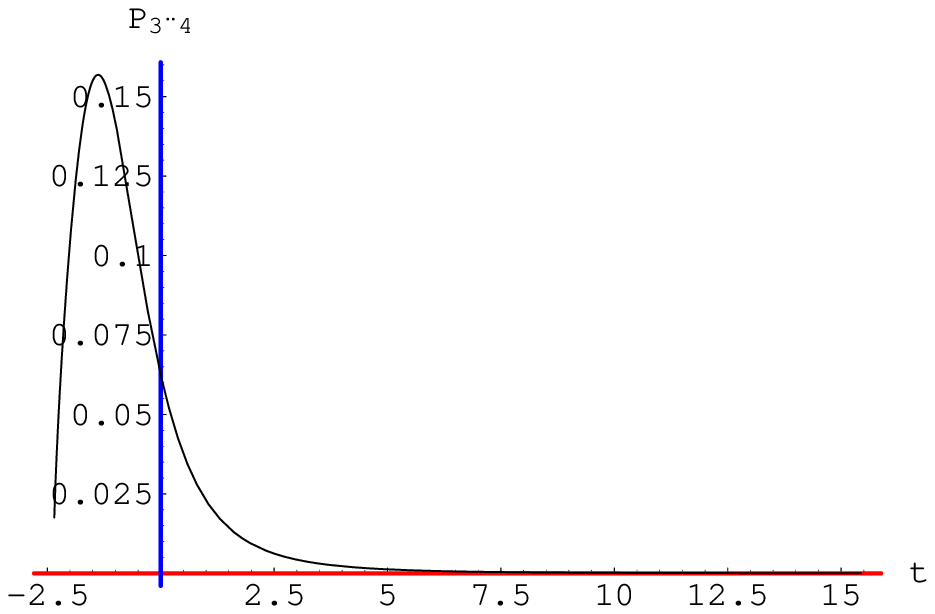}}
\epsfxsize =7cm
{\epsffile{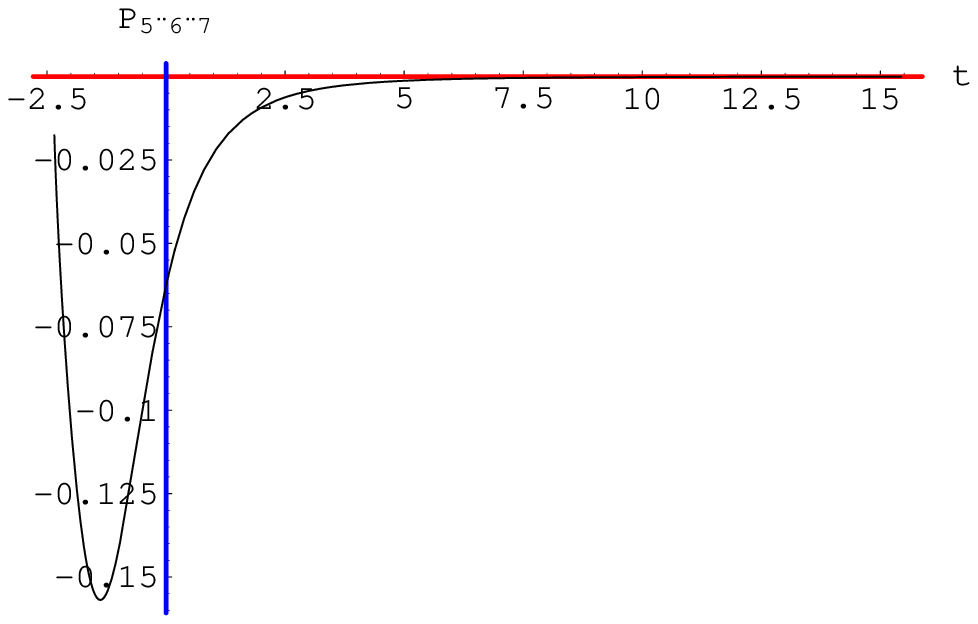}}
\epsfxsize =7cm
{\epsffile{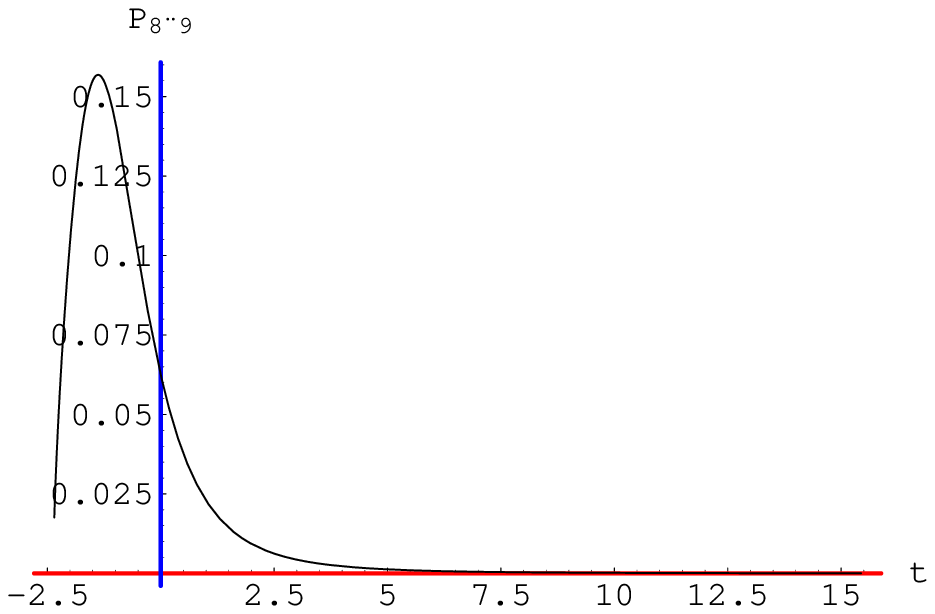}}
\caption{Plots of the pressure eigenvalues $P_{[\alpha]}$, $\alpha= 1|2 \, ,\, 3|4
\, , \, 5|6|7 \, , \,
8|9$ as functions of the cosmic time $t=\tau(T)$ in the case of the choice of
parameters $\omega =1$, $\kappa=0$
and for the $A_2$ solution with only the highest root switched on.
This case corresponds to a pure $D3$ brane system.
\label{pressuro1k0}}
\hskip 1.5cm \unitlength=1.1mm
\end{center}
\end{figure}
\fi
For a pure $D3$ brane system, namely for $\kappa=0$ and $\omega=1$
the energy density starts at zero, develops a maximum and then decays
again to zero. This can be seen in fig.(\ref{rhototo1k0}). The plot
of the pressures is displayed, for this case in
fig.(\ref{pressuro1k0}). In this case the pressure in the directions
transverse to the brane, i.e. 12567 is negative and it is just the opposite
of the pressure in the directions parallel to the brane, namely 3489.
This behavior causes the corresponding scale factors to suffer a minimum and a
maximum, respectively.
\subsubsection{Properties of the solution with all roots
switched on}
Let us now discuss the properties of the second solution where all
the roots have been excited. In section \ref{oxide2a2} we considered the oxidation
of such a sigma model solution and we constructed the corresponding $D=10$ supergravity
background  given by the metric
(\ref{a2secondtmetric}, \ref{a2secondscafacti}) and by the field strengths
(\ref{a2secondpform}). Looking at eq.s (\ref{a2secondpform}) we see
that the interpretation of the parameter $\omega$ is still the same
as it was before, namely it represents the magnetic charge of the
dyonic $D3$-brane. At $\omega=0$ the $D3$--brane disappears. Yet it
appears from eq.s (\ref{a2secondpform}) that there is no obvious
interpretation of the parameter $\kappa$ as a pure $D5$-brane charge.
Indeed there is no choice of $\kappa$ which suppresses both the NS
and the RR $3$--form field strengths.
\par
Following the same procedure as in the previous case we calculate the energy density
and the pressures
and we separate the contributions due to the $D3$--brane and to the
\textit{dilaton}--$D5$--brane system.
After straightforward but lengthy algebraic manipulations, implemented on a computer
with MATHEMATICA
 we obtain:
\begin{eqnarray}
\rho^{tot} & = &  \frac{e^
                     {t\,\left( \frac{\omega }{4} -
                                          {\sqrt{\frac{{\kappa }^2}{3} + {\omega
}^2}} \right) }\,}{192\,
             {\left( 1 + e^{t\,\omega } \right) }^{\frac{9}{4}}\,
             {\left( 1 + e^{t\,\omega } +
                                 e^{\frac{t\,\left( \kappa  + \omega  \right) }{2}}
\right) }^
                     {\frac{9}{4}}} \, \left [
                     {\left( 1 + e^{t\,\omega } \right) }^2\,
            \left( 4 + 8\,e^{t\,\omega } + 4\,e^{2\,t\,\omega } \right.\right.
\nonumber\\
  && \left. \left.+
                     20\,e^{\frac{t\,\left( \kappa  + \omega  \right) }{2}} +
                     e^{t\,\left( \kappa  + \omega  \right) } +
                     20\,e^{\frac{t\,\left( \kappa  + 3\,\omega  \right) }{2}}
\right) \,{\kappa }^2
\right. \nonumber\\
            && \left. - 6\,e^
             {\frac{t\,\left( \kappa  + \omega  \right) }{2}}\,
            \left( -1 + e^{2\,t\,\omega } \right) \,
            \left( 6 + 6\,e^{t\,\omega } +
                     e^{\frac{t\,\left( \kappa  + \omega  \right) }{2}} \right) \,
            \kappa \,\omega \right. \nonumber\\
            && \left. + 3\,\left( 16\,e^{t\,\omega } +
                     32\,e^{2\,t\,\omega } + 16\,e^{3\,t\,\omega } +
                     8\,e^{\frac{t\,\left( \kappa  + \omega  \right) }{2}} +
                     3\,e^{t\,\left( \kappa  + \omega  \right) } \right.\right.
\nonumber\\
             && \left.\left. +
                     26\,e^{t\,\left( \kappa  + 2\,\omega  \right) } +
                     40\,e^{\frac{t\,\left( \kappa  + 3\,\omega  \right) }{2}} +
                     3\,e^{t\,\left( \kappa  + 3\,\omega  \right) } +
                     40\,e^{\frac{t\,\left( \kappa  + 5\,\omega  \right) }{2}} +
                     8\,e^{\frac{t\,\left( \kappa  + 7\,\omega  \right) }{2}}
\right) \,
            {\omega }^2
                        \right] \nonumber\\
\rho^{d3} & = &   \frac{{\omega }^2}
  {4\,e^{t\,\left( \frac{-5\,\omega }{4} +
                                          {\sqrt{\frac{{\kappa }^2}{3} + {\omega
}^2}} \right) }\,
             {\left( 1 + e^{t\,\omega } \right) }^{\frac{5}{4}}\,
             {\left( 1 + e^{t\,\omega } +
                                 e^{\frac{t\,\left( \kappa  + \omega  \right) }{2}}
\right) }^
                     {\frac{5}{4}}}\nonumber\\
\rho^{d5}    & =&  \frac{e^
                     {t\,\left( \frac{\omega }{4} -
                                          {\sqrt{\frac{{\kappa }^2}{3} + {\omega
}^2}} \right) }\,}{192\,
             {\left( 1 + e^{t\,\omega } \right) }^{\frac{9}{4}}\,
             {\left( 1 + e^{t\,\omega } +
                                 e^{\frac{t\,\left( \kappa  + \omega  \right) }{2}}
\right) }^
                     {\frac{9}{4}}} \, \left [ {\left( 1 + e^{t\,\omega } \right) }^2\,
            \left( 4 + 8\,e^{t\,\omega } + 4\,e^{2\,t\,\omega } +
                     20\,e^{\frac{t\,\left( \kappa  + \omega  \right) }{2}} \right.
\right. \nonumber\\
                     && \left.\left. +
                     e^{t\,\left( \kappa  + \omega  \right) } +
                     20\,e^{\frac{t\,\left( \kappa  + 3\,\omega  \right) }{2}}
\right) \,{\kappa }^2
\right. \nonumber\\
             && \left. - 6\,e^
             {\frac{t\,\left( \kappa  + \omega  \right) }{2}}\,
            \left( -1 + e^{2\,t\,\omega } \right) \,
            \left( 6 + 6\,e^{t\,\omega } +
                     e^{\frac{t\,\left( \kappa  + \omega  \right) }{2}} \right) \,
            \kappa \,\omega \right. \nonumber\\
            &&\left. + 3\,\left( 8\,
                        e^{\frac{t\,\left( \kappa  + \omega  \right) }{2}} +
                     3\,e^{t\,\left( \kappa  + \omega  \right) } +
                     26\,e^{t\,\left( \kappa  + 2\,\omega  \right) } +
                     24\,e^{\frac{t\,\left( \kappa  + 3\,\omega  \right) }{2}}
\right.\right.\nonumber\\
                     && \left.\left. +
                     3\,e^{t\,\left( \kappa  + 3\,\omega  \right) } +
                     24\,e^{\frac{t\,\left( \kappa  + 5\,\omega  \right) }{2}} +
                     8\,e^{\frac{t\,\left( \kappa  + 7\,\omega  \right) }{2}}
\right) \,
            {\omega }^2\right]
\label{energydensitya2second}
\end{eqnarray}
We see from the above formulae that  the energy density
contributed by the  $D3$--brane system is proportional to $\omega^2$ as before and
vanishes
at $\omega=0$. However there is no choice of the parameter $\kappa$
which suppresses the \textit{dilaton}--$D5$--contribution leaving the
$D3$--contribution non--zero.
\par
The pressure eigenvalues can also be calculated just as in the previous example
but the resulting analytic formulae are quite messy and we do not
feel them worthy to be displayed. It is rather convenient to consider
a few more plots.
\par
Just as in the previous case we define the cosmic time through the
formula (\ref{cosmotime1}). In this case, however, the integral does
not lead to a closed formula in terms of special functions and we
just have an implicit definition:
\begin{equation}
 \tau(T) \, \equiv \, \int _{0}^{T} \, e^
                        {\frac{t\,\left( \frac{-\omega }{4} +
                                                      {\sqrt{\frac{{\kappa }^2}{3} +
{\omega }^2}}
                                                      \right) }{2}}\,
                     {\left( 1 + e^{t\,\omega } \right) }^{\frac{1}{8}}\,
                     {\left( 1 + e^{t\,\omega } +
                                          e^{\frac{t\,\left( \kappa  + \omega
\right) }
                                                      {2}} \right) }^{\frac{1}{8}}\,dt
\label{cosmictau2}
\end{equation}
Let us now observe from eq.s (\ref{a2secondscafacti}),
(\ref{a2secondpform}) that there are the following critical values of
the parameters:
\begin{description}
  \item[1] For $\omega=0$ and $\kappa \ne 0$ there is no $D3$--brane
  and there is just a $D$--string dual to a $D5$--brane.
  \item[2] For $\kappa = \pm \ft 3 2 \, \omega$ the scale factor in
  the directions 34 tends to a finite asymptotic value respectively
  at very early or very late times.
\end{description}
The plot of the scale factors for the choice $\omega=0$,
$\kappa=3/2$ is given in fig.(\ref{radii2o0k15})
\iffigs
\begin{figure}
\begin{center}
\epsfxsize =7cm
{\epsffile{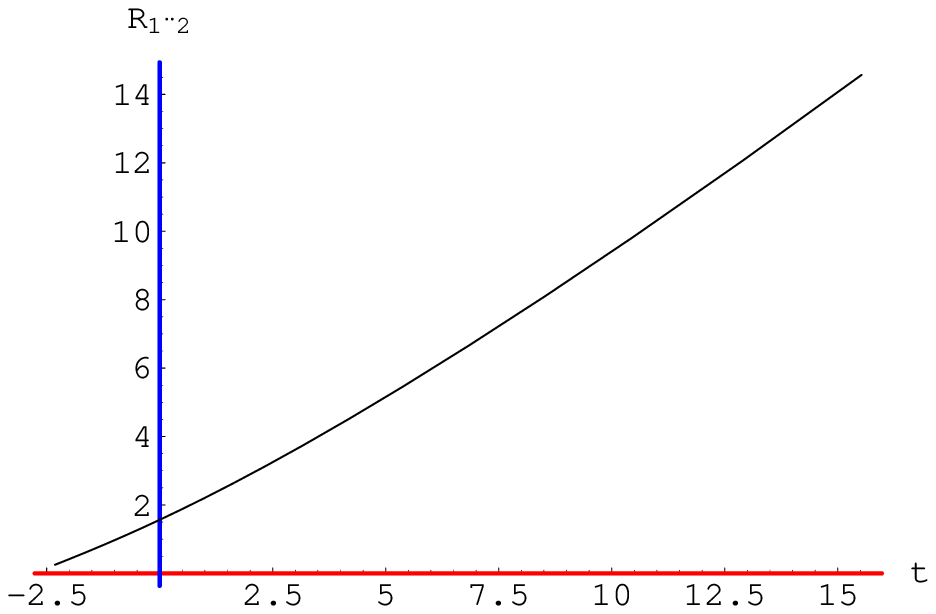}}
\epsfxsize =7cm
{\epsffile{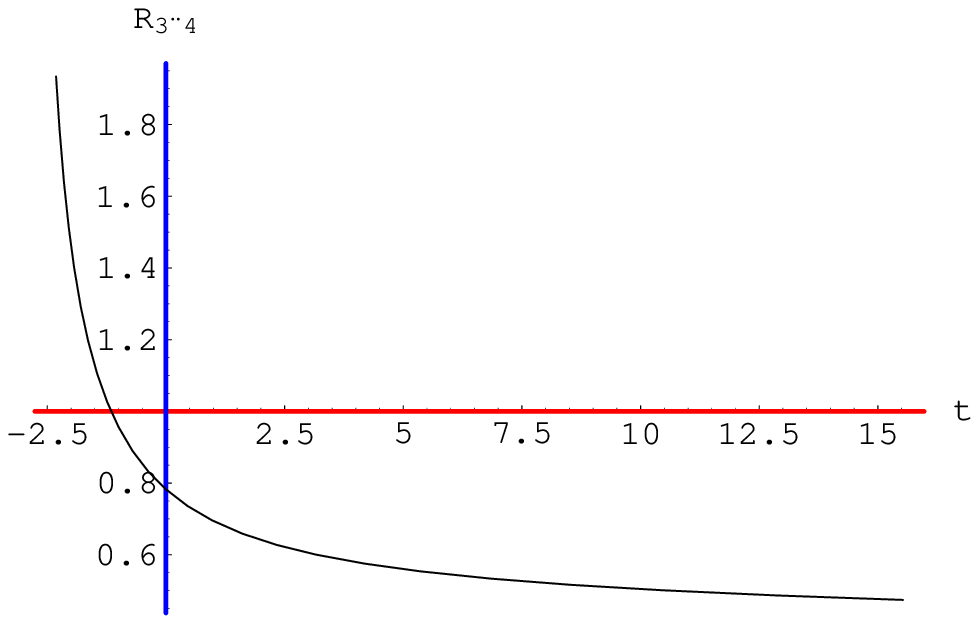}}
\epsfxsize =7cm
{\epsffile{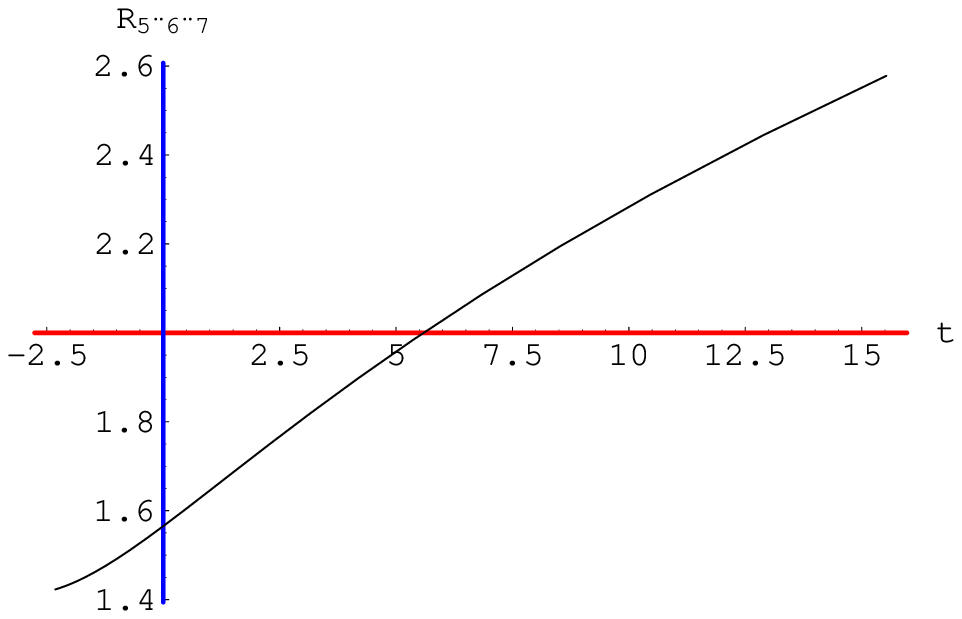}}
\epsfxsize =7cm
{\epsffile{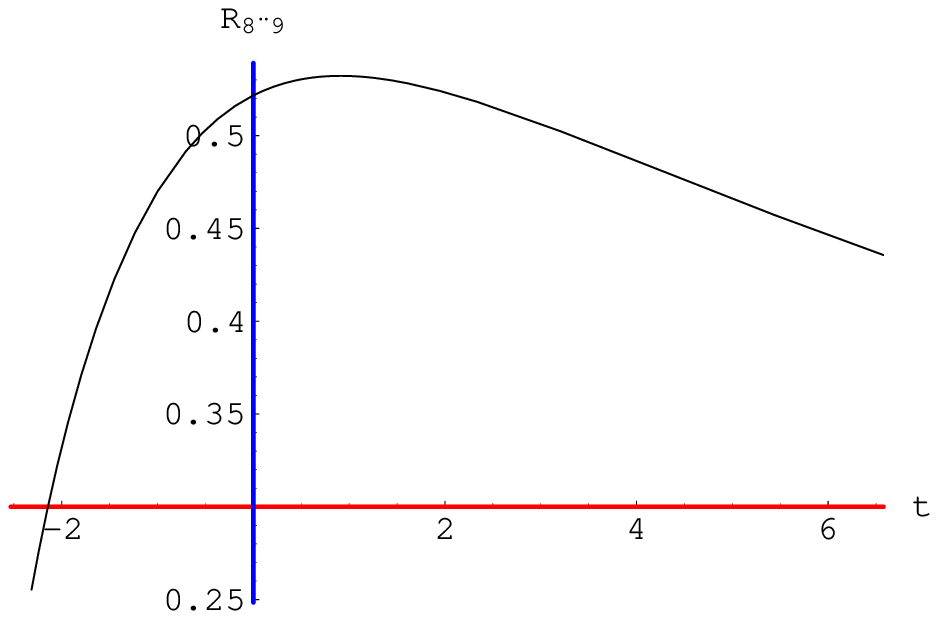}}
\caption{Plots of the scale factors $\overline{r}^2_{[\alpha]}$, $\alpha= 1|2 \, ,\,
3|4 \, , \, 5|6|7 \, , \,
8|9$ as functions of the cosmic time $t=\tau(T)$ with the parameter choice  $\omega
=0$, $\kappa=3/2$
and for the $A_2$ solution with all the  roots switched on.
\label{radii2o0k15}}
\hskip 1.5cm \unitlength=1.1mm
\end{center}
\end{figure}
\fi
\iffigs
\begin{figure}
\begin{center}
\epsfxsize =7cm
{\epsffile{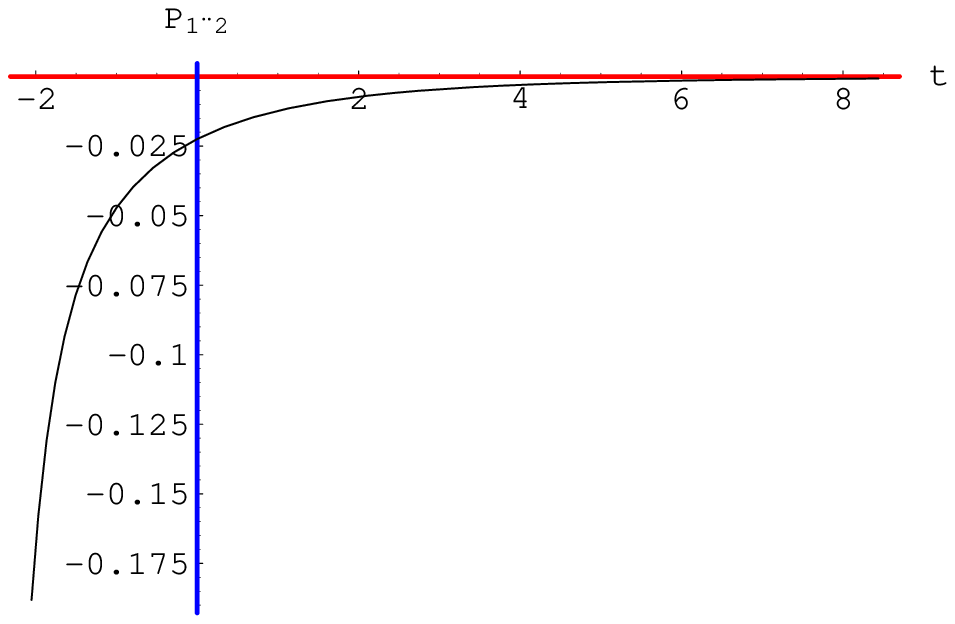}}
\epsfxsize =7cm
{\epsffile{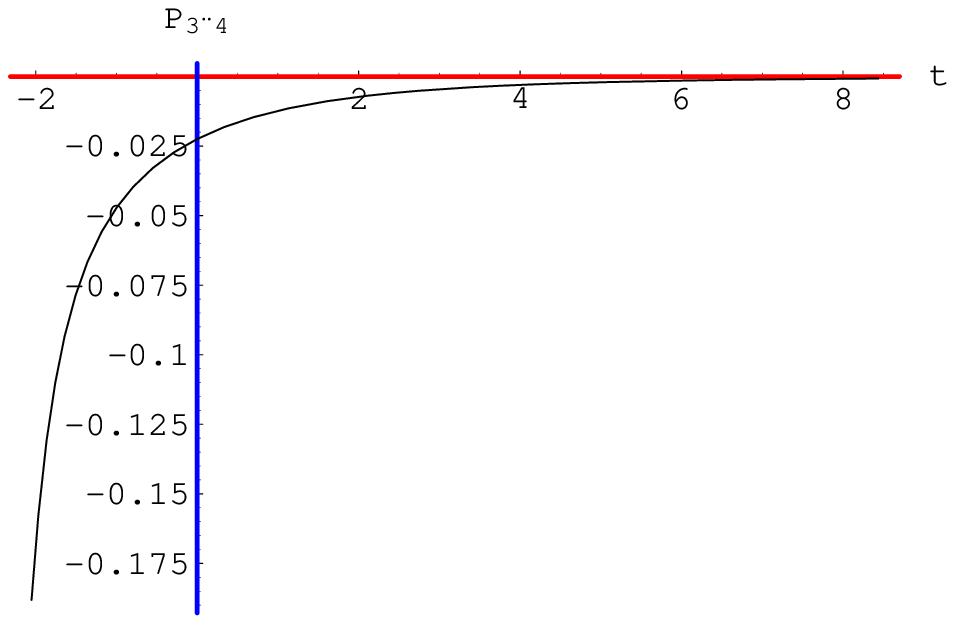}}
\epsfxsize =7cm
{\epsffile{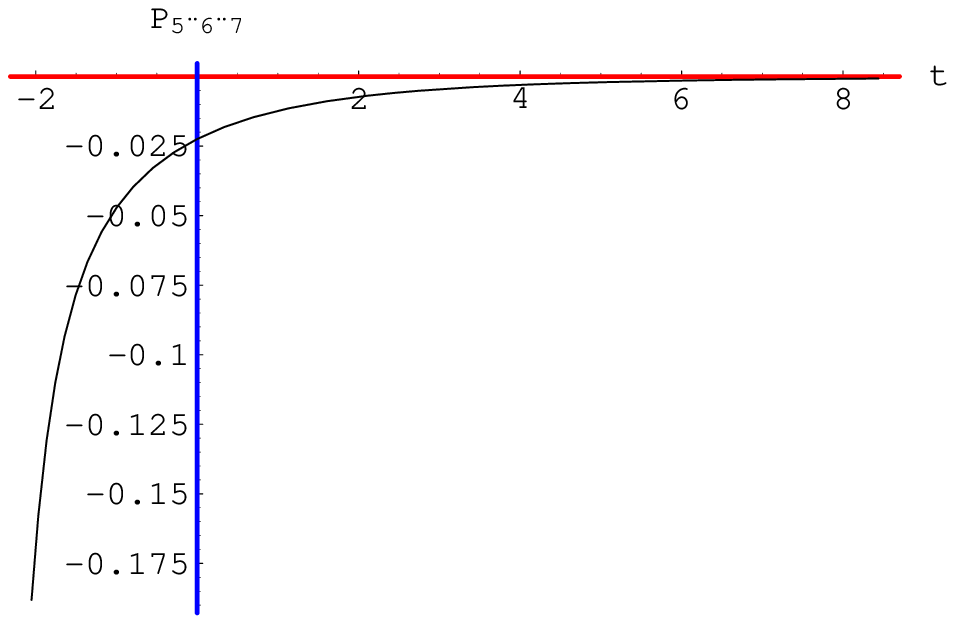}}
\epsfxsize =7cm
{\epsffile{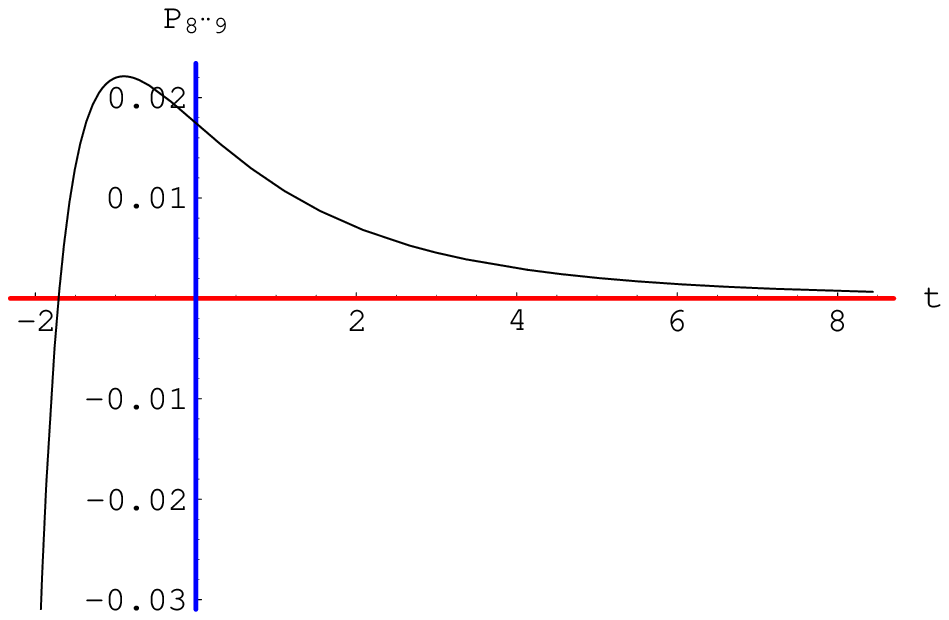}}
\caption{Plots of the pressure eigenvalues $\overline{P}_{[\alpha]}$, $\alpha= 1|2
\, ,\, 3|4 \, , \, 5|6|7 \, , \,
8|9$ as functions of the cosmic time $t=\tau(T)$ with the parameter choice  $\omega
=0$, $\kappa=3/2$
and for the $A_2$ solution with all the  roots switched on.
\label{press2o0k15}}
\hskip 1.5cm \unitlength=1.1mm
\end{center}
\end{figure}
\fi
As already stressed, this a pure $D$-string system and indeed the
billiard phenomenon occurs only in the directions 89 that correspond
to the euclidean $D$--string world--sheet. In all the other
directions there is a monotonous behavior of the scale factors.
The $D$-string nature of the solution is best appreciated by looking
at the behavior of the pressure eigenvalues, displayed in fig.(\ref{press2o0k15})
\par
As we see the positive bump in the pressure now occurs only in the
$D$--string directions 89, while in all the other directions the
pressure is the same and rises monotonously to zero from large
negative values. The pressure bump is in correspondence with the
billiard phenomenon. The energy density is instead a monotonously
decreasing function of time (see fig.(\ref{2o0k15rhotot})).
\iffigs
\begin{figure}
\begin{center}
\epsfxsize =7cm
{\epsffile{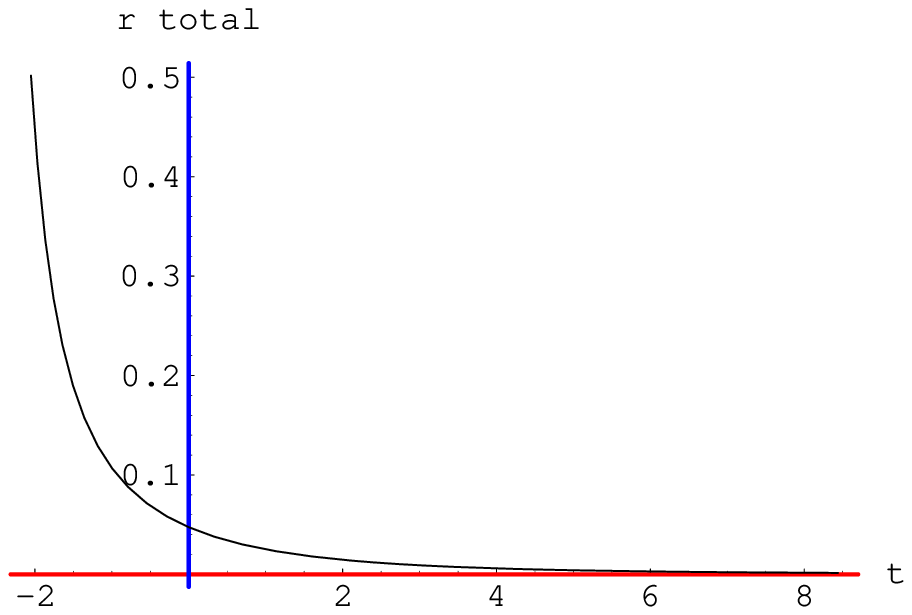}}
\caption{Plot of the  energy density as function of the cosmic time $t=\tau(T)$
with the parameter choice  $\omega =0$, $\kappa=3/2$
and for the $A_2$ solution with all the  roots switched on.
\label{2o0k15rhotot}}
\hskip 1.5cm \unitlength=1.1mm
\end{center}
\end{figure}
\fi
An intermediate case is provided by the parameter choice $\omega=1$,
$\kappa=0.8 < 3/2$. The plots of the scale factors are given in
fig.(\ref{radii2o1k08})
\iffigs
\begin{figure}
\begin{center}
\epsfxsize =7cm
{\epsffile{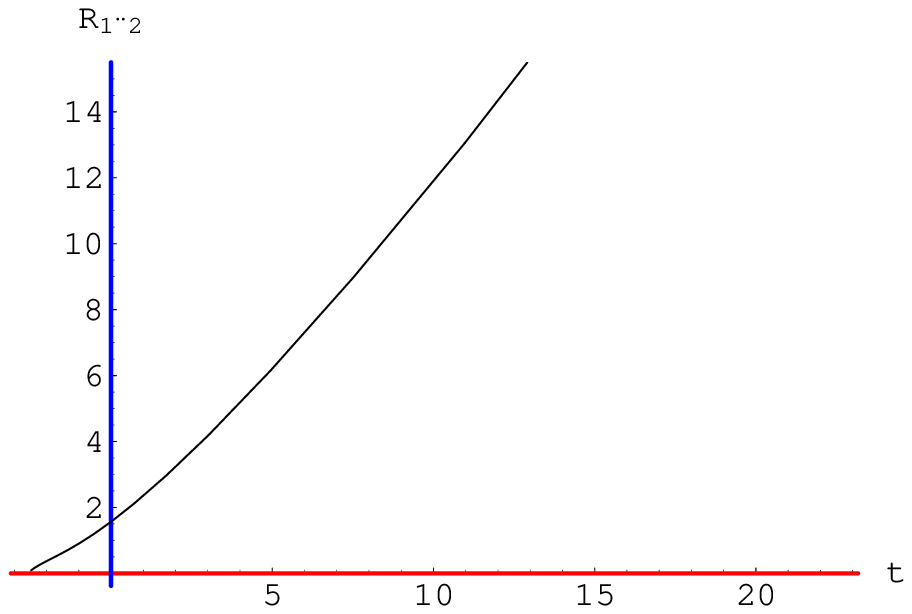}}
\epsfxsize =7cm
{\epsffile{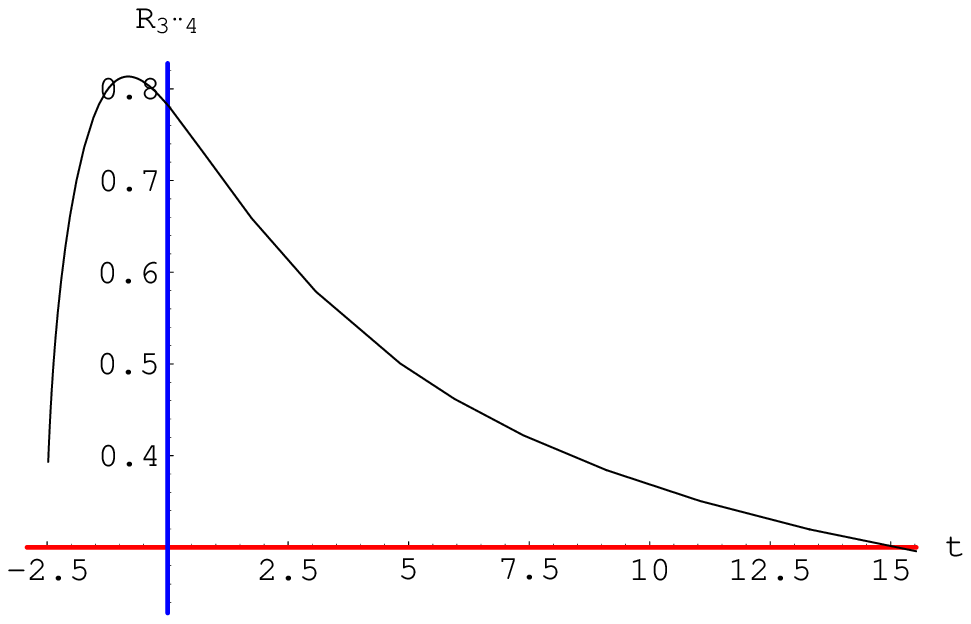}}
\epsfxsize =7cm
{\epsffile{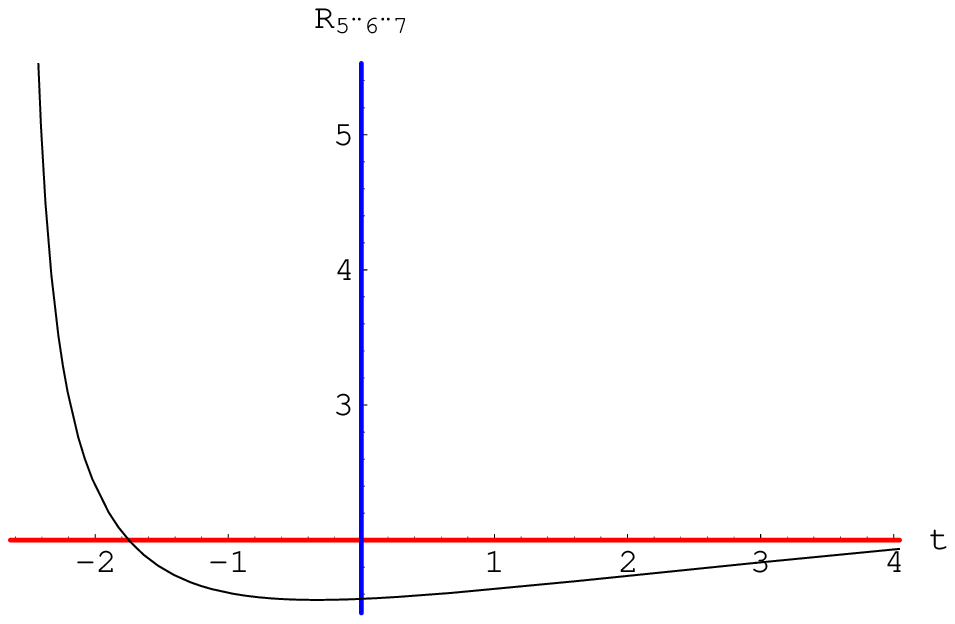}}
\epsfxsize =7cm
{\epsffile{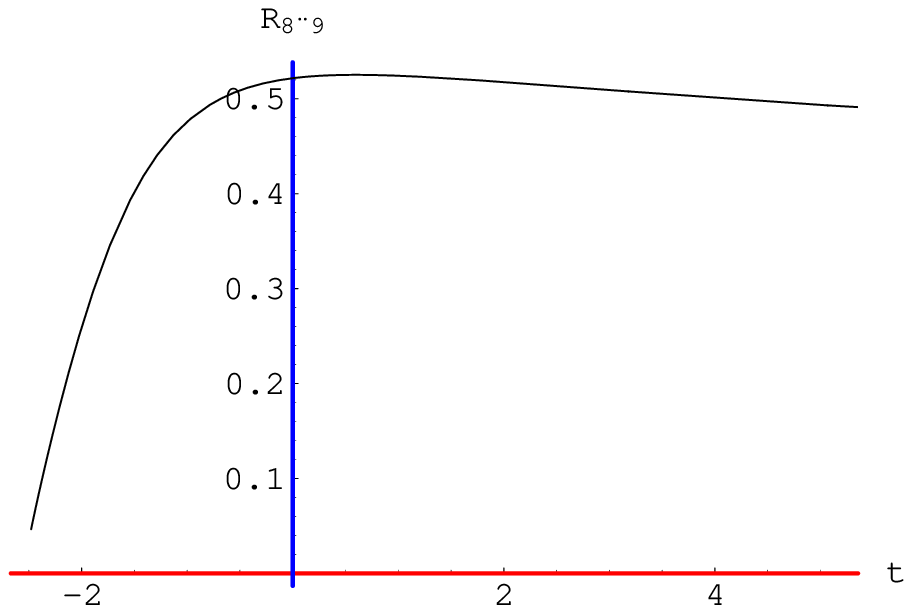}}
\caption{Plots of the scale factors $\overline{r}^2_{[\alpha]}$, $\alpha= 1|2 \, ,\,
3|4 \, , \, 5|6|7 \, , \,
8|9$ as functions of the cosmic time $t=\tau(T)$ with the parameter choice  $\omega
=1$, $\kappa=0.8$
and for the $A_2$ solution with all the  roots switched on.
\label{radii2o1k08}}
\hskip 1.5cm \unitlength=1.1mm
\end{center}
\end{figure}
\fi
The mixture of $D3$ and $D5$ systems is evident from the pictures.
Indeed we have now a billiard phenomenon in both the directions 34
and 89 as we expect from a $D3$--brane, but the maximum in 34 is much
sharper than in 89. The maximum in 89 is broader because it takes
contribution both from the $D3$ brane and from the $D$--string.
\iffigs
\begin{figure}
\begin{center}
\epsfxsize =7cm
{\epsffile{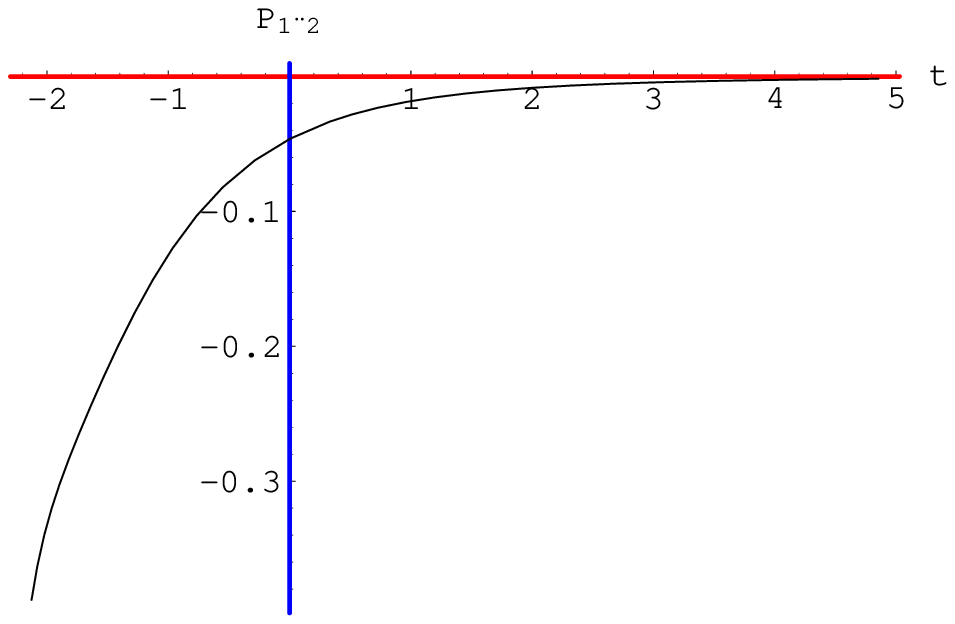}}
\epsfxsize =7cm
{\epsffile{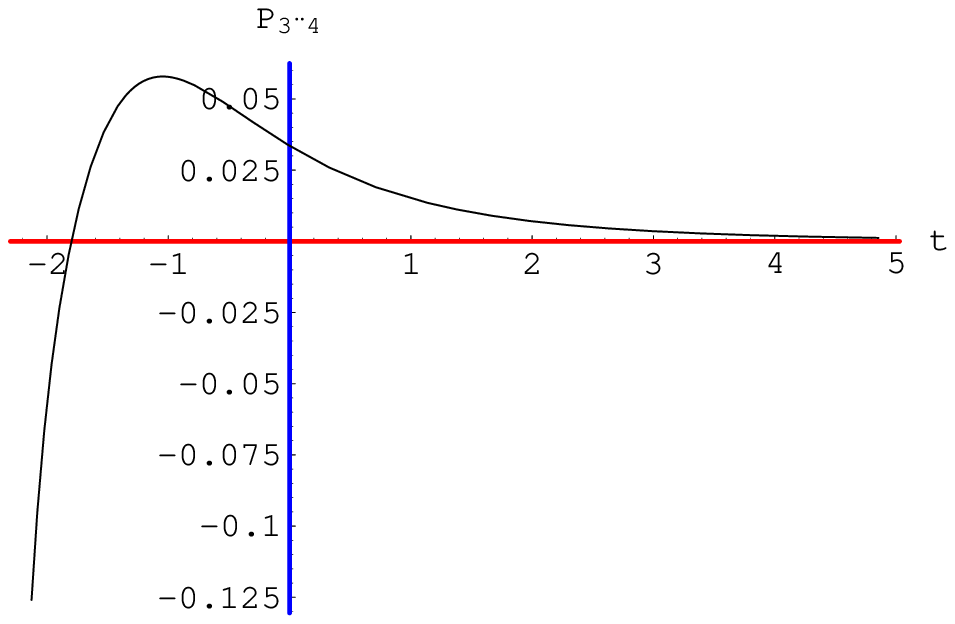}}
\epsfxsize =7cm
{\epsffile{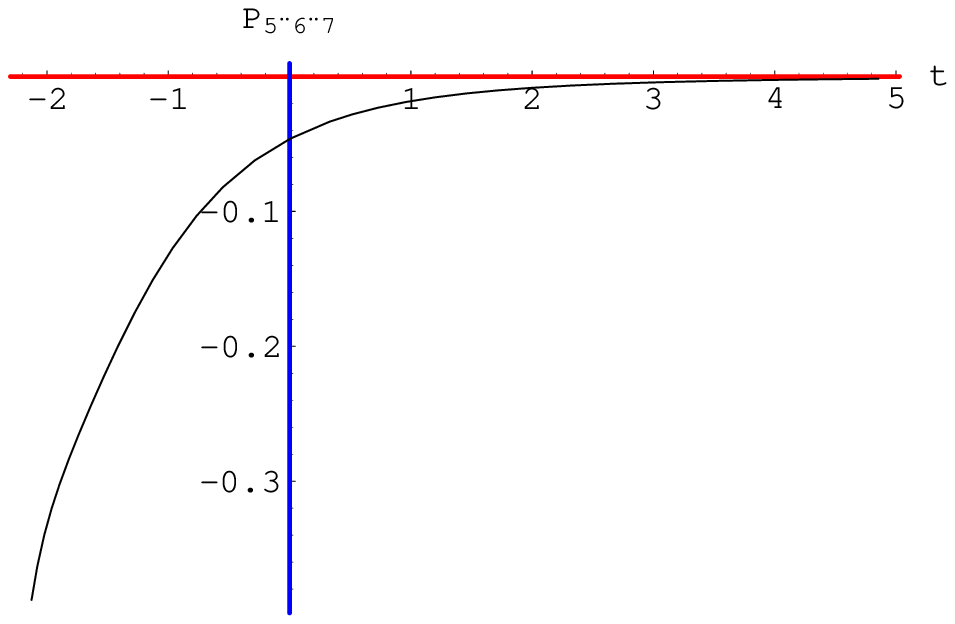}}
\epsfxsize =7cm
{\epsffile{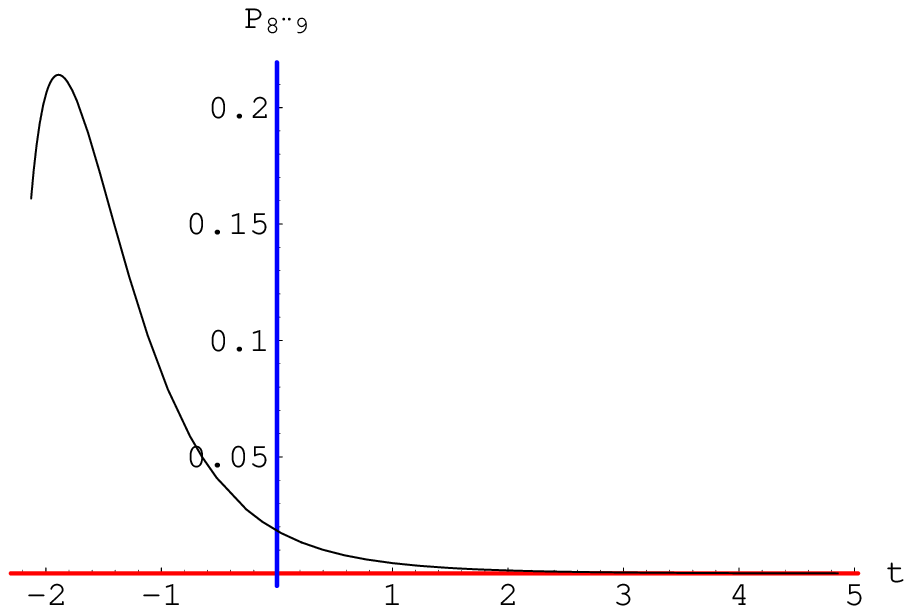}}
\caption{Plots of the pressure eigenvalues $\overline{P}_{[\alpha]}$, $\alpha= 1|2
\, ,\, 3|4 \, , \, 5|6|7 \, , \,
8|9$ as functions of the cosmic time $t=\tau(T)$ with the parameter choice  $\omega
=1$, $\kappa=0.8$
and for the $A_2$ solution with all the  roots switched on.
\label{press2o1k08}}
\hskip 1.5cm \unitlength=1.1mm
\end{center}
\end{figure}
\fi
The phenomenon is best appreciated by considering the plots of the
pressure eigenvalues (see figs. (\ref{press2o1k08}))
and of the energy
density (see figs.  (\ref{energ2o1k08})). In the pressure plots we see
that there is a positive bump both in the directions 34 and 89, yet
the bump in 89 is anticipated at earlier times and it is bigger
than the bump in 34, the reason being the cooperation between the
$D3$--brane and $D$-string contributions. Even more instructive is
the plot of the energy densities.
\iffigs
\begin{figure}
\begin{center}
\epsfxsize =7cm
{\epsffile{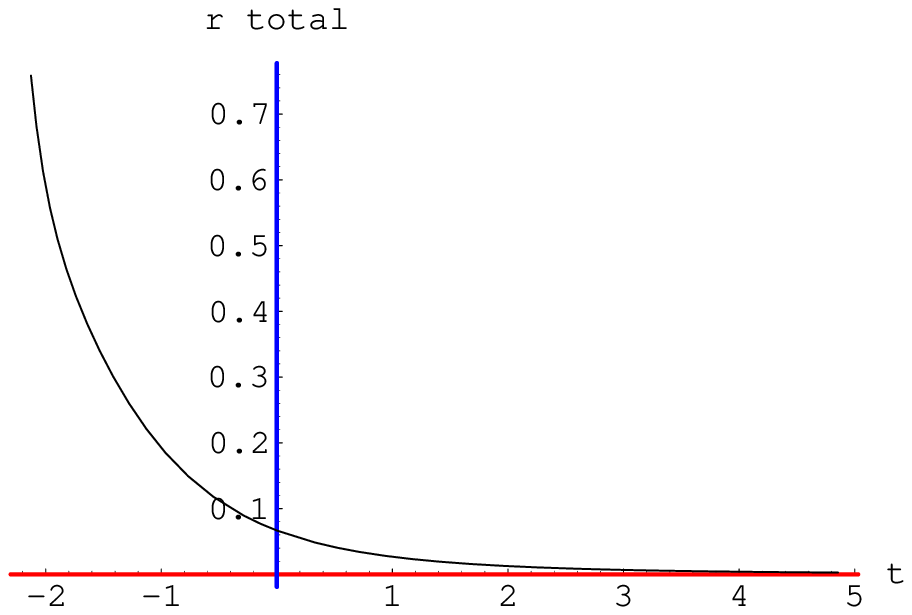}}
\epsfxsize =7cm
{\epsffile{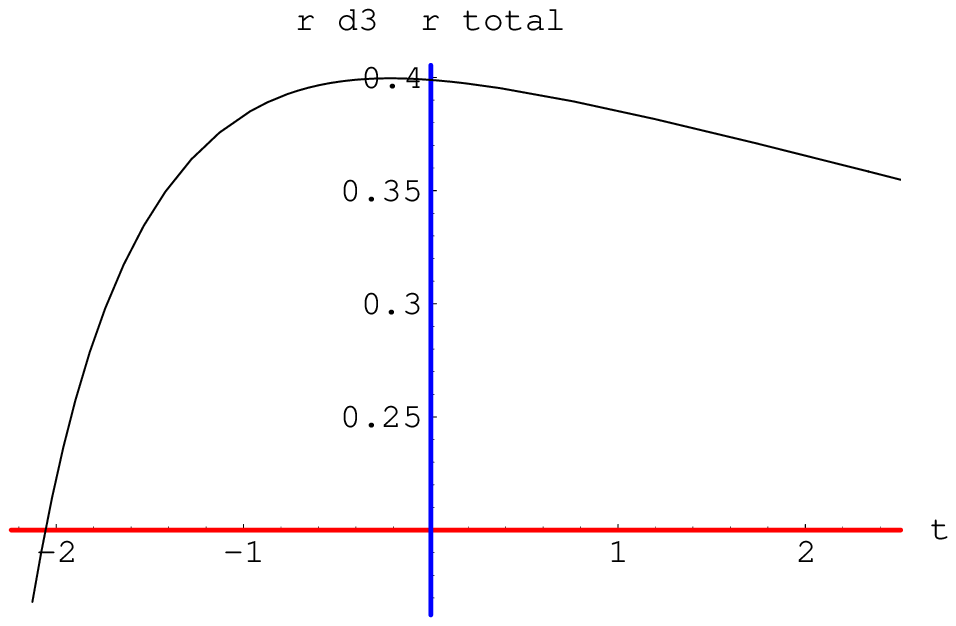}}
\epsfxsize =7cm
{\epsffile{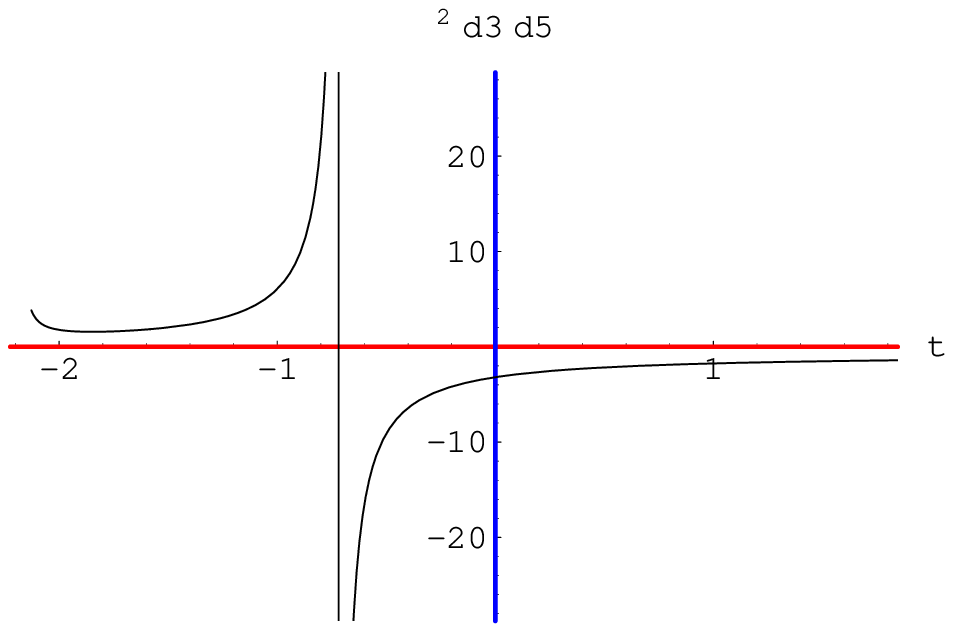}}
\epsfxsize =7cm
{\epsffile{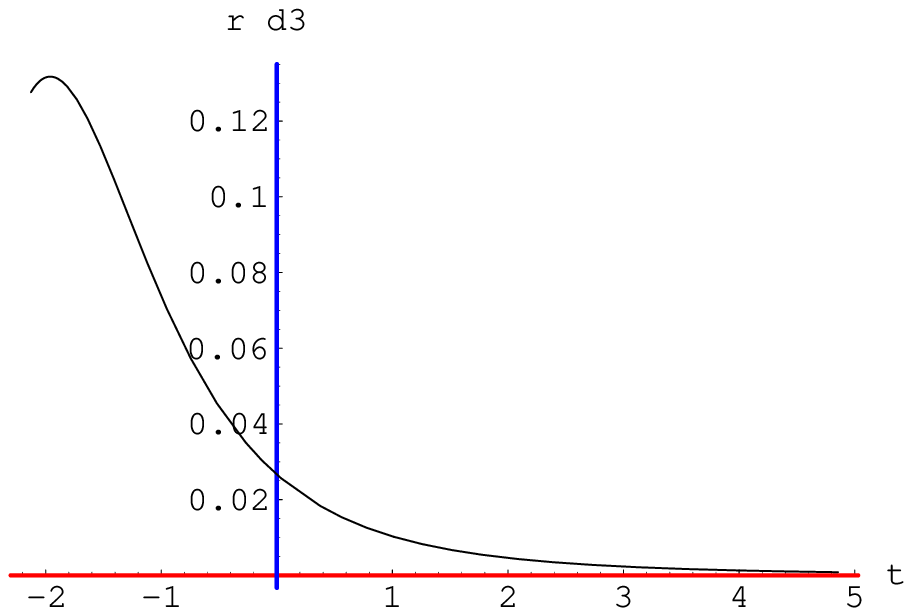}}
\epsfxsize =7cm
{\epsffile{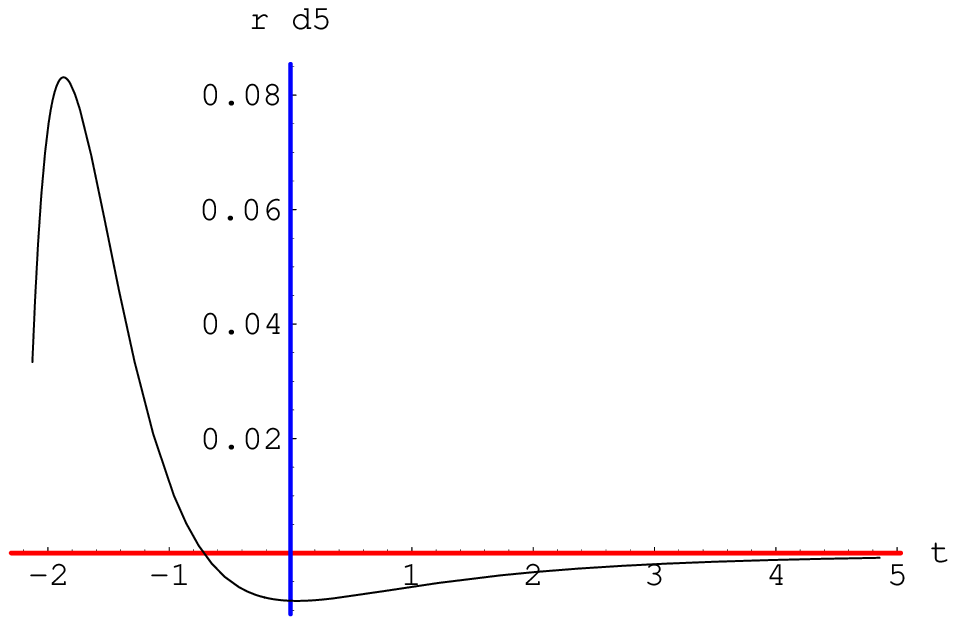}}
\caption{Plots of the energy densities as functions of the cosmic time $t=\tau(T)$
with the parameter
choice  $\omega =1$, $\kappa=0.8$
and for the $A_2$ solution with all the  roots switched on. The first
picture plots the behaviour of the total energy density. The second
plots the ratio of the $D3$--brane contribution to the energy density
with respect to the total density. The third plots the ratio of the
$D3$--brane contribution with respect to the contribution of the
dilaton $D5$--brane system. The fourth and the fifth picture plot the
energy density of the $D3$--brane and of the \textit{dilaton}--$D$--string systems,
respectively.
\label{energ2o1k08}}
\hskip 1.5cm \unitlength=1.1mm
\end{center}
\end{figure}
\fi
In fig.(\ref{energ2o1k08}) we see that the energy density of the
$D3$--brane has the usual positive bump, while the energy density of
the \textit{dilaton}--$D$--string system has a positive bump followed
by a smaller negative one, so that it passes through zero.

At the critical value $\kappa=\ft 32 \omega$ something very interesting occurs
in the behaviour of the scale factors.
\iffigs
\begin{figure}
\begin{center}
\epsfxsize =7cm
{\epsffile{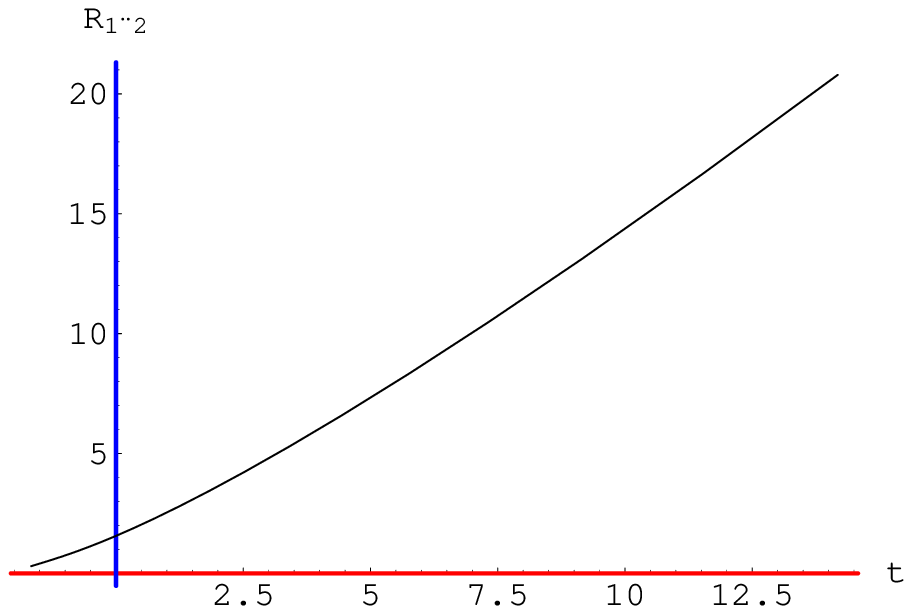}}
\epsfxsize =7cm
{\epsffile{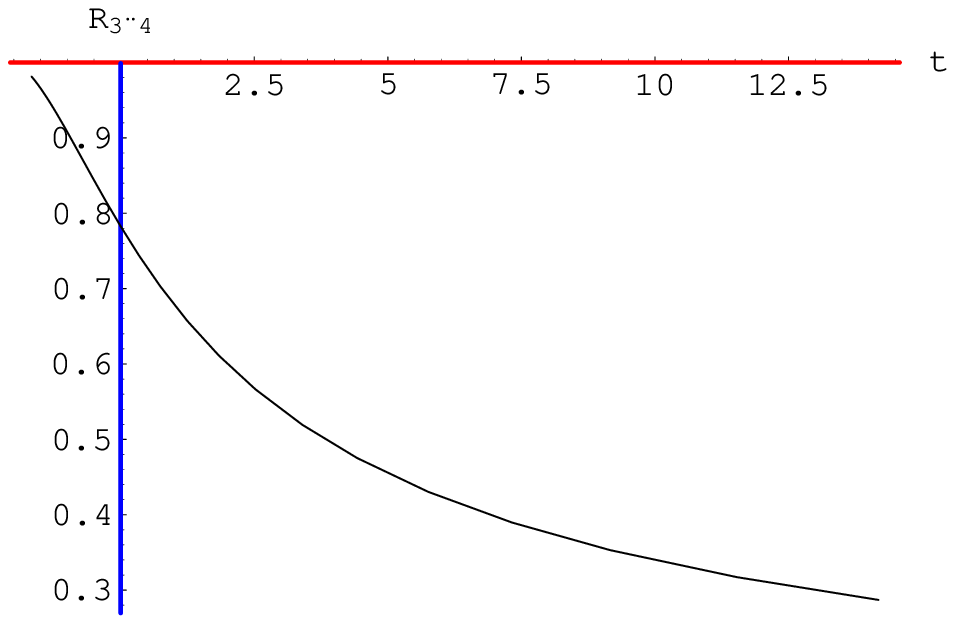}}
\epsfxsize =7cm
{\epsffile{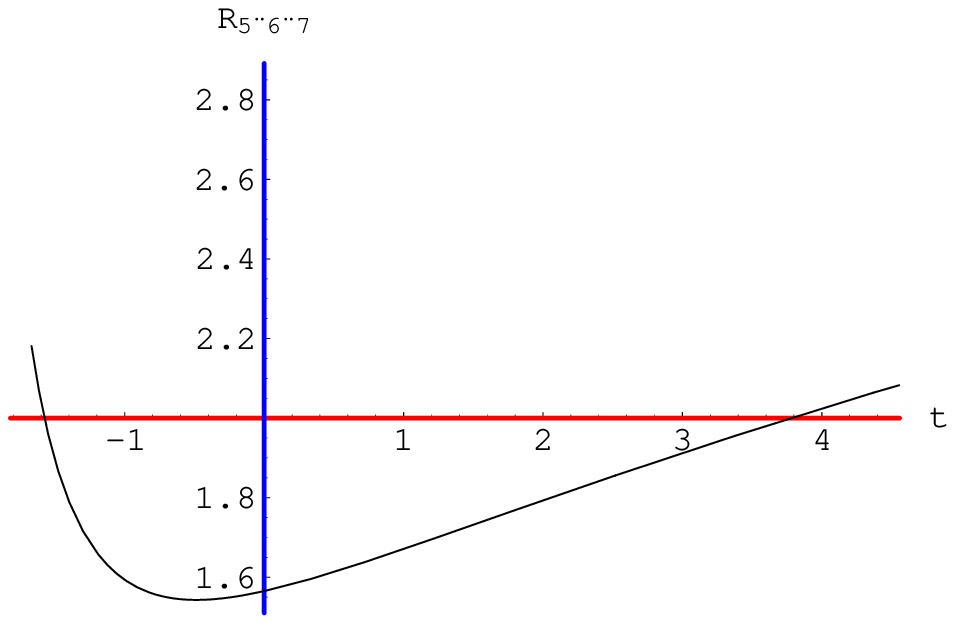}}
\epsfxsize =7cm
{\epsffile{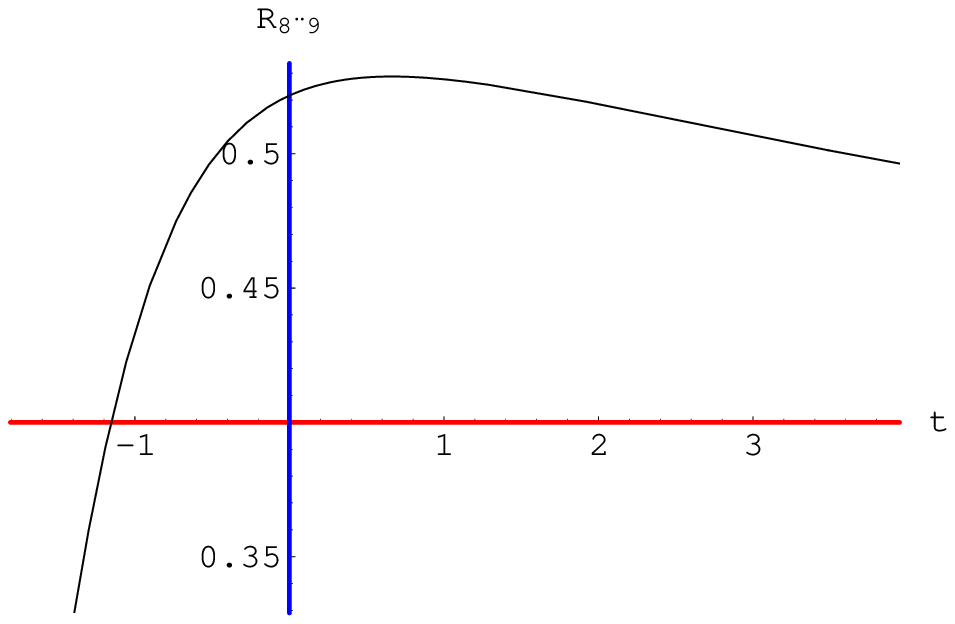}}
\caption{Plots of the scale factors $\overline{r}^2_{[\alpha]}$, $\alpha= 1|2 \, ,\,
3|4 \, , \, 5|6|7 \, , \,
8|9$ as functions of the cosmic time $t=\tau(T)$ with the critical choice of
parameters $\omega =1$, $\kappa=3/2$
and for the $A_2$ solution with all the  roots switched on.
\label{radiio1k15}}
\hskip 1.5cm \unitlength=1.1mm
\end{center}
\end{figure}
\fi
As we see from fig.(\ref{radiio1k15}), the scale factor in the
direction 34, rather than starting from zero as in all other cases
starts from a finite value and then always decreases without
suffering a billiard bump. The bump is only in the scale factor 89.
Essentially this means that the positive energy of the $D3$ brane and
the negative one of the $D$-string exactly compensate at the origin
of time for these critical value of the parameters.
\par
\subsubsection{Summarizing the above discussion and the cosmological billiard}
Summarizing what we have learned from the numerical analysis of
the type II B cosmological backgrounds obtained by a specific
oxidation of the $A_2$ sigma model solutions we can say what
follows.
\par
The expansion or contraction of the cosmological scale factors in the
diagonal metric is driven by the presence of euclidean $D$--branes
which behave like instantons (S-branes). Their energy--density and charge are
localized functions of time. Alternatively we see that these branes
contribute rather sharp bumps in the eigenvalues of the spatial part
of the stress--energy tensor which we have named \textit{pressures}.
Typically there are \textit{maxima} of these pressures in the space
directions parallell to the euclidean brane world--volume and
\textit{minima} of the same in the directions transverse to the
brane. These maxima and minima in the pressures correspond to maxima
and minima of the scale factors in the same directions. Such
inversions in the rate of expansion/contraction of the scale factors
is the \textit{cosmological billiard} phenomenon originally envisaged
by Damour et al. In the toy $A_2$ model we have presented, we observe
just one scattering, but this is due to the insufficient number of
branes (roots in the Lie algebra language) that we have excited.
Indeed it is like we had only one wall of a Weyl chamber. In
subsequent publications we plan to study the phenomenon in more
complex situations with more algebraic roots switched on. What is
relevant in our opinion is that we were able to see the postulated
\textit{bumping phenomenon} in the context of exact smooth solutions rather than in
asymptotic limiting regimes.
\section{Conclusions and Perspectives}
\label{inconcludo} The main purpose of our analysis was to develop
a convenient mathematical framework within three dimensional
(ungauged) maximal supergravity where to study homogeneous
cosmological solutions of type II A or II B theories. Our approach
exploits the correspondence between homogeneous time--dependent
solutions in ten and three dimensions. This mapping is realized
through \emph{toroidal dimensional reduction} from $D=10$ to $D=3$
or through \emph{oxidation} from $D=3$ to $D=10$. The starting
point of our study was the  ${\rm E}_{8(8)}$ orbit described by
three dimensional homogeneous solutions and we defined the precise
method for constructing a generic representative of the orbit
 from the generating solution which is
defined only by the radii of the internal seven--torus and the
dilaton. Exploiting the solvable Lie algebra (or Iwasawa)
representation of the scalar manifold in the three dimensional
theory  it was possible to control the ten dimensional
interpretation of the various bosonic fields. This allows for
instance to construct a ten dimensional solution characterized by
certain (off--diagonal) components of the metric or of  the tensor
fields by oxiding a three dimensional solution in which the scalar
fields associated with the corresponding ${\rm E}_{8(8)}$ roots
are switched on. As an example we have worked out in three
dimensions the general homogeneous time--dependent solution of an
$A_2$ model in which the scalar fields span a ${\rm
SL}(3,\mathbb{R})/{\rm SO}(3)$ submanifold of ${\rm E}_{8(8)}/{\rm
SO}(16)$. It was  shown that, depending on the embedding of ${\rm
SL}(3,\mathbb{R})$ within ${\rm E}_{8(8)}$ the ten dimensional
solution obtained upon oxidation of the three dimensional one can
have radically different physical interpretations in terms of ten
dimensional fields. We have then focused on one particular
embedding for which the axionic fields are interpreted as the
components $B_{34}$, $C_{89}$ and $C_{3489}$ of the type II B
tensor fields, and accomplished the oxidation of the solution to
ten dimensions. Its behavior, which has been described in detail
in the previous section, is characterized by an exchange of energy
between the tensor fields and the gravitational field which
results in consecutive phases of expansion and contraction of the
cosmological scale factors along the directions defined by the non
vanishing components of the tensor fields (the ``bouncing"
behavior of the scale factors in this kind of solutions was also
analyzed  by Emparan and Arriga in a different perspective
\cite{otherSbranes}). This background could be interpreted
microscopically in terms of a system of space--like or
\emph{S--branes} (or SD--branes)
\cite{Gutperle:2002ai,otherSbranes} along the directions $89$ and
$3489$, coupled to the Kalb--Ramond field.\par It is interesting
to make contact with the \emph{cosmological billiard} phenomenon
describing the behavior of solutions to Einstein equations near
space--like singularities. In this limit
 the evolution of the scale--factors/dilatons is described by a
null trajectory in a hyperbolic space which is reflected by
\emph{walls} or hyper--surfaces where the energy density of the
axionic fields diverges. Although ours is a different kind of
analysis which aims at the construction of exact smooth
cosmological solutions, we may retrieve  a similar
\emph{qualitative} description of the evolution of the
scale--factors/dilaton in relation to the evolution of the axionic
fields. In our formalism the logarithm $\sigma_i$ of the scale
factors associated with the internal directions together with the
ten dimensional dilaton $\phi$ are described by the vector $h(t)$
in the Euclidean eight--dimensional space of the ${\rm E}_{8(8)}$
Cartan subalgebra. The kinetic term of an axion $\chi $ associated
with the positive ${\rm E}_{8(8)}$ root $\alpha$ contains the
characteristic exponential factor $\exp{(-2\, \alpha\cdot h)}$.
The corresponding wall in the space of $h$ is defined by the
equation $ \alpha\cdot h= 0$ and the \emph{billiard} region by
$\alpha\cdot h\ge 0$. As it can be inferred from our solution, the
evolution of $h(t)$ is such that, if we denote by $h^\|$ the
component of $h$ along $\alpha$ and by $h^\perp$ its projection on
the hyperplane perpendicular to $\alpha$ (the wall), as the energy
density of $\chi$ reaches its maximum (this temporal region
corresponds to the \emph{thickness} of the S--branes in the $A_2$
solution) $h$ undergoes a reflection. This is most easily
illustrated for example in the $A_2$ solution with just one root
switched on (namely $\alpha[80]$). In this case the component of
$\dot{h}$ parallel to $\alpha[80]$ undergoes a  continuous sign
inversion from negative values to positive ones while
$|\dot{h}^\perp|$ is constant and proportional to the time
derivative of the dilaton :
\begin{eqnarray}
\dot{h}^\|\rightarrow -\dot{h}^\| \,\,\,;\,\,\,|\dot{h}^\perp|\,=\,
\frac{k}{2\sqrt{6}}\,\propto\,\dot{\phi}\,=\,\mbox{const.}
\end{eqnarray}
 The roots in the $A_2$ system are not enough to define
a finite volume billiard which would result in an oscillatory
behavior of the solution. Indeed, using for $h$ the
parametrization in terms of the variables $x,\,y$, namely
$h=\{x,x,y,y,y,-y,-y,y\}$, the billiard region is defined by the
dominant walls $\alpha[69]\cdot h=0,\,\alpha[15]\cdot h=0$:
\begin{eqnarray}
x&\ge& 0\,;\,\,\,y\le -\frac{x}{3}
\end{eqnarray}
which is an open region. This explains the Kasner--like (non--oscillatory) behavior of our solution for $t\rightarrow
+\infty$.\par As a possible direction for future investigations,
it would be interesting to extend the solution--constructing
technology developed in the present paper to one dimensional
theories, in which the symmetry group is ${\rm E}_{10(10)}$
\cite{Julia:1982gx}--\cite{e10} and the Cartan fields $h$ evolve
in a ten dimensional Lorentzian space. In this setting a system of
axions can be chosen so as to define through their roots a finite
billiard, yielding an oscillatory ten dimensional cosmological
solution with chaotic behavior.\par In perspective it would be
also interesting to analyze  more general ten dimensional
homogeneous solutions deriving from the coupling of gravity to
various other combinations of tensor fields and with a
non--diagonal metric.  We leave for instance to future
investigations the construction of backgrounds related to the
other classes of $A_2$ embeddings outlined in section
\ref{occidoa2}.
 These backgrounds can provide
  cosmological settings where to analyze interesting
\emph{braneworld} scenarios, in which the dynamics of the universe
expansion/contraction along directions parallel to the braneworld
(our four--dimensional universe) and transverse to it would in
general be different.
\par Of
course any microscopic interpretation of these homogeneous
backgrounds will bring about the issue of $\alpha^\prime $
corrections (as far as S--brane supergravity solutions are
concerned see \cite{andythermo,stro2,stro1}) and the possibility
of constructing exact string solutions. This is in general not the
case for standard cosmological backgrounds, on which stringy
corrections have a relevant effect. \par
 An other direction for future research would be to generalize our
solution--constructing technique to ansaetze which are
inomogeneous/anisotropic with respect to the three--dimensional
space--time and  to analyze the behavior of these solutions in
relation for instance to the singularity theorem discussed in
\cite{otherSbranes}.
\section{Acknowledgments}
M. T. would like to thank M. Bertolini and B. Stefanski for useful discussions.
\newpage
\appendix

\section{Listing of the $E_8$ positive roots}
\subsection{Listing according to height}
\vskip 0.1cm
In this listing we present the roots of the $E_8$ Lie algebra, giving their
definition both in terms
of the simple roots and in the eucledian basis. The notation $a_{i|j}$ is introduced to
denote the height of the root (i).
The number (j) is introduced to distinguish the roots of the same height

\begin{eqnarray*}
\begin{array}{lclclcl}
 {a_{1|1}} & = &{\alpha [} 1 ] & = &
 \{1,0,0,0,0,0,0,0\} & = &\{0,1,-1,0,0,0,0,0\}
\nonumber \\ {a_{1|2}} & = &{\alpha [} 2 ]& = &
 \{0,1,0,0,0,0,0,0\} & = &\{0,0,1,-1,0,0,0,0\}
\nonumber \\ {a_{1|3}} & = &{\alpha [} 3 ]& = &
 \{0,0,1,0,0,0,0,0\} & = &\{ 0,0,0,1,-1,0,0,0\}
\nonumber \\ {a_{1|4}} & = &{\alpha [} 4 ]& = &
 \{0,0,0,1,0,0,0,0\} & = &\{ 0,0,0,0,1,-1,0,0\}
\nonumber \\ {a_{1|5}} & = &{\alpha [} 5 ]& = &
 \{0,0,0,0,1,0,0,0\} & = &\{0,0,0,0,0,1,-1,0\}
\nonumber \\ {a_{1|6}}  & = & {\alpha [} 6 ]& = &
 \{0,0,0,0,0,1,0,0\} & = &\{ 0,0,0,0,0,1,1,0\}
\nonumber \\ {a_{1|7}} & = &{\alpha [} 7 ]& = &
\{0,0,0,0,0,0,1,0\}
& = &\big\{- \frac{1}{2}  ,
            - \frac{1}{2}  ,
            - \frac{1}{2}  ,
            - \frac{1}{2}  ,
            - \frac{1}{2}  ,
            - \frac{1}{2}  ,
            - \frac{1}{2}  ,
            - \frac{1}{2}\big\}
\nonumber \\ {a_{1|8}} & = &{\alpha [} 8 ]& = & \{ 0,0,0,0,0,0,0,1 \}
& =
&\big\{ 1,-1,0,0,0,0,0,0\big\}\
\nonumber \\
 {a_{2|1}} & = &{\alpha [} 9 ]& = &
 \{1,1,0,0,0,0,0,0\} & = &\{ 0,1,0,-1,0,0,0,0\}
\nonumber \\ {a_{2|2}} & = &{\alpha [} 10 ]& = &
\{1,0,0,0,0,0,0,1\}
& = &\big\{1,0,-1,0,0,0,0,0\big\}
\nonumber \\ {a_{2|3}} & = &{\alpha [} 11 ]& = &
 \{0,1,1,0,0,0,0,0\} & = &\{0,0,1,0,-1,0,0,0\}
\nonumber \\ {a_{2|4}} & = &{\alpha [} 12 ]& = &
 \{0,0,1,1,0,0,0,0\} & = &\{ 0,0,0,1,0,-1,0,0\}
\nonumber \\ {a_{2|5}} & = &{\alpha [} 13 ]& = &
 \{0,0,0,1,1,0,0,0\} & = &\{0,0,0,0,1,0,-1,0\}
\nonumber \\ {a_{2|6}} & = &{\alpha [} 14 ]& = &
 \{0,0,0,1,0,1,0,0\} & = &\{ 0,0,0,0,1,0,1,0\}
\nonumber \\ {a_{2|7}} & = &{\alpha [} 15 ]& = &
\{0,0,0,0,0,1,1,0\}
& = &\big\{-\frac{1}{2}  ,
            -\frac{1}{2} ,
            -\frac{1}{2} ,
            -\frac{1}{2} ,
            -\frac{1}{2} ,\frac{1}{2},\frac{1}{2},
            -\frac{1}{2}\big\} \
\nonumber \\
 {a_{3|1}} & = &{\alpha [} 16 ]& = &
 \{1,1,1,0,0,0,0,0\} & = &\{0,1,0,0,-1,0,0,0\}
\nonumber \\ {a_{3|2}} & = &{\alpha [} 17 ]& = &
\{1,1,0,0,0,0,0,1\}
& = &\big\{1,0,0,-1,0,0,0,0\big\}
\nonumber \\ {a_{3|3}} & = &{\alpha [} 18 ]& = &
 \{0,1,1,1,0,0,0,0\} & = &\{ 0,0,1,0,0,-1,0,0\}
\nonumber \\ {a_{3|4}} & = &{\alpha [} 19 ]& = &
 \{0,0,1,1,1,0,0,0\} & = &\{ 0,0,0,1,0,0,-1,0\}
\nonumber \\ {a_{3|5}} & = &{\alpha [} 20 ]& = &
 \{0,0,1,1,0,1,0,0\} & = &\{ 0,0,0,1,0,0,1,0\}
\nonumber \\ {a_{3|6}} & = &{\alpha [} 21 ]& = &
\{0,0,0,1,0,1,1,0\}
& = &\big\{ -\frac{1}{2} ,
            -\frac{1}{2} ,
            -\frac{1}{2} ,
            -\frac{1}{2} ,\frac{1}{2},
            -\frac{1}{2} ,\frac{1}{2},
            -\frac{1}{2}\big\}
\nonumber \\ {a_{3|7}} & = &{\alpha [} 22 ]& = &
 \{0,0,0,1,1,1,0,0\} & = &\{0,0,0,0,1,1,0,0\} \
\nonumber \\
 {a_{4|1}} & = &{\alpha [} 23 ]& = &
 \{1,1,1,1,0,0,0,0\} & = &\{0,1,0,0,0,-1,0,0\}
\nonumber \\ {a_{4|2}} & = &{\alpha [} 24 ]& = &
 \{0,1,1,1,1,0,0,0\} & = &\{0,0,1,0,0,0,-1,0\}
\nonumber \\ {a_{4|3}} & = &{\alpha [} 25 ]& = &
 \{0,1,1,1,0,1,0,0\} & = &\{ 0,0,1,0,0,0,1,0\}
\nonumber \\ {a_{4|4}} & = &{\alpha [} 26 ]& = &
\{1,1,1,0,0,0,0,1\}
& = &\big\{ 1,0,0,0,-1,0,0,0\big\}
\nonumber
\\
 {a_{4|5}} & = &{\alpha [} 27 ]& = &
\{0,0,1,1,0,1,1,0\}
& = &\big\{-\frac{1}{2} ,
            -\frac{1}{2} ,
            -\frac{1}{2} ,\frac{1}{2},
            -\frac{1}{2} ,
            -\frac{1}{2} ,\frac{1}{2},
            -\frac{1}{2} \big\}
\nonumber
\\
 {a_{4|6}} & = &{\alpha [} 28 ]& = &
 \{0,0,1,1,1,1,0,0\} & = &\{0,0,0,1,0,1,0,0\}
\nonumber
\end{array}
\end{eqnarray*}
\begin{eqnarray*}
\begin{array}{lclclcl}
 {a_{4|7}} & = &{\alpha [} 29 ]& = &
\{0,0,0,1,1,1,1,0\}
& = &\big\{-\frac{1}{2} ,
            -\frac{1}{2} ,
            -\frac{1}{2} ,
            -\frac{1}{2} ,\frac{1}{2},\frac{1}{2},
            -\frac{1}{2} ,
            -\frac{1}{2}\big\} \nonumber
 \\
 {a_{5|1}} & = &{\alpha [} 30 ]& = &
 \{1,1,1,1,1,0,0,0\} & = &\{ 0,1,0,0,0,0,-1,0\}
\nonumber \\ {a_{5|2}} & = &{\alpha [} 31 ]& = &
 \{1,1,1,1,0,1,0,0\} & = &\{0,1,0,0,0,0,1,0\}
\nonumber \\ {a_{5|3}} & = &{\alpha [} 32 ]& = &
\{0,1,1,1,0,1,1,0\}
& = &\big\{-\frac{1}{2} ,
            -\frac{1}{2} ,\frac{1}{2},
            -\frac{1}{2} ,
            -\frac{1}{2} ,
            -\frac{1}{2} ,\frac{1}{2},
            -\frac{1}{2} \big\}
\nonumber \\ {a_{5|4}} & = &{\alpha [} 33 ]& = &
 \{0,1,1,1,1,1,0,0\} & = &\{ 0,0,1,0,0,1,0,0\}
  \nonumber \\
 {a_{5|5}} & = &{\alpha [} 34 ]& = &
\{0,0,1,1,1,1,1,0\}
& = &\big\{-\frac{1}{2} ,
            -\frac{1}{2} ,
            -\frac{1}{2} ,\frac{1}{2},
            -\frac{1}{2} ,\frac{1}{2},
            -\frac{1}{2} ,
            -\frac{1}{2}\big\}
\nonumber \\ {a_{5|6}} & = &{\alpha [} 35 ]& = &
\{1,1,1,1,0,0,0,1\}
& = &\big\{1,0,0,0,0,-1,0,0\big\}
\nonumber \\ {a_{5|7}} & = &{\alpha [} 36 ]& = &
 \{0,0,1,2,1,1,0,0\} & = &\{0,0,0,1,1,0,0,0\} \nonumber \\
 {a_{6|1}} & = &{\alpha [} 37 ]& = &
\{1,1,1,1,0,1,1,0\}
& = &\big\{-\frac{1}{2} ,\frac{1}{2},
            -\frac{1}{2} ,
            -\frac{1}{2} ,
            -\frac{1}{2} ,
            -\frac{1}{2} ,\frac{1}{2},
            -\frac{1}{2}\big\}
\nonumber \\ {a_{6|2}} & = &{\alpha [} 38 ]& = &
 \{1,1,1,1,1,1,0,0\} & = &\{0,1,0,0,0,1,0,0\}
\nonumber \\ {a_{6|3}} & = &{\alpha [} 39 ]& = &
\{0,1,1,1,1,1,1,0\}
& = &\big\{-\frac{1}{2} ,
            -\frac{1}{2} ,\frac{1}{2},
            -\frac{1}{2} ,
            -\frac{1}{2} ,\frac{1}{2},
            -\frac{1}{2} ,
            -\frac{1}{2}\big\}
\nonumber \\ {a_{6|4}} & = &{\alpha [} 40 ]& = &
 \{0,1,1,2,1,1,0,0\} & = &\{ 0,0,1,0,1,0,0,0\}
\nonumber \\ {a_{6|5}} & = &{\alpha [} 41 ]& = &
\{0,0,1,2,1,1,1,0\}
& = &\big\{-\frac{1}{2} ,
            -\frac{1}{2} ,
            -\frac{1}{2} ,\frac{1}{2},\frac{1}{2},
            -\frac{1}{2} ,
            -\frac{1}{2} ,
            -\frac{1}{2}\big\}
\nonumber \\ {a_{6|6}} & = &{\alpha [} 42 ]& = &
\{1,1,1,1,1,0,0,1\}
& = &\big\{1,0,0,0,0,0,-1,0\big\}
\nonumber \\ {a_{6|7}} & = &{\alpha [} 43 ]& = &
\{1,1,1,1,0,1,0,1\}
& = &\big\{1,0,0,0,0,0,1,0\big\}\nonumber \\
 {a_{7|1}} & = &{\alpha [} 44 ]& = &
\{1,1,1,1,1,1,1,0\}
& = &\big\{-\frac{1}{2} ,\frac{1}{2},
            -\frac{1}{2} ,
            -\frac{1}{2} ,
            -\frac{1}{2} ,\frac{1}{2},
            -\frac{1}{2} ,
            -\frac{1}{2}\big\}
\nonumber \\ {a_{7|2}} & = &{\alpha [} 45 ]& = &
 \{1,1,1,2,1,1,0,0\} & = &\{ 0,1,0,0,1,0,0,0\}
\nonumber \\ {a_{7|3}} & = &{\alpha [} 46 ]& = &
\{0,1,1,2,1,1,1,0\}
& = &\big\{ -\frac{1}{2} ,
            -\frac{1}{2} ,\frac{1}{2},
            -\frac{1}{2} ,\frac{1}{2},
            -\frac{1}{2} ,
            -\frac{1}{2} ,
            -\frac{1}{2}\big\}
\nonumber \\ {a_{7|4}} & = &{\alpha [} 47 ]& = &
 \{0,1,2,2,1,1,0,0\} & = &\{0,0,1,1,0,0,0,0\}
\nonumber \\ {a_{7|5}} & = &{\alpha [} 48 ]& = &
\{1,1,1,1,1,1,0,1\}
& = &\big\{1,0,0,0,0,1,0,0\big\}
\nonumber \\ {a_{7|6}} & = &{\alpha [} 49 ]& = &
\{0,0,1,2,1,2,1,0\}
& = &\big\{-\frac{1}{2} ,
            -\frac{1}{2} ,
            -\frac{1}{2} ,
            \frac{1}{2},\frac{1}{2},
            \frac{1}{2},\frac{1}{2},
            -\frac{1}{2}\big\}
\nonumber \\ {a_{7|7}} & = &{\alpha [} 50 ]& = &
\{1,1,1,1,0,1,1,1\}
& = &\big\{\frac{1}{2},-\frac{1}{2} ,
            -\frac{1}{2} ,
            -\frac{1}{2} ,
            -\frac{1}{2} ,
            -\frac{1}{2} ,\frac{1}{2},
            -\frac{1}{2}\big\} \nonumber
\\
 {a_{8|1}} & = &{\alpha [} 51 ]& = &
\{1,1,1,2,1,1,1,0\}
& = &\big\{-\frac{1}{2} ,\frac{1}{2},
            -\frac{1}{2} ,
            -\frac{1}{2} ,\frac{1}{2},
            -\frac{1}{2} ,
            -\frac{1}{2} ,
            -\frac{1}{2}\big\}
\nonumber \\ {a_{8|2}} & = &{\alpha [} 52 ]& = &
 \{1,1,2,2,1,1,0,0\} & = &\{ 0,1,0,1,0,0,0,0\}
\nonumber \\ {a_{8|3}} & = &{\alpha [} 53 ]& = &
\{0,1,1,2,1,2,1,0\}
& = &\big\{-\frac{1}{2} ,
            -\frac{1}{2} ,\frac{1}{2},
            -\frac{1}{2} ,\frac{1}{2},\frac{1}{2},
            \frac{1}{2},-\frac{1}{2} \big\}
\nonumber \\ {a_{8|4}} & = &{\alpha [} 54 ]& = &
\{0,1,2,2,1,1,1,0\}
& = &\big\{-\frac{1}{2} ,
            -\frac{1}{2} ,\frac{1}{2},\frac{1}{2},
            -\frac{1}{2} ,
            -\frac{1}{2} ,
            -\frac{1}{2} ,
            -\frac{1}{2}\big\}
\nonumber \\ {a_{8|5}} & = &{\alpha [} 55 ]& = &
\{1,1,1,2,1,1,0,1\}
& = &\big\{1,0,0,0,1,0,0,0\big\}
\nonumber \\ {a_{8|6}} & = &{\alpha [} 56 ]& = &
\{1,1,1,1,1,1,1,1\}
& = &\big\{\frac{1}{2},-\frac{1}{2} ,
            -\frac{1}{2} ,
            -\frac{1}{2} ,
            -\frac{1}{2} ,\frac{1}{2},
            -\frac{1}{2} ,
            -\frac{1}{2}\big\}\nonumber
\\
 {a_{9|1}} & = &{\alpha [} 57 ]& = &
\{1,1,1,2,1,2,1,0\}
& = &\big\{-\frac{1}{2} ,\frac{1}{2},
            -\frac{1}{2} ,
            -\frac{1}{2} ,\frac{1}{2},\frac{1}{2},
            \frac{1}{2},-\frac{1}{2}\big\}
\nonumber \\ {a_{9|2}} & = &{\alpha [} 58 ]& = &
\{1,1,2,2,1,1,1,0\}
& = &\big\{-\frac{1}{2} ,\frac{1}{2},
            -\frac{1}{2} ,\frac{1}{2},
            -\frac{1}{2} ,
            -\frac{1}{2} ,
            -\frac{1}{2} ,
            -\frac{1}{2}\big\}
\nonumber \\ {a_{9|3}} & = &{\alpha [} 59 ]& = &
 \{1,2,2,2,1,1,0,0\} & = &\{0,1,1,0,0,0,0,0\}
\nonumber \\ {a_{9|4}} & = &{\alpha [} 60 ]& = &
\{0,1,2,2,1,2,1,0\}
& = &\big\{-\frac{1}{2} ,
            -\frac{1}{2} ,\frac{1}{2},\frac{1}{2},
            -\frac{1}{2} ,\frac{1}{2},\frac{1}{2},
            -\frac{1}{2}\big\}
\nonumber \\ {a_{9|5}} & = &{\alpha [} 61 ]& = &
\{1,1,2,2,1,1,0,1\}
& = &\big\{1,0,0,1,0,0,0,0\big\}
\nonumber \\ {a_{9|6}} & = &{\alpha [} 62 ]& = &
\{1,1,1,2,1,1,1,1\}
& = &\big\{\frac{1}{2},-\frac{1}{2} ,
            -\frac{1}{2} ,
            -\frac{1}{2} ,\frac{1}{2},
            -\frac{1}{2} ,
            -\frac{1}{2} ,
            -\frac{1}{2}\big\}
\end{array}
\end{eqnarray*}

\begin{eqnarray*}
\begin{array}{lclclcl}
 {a_{10|1}} & = &{\alpha [} 63 ]& = &
\{1,1,2,2,1,2,1,0\}
& = &\big\{ -\frac{1}{2} ,\frac{1}{2},
            -\frac{1}{2} ,\frac{1}{2},
            -\frac{1}{2} ,\frac{1}{2},\frac{1}{2},
            -\frac{1}{2}\big\}
\nonumber \\ {a_{10|2}} & = &{\alpha [} 64 ]& = &
\{1,2,2,2,1,1,1,0\}
& = &\big\{-\frac{1}{2} ,\frac{1}{2},\frac{1}{2},
            -\frac{1}{2} ,
            -\frac{1}{2} ,
            -\frac{1}{2} ,
            -\frac{1}{2} ,
            -\frac{1}{2} \big\}
\nonumber \\ {a_{10|3}} & = &{\alpha [} 65 ]& = &
\{1,2,2,2,1,1,0,1\}
& = &\big\{1,0,1,0,0,0,0,0\big\}
\nonumber \\ {a_{10|4}} & = &{\alpha [} 66 ]& = &
\{1,1,2,2,1,1,1,1\}
& = &\big\{\frac{1}{2},-\frac{1}{2} ,
            -\frac{1}{2} ,\frac{1}{2},
            -\frac{1}{2} ,
            -\frac{1}{2} ,
            -\frac{1}{2} ,
            -\frac{1}{2}\big\}
\nonumber \\ {a_{10|5}} & = &{\alpha [} 67 ]& = &
\{0,1,2,3,1,2,1,0\}
& = &\big\{-\frac{1}{2} ,
            -\frac{1}{2} ,\frac{1}{2},\frac{1}{2},
            \frac{1}{2},-\frac{1}{2} ,\frac{1}{2},
            -\frac{1}{2}\big\}
\nonumber \\ {a_{10|6}} & = &{\alpha [} 68 ]& = &
\{1,1,1,2,1,2,1,1\}
& = &\big\{\frac{1}{2},-\frac{1}{2} ,
            -\frac{1}{2} ,
            -\frac{1}{2} ,\frac{1}{2},\frac{1}{2},
            \frac{1}{2},-\frac{1}{2} \big\}\nonumber
\\
 {a_{11|1}} & = &{\alpha [} 69 ]& = &
\{2,2,2,2,1,1,0,1\}
& = &\big\{1,1,0,0,0,0,0,0\big\}
\nonumber \\ {a_{11|2}} & = &{\alpha [} 70 ]& = &
\{1,1,2,3,1,2,1,0\}
& = &\big\{-\frac{1}{2} ,\frac{1}{2},
            -\frac{1}{2} ,\frac{1}{2},\frac{1}{2},
            -\frac{1}{2} ,\frac{1}{2},
            -\frac{1}{2}\big\}
\nonumber \\ {a_{11|3}} & = &{\alpha [} 71 ]& = &
\{1,2,2,2,1,2,1,0\}
& = &\big\{-\frac{1}{2} ,\frac{1}{2},\frac{1}{2},
            -\frac{1}{2} ,
            -\frac{1}{2} ,\frac{1}{2},\frac{1}{2},
            -\frac{1}{2}\big\}
\nonumber \\ {a_{11|4}} & = &{\alpha [} 72 ]& = &
\{1,2,2,2,1,1,1,1\}
& = &\big\{\frac{1}{2},-\frac{1}{2} ,\frac{1}{2},
            -\frac{1}{2} ,
            -\frac{1}{2} ,
            -\frac{1}{2} ,
            -\frac{1}{2} ,
            -\frac{1}{2}\big\}
\nonumber \\ {a_{11|5}} & = &{\alpha [} 73 ]& = &
\{1,1,2,2,1,2,1,1\}
& = &\big\{\frac{1}{2},-\frac{1}{2} ,
            -\frac{1}{2} ,\frac{1}{2},
            -\frac{1}{2} ,\frac{1}{2},\frac{1}{2},
            -\frac{1}{2}\big\}
\nonumber \\ {a_{11|6}} & = &{\alpha [} 74 ]& = &
\{0,1,2,3,2,2,1,0\}
& = &\big\{-\frac{1}{2} ,
            -\frac{1}{2} ,\frac{1}{2},\frac{1}{2},
            \frac{1}{2},\frac{1}{2},-\frac{1}{2} ,
            -\frac{1}{2}\big\}
\nonumber
\\
 {a_{12|1}} & = &{\alpha [} 75 ]& = &
\{2,2,2,2,1,1,1,1\}
& = &\big\{\frac{1}{2},\frac{1}{2},-\frac{1}{2} ,
            -\frac{1}{2} ,
            -\frac{1}{2} ,
            -\frac{1}{2} ,
            -\frac{1}{2} ,
            -\frac{1}{2} \big\}
\nonumber \\ {a_{12|2}} & = &{\alpha [} 76 ]& = &
\{1,1,2,3,2,2,1,0\}
& = &\big\{-\frac{1}{2} ,\frac{1}{2},
            -\frac{1}{2} ,\frac{1}{2},\frac{1}{2},
            \frac{1}{2},-\frac{1}{2} ,
            -\frac{1}{2}\big\}
\nonumber \\ {a_{12|3}} & = &{\alpha [} 77 ]& = &
\{1,2,2,3,1,2,1,0\}
& = &\big\{-\frac{1}{2} ,\frac{1}{2},\frac{1}{2},
            -\frac{1}{2} ,\frac{1}{2},
            -\frac{1}{2} ,\frac{1}{2},
            -\frac{1}{2} \big\}
\nonumber \\ {a_{12|4}} & = &{\alpha [} 78 ]& = &
\{1,2,2,2,1,2,1,1\}
& = &\big\{\frac{1}{2},-\frac{1}{2} ,\frac{1}{2},
            -\frac{1}{2} ,
            -\frac{1}{2} ,\frac{1}{2},\frac{1}{2},
            -\frac{1}{2}\big\}
\nonumber \\ {a_{12|5}} & = &{\alpha [} 79 ]& = &
\{1,1,2,3,1,2,1,1\}
& = &\big\{\frac{1}{2},-\frac{1}{2} ,
            -\frac{1}{2} ,\frac{1}{2},\frac{1}{2},
            -\frac{1}{2} ,\frac{1}{2},
            -\frac{1}{2}\big\}
\nonumber
\\
 {a_{13|1}} & = &{\alpha [} 80 ]& = &
\{2,2,2,2,1,2,1,1\}
& = &\big\{\frac{1}{2},\frac{1}{2},-\frac{1}{2} ,
            -\frac{1}{2} ,
            -\frac{1}{2} ,\frac{1}{2},\frac{1}{2},
            -\frac{1}{2} \big\}
\nonumber \\ {a_{13|2}} & = &{\alpha [} 81 ]& = &
\{1,2,2,3,2,2,1,0\}
& = &\big\{-\frac{1}{2} ,\frac{1}{2},\frac{1}{2},
            -\frac{1}{2} ,\frac{1}{2},\frac{1}{2},
            -\frac{1}{2} ,
            -\frac{1}{2}\big\}
\nonumber \\ {a_{13|3}} & = &{\alpha [} 82 ]& = &
\{1,2,2,3,1,2,1,1\}
& = &\big\{\frac{1}{2},-\frac{1}{2} ,\frac{1}{2},
            -\frac{1}{2} ,\frac{1}{2},
            -\frac{1}{2} ,\frac{1}{2},
            -\frac{1}{2}\big\}
\nonumber \\ {a_{13|4}} & = &{\alpha [} 83 ]& = &
\{1,2,3,3,1,2,1,0\}
& = &\big\{-\frac{1}{2} ,\frac{1}{2},\frac{1}{2},
            \frac{1}{2},-\frac{1}{2} ,
            -\frac{1}{2} ,\frac{1}{2},
            -\frac{1}{2}\big\}
\nonumber \\ {a_{13|5}} & = &{\alpha [} 84 ]& = &
\{1,1,2,3,2,2,1,1\}
& = &\big\{\frac{1}{2},-\frac{1}{2} ,
            -\frac{1}{2} ,\frac{1}{2},\frac{1}{2},
            \frac{1}{2},-\frac{1}{2} ,
            -\frac{1}{2}\big\}
\nonumber
\\
 {a_{14|1}} & = &{\alpha [} 85 ]& = &
\{2,2,2,3,1,2,1,1\}
& = &\big\{\frac{1}{2},\frac{1}{2},-\frac{1}{2} ,
            -\frac{1}{2} ,\frac{1}{2},
            -\frac{1}{2} ,\frac{1}{2},
            -\frac{1}{2}\big\}
\nonumber \\ {a_{14|2}} & = &{\alpha [} 86 ]& = &
\{1,2,2,3,2,2,1,1\}
& = &\big\{\frac{1}{2},-\frac{1}{2} ,\frac{1}{2},
            -\frac{1}{2} ,\frac{1}{2},\frac{1}{2},
            -\frac{1}{2} ,
            -\frac{1}{2}\big\}
\nonumber \\ {a_{14|3}} & = &{\alpha [} 87 ]& = &
\{1,2,3,3,2,2,1,0\}
& = &\big\{-\frac{1}{2} ,\frac{1}{2},\frac{1}{2},
            \frac{1}{2},-\frac{1}{2} ,\frac{1}{2},
            -\frac{1}{2} ,
            -\frac{1}{2}\big\}
\nonumber \\ {a_{14|4}} & = &{\alpha [} 88 ]& = &
\{1,2,3,3,1,2,1,1\}
& = &\big\{ \frac{1}{2},-\frac{1}{2} ,\frac{1}{2},
            \frac{1}{2},-\frac{1}{2} ,
            -\frac{1}{2} ,\frac{1}{2},
            -\frac{1}{2} \big\}
\nonumber
\\
 {a_{15|1}} & = &{\alpha [} 89 ]& = &
\{2,2,2,3,2,2,1,1\}
& = &\big\{\frac{1}{2},\frac{1}{2},-\frac{1}{2} ,
            -\frac{1}{2} ,\frac{1}{2},\frac{1}{2},
            -\frac{1}{2} ,
            -\frac{1}{2}\big\}
\nonumber \\ {a_{15|2}} & = &{\alpha [} 90 ]& = &
\{2,2,3,3,1,2,1,1\}
& = &\big\{\frac{1}{2},\frac{1}{2},-\frac{1}{2} ,
            \frac{1}{2},-\frac{1}{2} ,
            -\frac{1}{2} ,\frac{1}{2},
            -\frac{1}{2}\big\}
\nonumber \\ {a_{15|3}} & = &{\alpha [} 91 ]& = &
\{1,2,3,3,2,2,1,1\}
& = &\big\{\frac{1}{2},-\frac{1}{2} ,\frac{1}{2},
            \frac{1}{2},-\frac{1}{2} ,\frac{1}{2},
            -\frac{1}{2} ,
            -\frac{1}{2}\big\}
\nonumber \\ {a_{15|4}} & = &{\alpha [} 92 ]& = &
\{1,2,3,4,2,2,1,0\}
& = &\big\{-\frac{1}{2} ,\frac{1}{2},\frac{1}{2},
            \frac{1}{2},\frac{1}{2},-\frac{1}{2} ,
            -\frac{1}{2} ,
            -\frac{1}{2}\big\}
\nonumber
\\
 {a_{16|1}} & = &{\alpha [} 93 ]& = &
\{2,2,3,3,2,2,1,1\}
& = &\big\{\frac{1}{2},\frac{1}{2},-\frac{1}{2} ,
            \frac{1}{2},-\frac{1}{2} ,\frac{1}{2},
            -\frac{1}{2} ,
            -\frac{1}{2} \big\}
\nonumber \\ {a_{16|2}} & = &{\alpha [} 94 ]& = &
\{2,3,3,3,1,2,1,1\}
& = &\big\{\frac{1}{2},\frac{1}{2},\frac{1}{2},
            -\frac{1}{2} ,
            -\frac{1}{2} ,
            -\frac{1}{2},\frac{1}{2},
            -\frac{1}{2}\big\}
\nonumber \\ {a_{16|3}} & = &{\alpha [} 95 ]& = &
\{1,2,3,4,2,2,1,1\}
& = &\big\{\frac{1}{2},-\frac{1}{2} ,\frac{1}{2},
            \frac{1}{2},\frac{1}{2},-\frac{1}{2} ,
            -\frac{1}{2} ,
            -\frac{1}{2} \big\}
\nonumber \\ {a_{16|4}} & = &{\alpha [} 96 ]& = &
\{1,2,3,4,2,3,1,0\}
& = &\big\{-\frac{1}{2} ,\frac{1}{2},\frac{1}{2},
            \frac{1}{2},\frac{1}{2},\frac{1}{2},\frac{1}{2},
            -\frac{1}{2}\big\}
\
\end{array}
\end{eqnarray*}

\begin{eqnarray*}
\begin{array}{lclclcl}
 {a_{17|1}} & = &{\alpha [} 97 ]& = &
\{2,2,3,4,2,2,1,1\}
& = &\big\{\frac{1}{2},\frac{1}{2},-\frac{1}{2} ,
            \frac{1}{2},\frac{1}{2},-\frac{1}{2} ,
            -\frac{1}{2} ,
            -\frac{1}{2} \big\}
\nonumber \\ {a_{17|2}} & = &{\alpha [} 98 ]& = &
\{2,3,3,3,2,2,1,1\}
& = &\big\{\frac{1}{2},\frac{1}{2},\frac{1}{2},
            -\frac{1}{2} ,
            -\frac{1}{2} ,\frac{1}{2},
            -\frac{1}{2} ,
            -\frac{1}{2}\big\}
\nonumber \\ {a_{17|3}} & = &{\alpha [} 99 ]& = &
\{1,2,3,4,2,3,1,1\}
& = &\big\{\frac{1}{2},-\frac{1}{2} ,\frac{1}{2},
            \frac{1}{2},\frac{1}{2},\frac{1}{2},\frac{1}{2},
            -\frac{1}{2}\big\}
\nonumber \\ {a_{17|4}} & = &{\alpha [} 100 ]& = &
 \{1,2,3,4,2,3,2,0\} & = &\big\{-1,0,0,0,0,0,0,-1\big\}
\nonumber
\\
 {a_{18|1}} & = &{\alpha [} 101 ]& = &
\{2,2,3,4,2,3,1,1\}
& = &\big\{ \frac{1}{2},\frac{1}{2},-\frac{1}{2} ,
            \frac{1}{2},\frac{1}{2},\frac{1}{2},\frac{1}{2},
            -\frac{1}{2}\big\}
\nonumber \\ {a_{18|2}} & = &{\alpha [} 102 ]& = &
\{2,3,3,4,2,2,1,1\}
& = &\big\{\frac{1}{2},\frac{1}{2},\frac{1}{2},
            -\frac{1}{2} ,\frac{1}{2},
            -\frac{1}{2} ,
            -\frac{1}{2} ,
            -\frac{1}{2}\big\}
\nonumber \\ {a_{18|3}} & = &{\alpha [} 103 ]& = &
\{1,2,3,4,2,3,2,1\}
& = &\big\{0,-1,0,0,0,0,0,-1\big\}
\nonumber
\\
 {a_{19|1}} & = &{\alpha [} 104 ]& = &
\{2,2,3,4,2,3,2,1\}
& = &\big\{0,0,-1,0,0,0,0,-1\big\}
\nonumber \\ {a_{19|2}} & = &{\alpha [} 105 ]& = &
\{2,3,3,4,2,3,1,1\}
& = &\big\{\frac{1}{2},\frac{1}{2},\frac{1}{2},
            -\frac{1}{2} ,\frac{1}{2},\frac{1}{2},
            \frac{1}{2},-\frac{1}{2}\big\}
\nonumber \\ {a_{19|3}} & = &{\alpha [} 106 ]& = &
\{2,3,4,4,2,2,1,1\}
& = &\big\{\frac{1}{2},\frac{1}{2},\frac{1}{2},\frac{1}{2},
            -\frac{1}{2} ,
            -\frac{1}{2} ,
            -\frac{1}{2} ,
            -\frac{1}{2}\big\}
\nonumber
\\
 {a_{20|1}} & = &{\alpha [} 107 ]& = &
\{2,3,3,4,2,3,2,1\}
& = &\big\{0,0,0,-1,0,0,0,-1\big\}
\nonumber \\ {a_{20|2}} & = &{\alpha [} 108 ]& = &
\{2,3,4,4,2,3,1,1\}
& = &\big\{\frac{1}{2},\frac{1}{2},\frac{1}{2},\frac{1}{2},
            -\frac{1}{2} ,\frac{1}{2},\frac{1}{2},
            -\frac{1}{2}\big\}
\nonumber
\\
 {a_{21|1}} & = &{\alpha [} 109 ]& = &
\{2,3,4,4,2,3,2,1\}
& = &\big\{ 0,0,0,0,-1,0,0,-1\big\}
\nonumber \\ {a_{21|2}} & = &{\alpha [} 110 ]& = &
\{2,3,4,5,2,3,1,1\}
& = &\big\{\frac{1}{2},\frac{1}{2},\frac{1}{2},\frac{1}{2},
            \frac{1}{2},-\frac{1}{2} ,\frac{1}{2},
            -\frac{1}{2} \big\}
\nonumber
\\
 {a_{22|1}} & = &{\alpha [} 111 ]& = &
\{2,3,4,5,2,3,2,1\}
& = &\big\{0,0,0,0,0,-1,0,-1\big\}
\nonumber \\ {a_{22|2}} & = &{\alpha [} 112 ]& = &
\{2,3,4,5,3,3,1,1\}
& = &\big\{ \frac{1}{2},\frac{1}{2},\frac{1}{2},\frac{1}{2},
            \frac{1}{2},\frac{1}{2},-\frac{1}{2} ,
            -\frac{1}{2}\big\}
\nonumber
\\
{a_{23|1}} & = &{\alpha [} 113 ]& = &
\{2,3,4,5,3,3,2,1\}
& = &\big\{0,0,0,0,0,0,-1,-1\}
\nonumber \\ {a_{23|2}} & = &{\alpha [} 114 ]& = &
\{2,3,4,5,2,4,2,1\}
& = &\big\{0,0,0,0,0,0,1,-1\big\}
\nonumber
\\
 {a_{24|1}} & = &{\alpha [} 115 ]& = &
\{2,3,4,5,3,4,2,1\}
& = &\big\{0,0,0,0,0,1,0,-1\big\}
\nonumber
\\
 {a_{25|1}} & = &{\alpha [} 116 ]& = &
\{2,3,4,6,3,4,2,1\}
& = &\big\{0,0,0,0,1,0,0,-1\big\}
\nonumber
\\
 {a_{26|1}} & = &{\alpha [} 117 ]& = &
\{2,3,5,6,3,4,2,1\}
& = &\big\{0,0,0,1,0,0,0,-1\big\}
\nonumber
\\
 {a_{27|1}} & = &{\alpha [} 118 ]& = &
\{2,4,5,6,3,4,2,1\}
& = &\big\{ 0,0,1,0,0,0,0,-1\big\}
\nonumber
\\
 {a_{28|1}} & = &{\alpha [} 119 ]& = &
\{3,4,5,6,3,4,2,1\}
& = &\big\{0,1,0,0,0,0,0,-1\big\}
\nonumber
\\
 {a_{29|1}} & = &{\alpha [} 120 ]& = &
 \{3,4,5,6,3,4,2,2\} & = &\big\{1,0,0,0,0,0,0,-1\big\}
\
\end{array}
\end{eqnarray*}
\subsection{Listing the  roots according to the dimensional filtration}
\vskip 0.1 cm
\noindent {{Roots in D[} 1 ] and D[ 2 ]}
\begin{eqnarray*}
\begin{array}{lcccl|cc}
\hline
\mbox{label}& \null & \mbox{root number} &\null & \mbox{Dynkin label} &
\mbox{type II B }&\mbox{type II A } \\
\hline
{d_{1|1}} & = &  {\alpha [} 7 ]& = &
 \{0,0,0,0,0,0,1,0\}  \nonumber & \rho & C_{9}\\
{d_{2|1}} & = &  {\alpha [} 6 ]& = &
 \{0,0,0,0,0,1,0,0\}  \nonumber & B_{8\,9} & \gamma_{8}^{\phantom{8}\,9}\\
{d_{2|2}} & = &  {\alpha [} 5 ]& = &
 \{0,0,0,0,1,0,0,0\}  \nonumber & \gamma_{8}{}^{9} & \gamma_{8}{}^{9} \\
{d_{2|3}} & = &  {\alpha [} 15 ]& = &
 \{0,0,0,0,0,1,1,0\}& C_{8\,9} & C_{8}  \nonumber \\
\end{array}
\end{eqnarray*}
{{Roots in D[} 3 ], D[ 4 ] and D[ 5 ]}
\begin{eqnarray*}
\begin{array}{lcccl|cc}
\hline \mbox{label}& \null & \mbox{root number} &\null &
\mbox{Dynkin label} & \mbox{type II B }&\mbox{type II A }
\\\hline {d_{3|1}} & = &  {\alpha [} 22 ]& = &
 \{0,0,0,1,1,1,0,0\}  \nonumber & B_{7\,8} & \gamma_{7}^{\phantom{7}\, 8} \\
{d_{3|2}} & = &  {\alpha [} 4 ]& = &
 \{0,0,0,1,0,0,0,0\}  \nonumber &  \gamma_{7}{}^{8}&\gamma_{7}{}^{8}\\
{d_{3|3}} & = &  {\alpha [} 14 ]& = &
 \{0,0,0,1,0,1,0,0\}  \nonumber & B_{7\,9} & \gamma_{7}{}^{9}\\
{d_{3|4}} & = &  {\alpha [} 13 ]& = &
 \{0,0,0,1,1,0,0,0\}  \nonumber &  \gamma_{7}{}^{9} & B_{7\,9} \\
{d_{3|5}} & = &  {\alpha [} 29 ]& = &
 \{0,0,0,1,1,1,1,0\}  \nonumber& C_{7\,8} & C_{7\,8\,9} \\
{d_{3|6}} & = &  {\alpha [} 21 ]& = &
 \{0,0,0,1,0,1,1,0\}  \nonumber & C_{7\,9} & C_{7} \\
{d_{4|1}} & = &  {\alpha [} 36 ]& = &
 \{0,0,1,2,1,1,0,0\}  \nonumber &B_{6\,7} & B_{6\,7} \\
{d_{4|2}} & = &  {\alpha [} 3 ]& = &
 \{0,0,1,0,0,0,0,0\}  \nonumber &\gamma_{6}{}^{7}  & \gamma_{6}{}^{7} \\
{d_{4|3}} & = &  {\alpha [} 28 ]& = &
 \{0,0,1,1,1,1,0,0\}  \nonumber &B_{6\,8} & B_{6\,8}\\
{d_{4|4}} & = &  {\alpha [} 12 ]& = &
 \{0,0,1,1,0,0,0,0\}  \nonumber &\gamma_{6}{}^{8}  &\gamma_{6}{}^{8} \\
{d_{4|5}} & = &  {\alpha [} 20 ]& = &
 \{0,0,1,1,0,1,0,0\}  \nonumber &B_{6\,9} &\gamma_{6}{}^{9} \\
{d_{4|6}} & = &  {\alpha [} 19 ]& = &
 \{0,0,1,1,1,0,0,0\}  \nonumber &\gamma_{6}{}^{9} &B_{6\,9} \\
{d_{4|7}} & = &  {\alpha [} 41 ]& = &
 \{0,0,1,2,1,1,1,0\}  \nonumber &C_{6\,7} & C_{6\,7\,9} \\
{d_{4|8}} & = &  {\alpha [} 34 ]& = &
 \{0,0,1,1,1,1,1,0\}  \nonumber &C_{6\,8} & C_{6\,8\,9}\\
{d_{4|9}} & = &  {\alpha [} 27 ]& = &
 \{0,0,1,1,0,1,1,0\}  \nonumber &C_{6\,9} & C_{6}\\
{d_{4|10}} & = &  {\alpha [} 49 ]& = &
 \{0,0,1,2,1,2,1,0\}  \nonumber&C_{6\,7\,8\,9} & C_{6\,7\,8} \\
{d_{5|1}} & = &  {\alpha [} 47 ]& = &
 \{0,1,2,2,1,1,0,0\}  \nonumber  &B_{5\,6} & B_{5\,6}\\
{d_{5|2}} & = &  {\alpha [} 2 ]& = &
 \{0,1,0,0,0,0,0,0\}  \nonumber &\gamma_{5}{}^{6} & \gamma_{5}{}^{6} \\
{d_{5|3}} & = &  {\alpha [} 40 ]& = &
 \{0,1,1,2,1,1,0,0\}  \nonumber &B_{5\,7} & B_{5\,7} \\
{d_{5|4}} & = &  {\alpha [} 11 ]& = &
 \{0,1,1,0,0,0,0,0\}  \nonumber&\gamma_{5}{}^{7} &\gamma_{5}{}^{7} \\
{d_{5|5}} & = &  {\alpha [} 33 ]& = &
 \{0,1,1,1,1,1,0,0\}  \nonumber &B_{5\,8} & B_{5\,8}\\
{d_{5|6}} & = &  {\alpha [} 18 ]& = &
 \{0,1,1,1,0,0,0,0\}  \nonumber &\gamma_{5}{}^{8}  &\gamma_{5}{}^{8}  \\
{d_{5|7}} & = &  {\alpha [} 25 ]& = &
 \{0,1,1,1,0,1,0,0\}  \nonumber &B_{5\,9} &\gamma_{5}{}^{9}\\
{d_{5|8}} & = &  {\alpha [} 24 ]& = &
 \{0,1,1,1,1,0,0,0\}  \nonumber &\gamma_{5}{}^{9} &B_{5\,9}\\
{d_{5|9}} & = &  {\alpha [} 54 ]& = &
 \{0,1,2,2,1,1,1,0\}  \nonumber &C_{5\,6} & C_{5\,6\,9} \\
{d_{5|10}} & = &  {\alpha [} 46 ]& = &
 \{0,1,1,2,1,1,1,0\}  \nonumber &C_{5\,7} & C_{5\,7\,9}\\
{d_{5|11}} & = &  {\alpha [} 39 ]& = &
 \{0,1,1,1,1,1,1,0\}  \nonumber&C_{5\,8} & C_{5\,8\,9} \\
{d_{5|12}} & = &  {\alpha [} 32 ]& = &
 \{0,1,1,1,0,1,1,0\}  \nonumber &C_{5\,9} & C_{5} \\
{d_{5|13}} & = &  {\alpha [} 53 ]& = &
 \{0,1,1,2,1,2,1,0\}  \nonumber &C_{5\,7\,8\,9} & C_{5\,7\,8}\\
{d_{5|14}} & = &  {\alpha [} 60 ]& = &
 \{0,1,2,2,1,2,1,0\}  \nonumber &C_{5\,6\,8\,9} & C_{5\,6\,8}\\
{d_{5|15}} & = &  {\alpha [} 67 ]& = &
 \{0,1,2,3,1,2,1,0\}  \nonumber &C_{5\,6\,7\,9} & C_{5\,6\,7}\\
{d_{5|16}} & = &  {\alpha [} 74 ]& = &
 \{0,1,2,3,2,2,1,0\}  \nonumber &C_{5\,6\,7\,8} & C_{3\, 4\,\mu}\\
\end{array}
\end{eqnarray*}
{{Roots in D[} 6 ]}
\begin{eqnarray*}
\begin{array}{lcccl|cc}
\hline \mbox{label}& \null & \mbox{root number} &\null &
\mbox{Dynkin label} & \mbox{type II B }&\mbox{type II A }
 \\\hline
{d_{6|1}} & = &  {\alpha [} 59 ]& = &
 \{1,2,2,2,1,1,0,0\}  \nonumber &B_{4\,5} & B_{4\,5} \\
{d_{6|2}} & = &  {\alpha [} 1 ]& = &
 \{1,0,0,0,0,0,0,0\}  \nonumber &\gamma_{4}{}^{5} &\gamma_{4}{}^{5}\\
{d_{6|3}} & = &  {\alpha [} 52 ]& = &
 \{1,1,2,2,1,1,0,0\}  \nonumber &B_{4\,6} & B_{4\,6}\\
{d_{6|4}} & = &  {\alpha [} 9 ]& = &
 \{1,1,0,0,0,0,0,0\}  \nonumber &\gamma_{4}{}^{6}&\gamma_{4}{}^{6}\\
{d_{6|5}} & = &  {\alpha [} 45 ]& = &
 \{1,1,1,2,1,1,0,0\}  \nonumber &B_{4\,7} & B_{4\,7}\\
{d_{6|6}} & = &  {\alpha [} 16 ]& = &
 \{1,1,1,0,0,0,0,0\}  \nonumber &\gamma_{4}{}^{7} &\gamma_{4}{}^{7}\\\
{d_{6|7}} & = &  {\alpha [} 38 ]& = &
 \{1,1,1,1,1,1,0,0\}  \nonumber &B_{4\,8} & B_{4\,8}\\
{d_{6|8}} & = &  {\alpha [} 23 ]& = &
 \{1,1,1,1,0,0,0,0\}  \nonumber &\gamma_{4}{}^{8}&\gamma_{4}{}^{8}\\
{d_{6|9}} & = &  {\alpha [} 31 ]& = &
 \{1,1,1,1,0,1,0,0\}  \nonumber &B_{4\,9} & \gamma_{4}{}^{9}\\
{d_{6|10}} & = &  {\alpha [} 30 ]& = &
 \{1,1,1,1,1,0,0,0\}  \nonumber &\gamma_{4}{}^{9} &B_{4\,9}\\
{d_{6|11}} & = &  {\alpha [} 100 ]& = &
 \{1,2,3,4,2,3,2,0\}  \nonumber  &B_{3\,\mu} & B_{3\,\mu}\\
{d_{6|12}} & = &  {\alpha [} 96 ]& = &
 \{1,2,3,4,2,3,1,0\}  \nonumber &C_{3\,\mu} & C_{3\,9\,\mu}\\
{d_{6|13}} & = &  {\alpha [} 64 ]& = &
 \{1,2,2,2,1,1,1,0\}  \nonumber&C_{4\,5} & C_{4\,5\,9} \\
{d_{6|14}} & = &  {\alpha [} 58 ]& = &
 \{1,1,2,2,1,1,1,0\}  \nonumber&C_{4\,6} & C_{4\,6\,9} \\
{d_{6|15}} & = &  {\alpha [} 51 ]& = &
 \{1,1,1,2,1,1,1,0\}  \nonumber&C_{4\,7} & C_{4\,7\,9} \\
{d_{6|16}} & = &  {\alpha [} 44 ]& = &
 \{1,1,1,1,1,1,1,0\}  \nonumber&C_{4\,8} & C_{4\,8\,9} \\
{d_{6|17}} & = &  {\alpha [} 37 ]& = &
 \{1,1,1,1,0,1,1,0\}  \nonumber &C_{4\,9} & C_{4}\\
{d_{6|18}} & = &  {\alpha [} 57 ]& = &
 \{1,1,1,2,1,2,1,0\}  \nonumber&C_{4\,7\,8\,9} & C_{4\,7\,8}  \\
{d_{6|19}} & = &  {\alpha [} 63 ]& = &
 \{1,1,2,2,1,2,1,0\}  \nonumber&C_{4\,6\,8\,9} & C_{4\,6\,8} \\
{d_{6|20}} & = &  {\alpha [} 70 ]& = &
 \{1,1,2,3,1,2,1,0\}  \nonumber &C_{4\,6\,7\,9} & C_{4\,6\,7}\\
{d_{6|21}} & = &  {\alpha [} 76 ]& = &
 \{1,1,2,3,2,2,1,0\}  \nonumber &C_{4\,6\,7\,8} & C_{3\,5\,\mu}\\
{d_{6|22}} & = &  {\alpha [} 71 ]& = &
 \{1,2,2,2,1,2,1,0\}  \nonumber &C_{4\,5\,8\,9} & C_{4\,5\,8}\\
{d_{6|23}} & = &  {\alpha [} 77 ]& = &
 \{1,2,2,3,1,2,1,0\}  \nonumber &C_{4\,5\,7\,9} & C_{4\,5\,7}\\
{d_{6|24}} & = &  {\alpha [} 81 ]& = &
 \{1,2,2,3,2,2,1,0\}  \nonumber &C_{4\,5\,7\,8} & C_{3\,6\,\mu}\\
{d_{6|25}} & = &  {\alpha [} 83 ]& = &
 \{1,2,3,3,1,2,1,0\}  \nonumber &C_{4\,5\,6\,9} & C_{4\,5\,6}\\
{d_{6|26}} & = &  {\alpha [} 87 ]& = &
 \{1,2,3,3,2,2,1,0\}  \nonumber &C_{4\,5\,6\,8} & C_{3\,7\,\mu}\\
{d_{6|27}} & = &  {\alpha [} 92 ]& = &
 \{1,2,3,4,2,2,1,0\}  \nonumber &C_{4\,5\,6\,7} & C_{3\,8\,\mu}\\
\end{array}
\end{eqnarray*}
\newpage
{{Roots in D[} 7 ]}\nonumber \\
{Electric with respect to the electric subgroup $\mathrm{SL(8)} \subset
\mathrm{E_{7(7)}} \subset\mathrm{ E_{8(8)}}$}
\begin{eqnarray*}
\begin{array}{lclclcl|cc}
\hline
\mbox{label}& \null & \mbox{root $\#$} &\null & \mbox{Dynkin label}
&\null & \mbox{q-vector} &\mbox{type II B}&\mbox{type II A} \\
\hline
{d_{7|1}} & = &  {\alpha [} 50 ]& = &
 \{1,1,1,1,0,1,1,1\} & \Rightarrow &  \{1,1,1,1,0,1,1\} \nonumber& C_{3\,9}&C_{3} \\
{d_{7|2}} & = &  {\alpha [} 99 ]& = &
 \{1,2,3,4,2,3,1,1\} & \Rightarrow &  \{1,2,3,4,2,3,1\} &  C_{4\,\mu}&C_{4\,9\,\mu}
\nonumber \\
{d_{7|3}} & = &  {\alpha [} 101 ]& = &
 \{2,2,3,4,2,3,1,1\} & \Rightarrow &  \{2,2,3,4,2,3,1\} &  C_{5\,\mu}&C_{5\,9\,\mu}
\nonumber \\
{d_{7|4}} & = &  {\alpha [} 105 ]& = &
 \{2,3,3,4,2,3,1,1\} & \Rightarrow &  \{2,3,3,4,2,3,1\} & C_{6\,\mu}&C_{6\,9\,\mu}
\nonumber \\
{d_{7|5}} & = &  {\alpha [} 108 ]& = &
 \{2,3,4,4,2,3,1,1\} & \Rightarrow &  \{2,3,4,4,2,3,1\} & C_{7\,\mu}&C_{7\,9\,\mu}
\nonumber \\
{d_{7|6}} & = &  {\alpha [} 110 ]& = &
 \{2,3,4,5,2,3,1,1\} & \Rightarrow &  \{2,3,4,5,2,3,1\} & C_{8\,\mu}&C_{8\,9\,\mu}
\nonumber \\
{d_{7|7}} & = &  {\alpha [} 84 ]& = &
 \{1,1,2,3,2,2,1,1\} & \Rightarrow &  \{1,1,2,3,2,2,1\} &  C_{3\,6\,7\,8}&
C_{4\,5\,\mu} \nonumber \\
{d_{7|8}} & = &  {\alpha [} 86 ]& = &
 \{1,2,2,3,2,2,1,1\} & \Rightarrow &  \{1,2,2,3,2,2,1\} & C_{3\,5\,7\,8}&
C_{4\,6\,\mu} \nonumber \\
{d_{7|9}} & = &  {\alpha [} 91 ]& = &
 \{1,2,3,3,2,2,1,1\} & \Rightarrow &  \{1,2,3,3,2,2,1\} & C_{3\,5\,6\,8}&
C_{4\,7\,\mu}\nonumber \\
{d_{7|10}} & = &  {\alpha [} 95 ]& = &
 \{1,2,3,4,2,2,1,1\} & \Rightarrow &  \{1,2,3,4,2,2,1\} &  C_{3\,5\,6\,7}&
C_{4\,8\,\mu} \nonumber \\
{d_{7|11}} & = &  {\alpha [} 97 ]& = &
 \{2,2,3,4,2,2,1,1\} & \Rightarrow &  \{2,2,3,4,2,2,1\} & C_{3\,4\,6\,7}&
C_{5\,8\,\mu}\nonumber \\
{d_{7|12}} & = &  {\alpha [} 102 ]& = &
 \{2,3,3,4,2,2,1,1\} & \Rightarrow &  \{2,3,3,4,2,2,1\} & C_{3\,4\,5\,7}&
C_{6\,8\,\mu} \nonumber \\
{d_{7|13}} & = &  {\alpha [} 106 ]& = &
 \{2,3,4,4,2,2,1,1\} & \Rightarrow &  \{2,3,4,4,2,2,1\} & C_{3\,4\,5\,6}&
C_{7\,8\,\mu}  \nonumber \\
{d_{7|14}} & = &  {\alpha [} 93 ]& = &
 \{2,2,3,3,2,2,1,1\} & \Rightarrow &  \{2,2,3,3,2,2,1\} & C_{3\,4\,6\,8}&
C_{5\,7\,\mu}  \nonumber \\
{d_{7|15}} & = &  {\alpha [} 89 ]& = &
 \{2,2,2,3,2,2,1,1\} & \Rightarrow &  \{2,2,2,3,2,2,1\} & C_{3\,4\,7\,8}&
C_{5\,6\,\mu}\nonumber \\
{d_{7|16}} & = &  {\alpha [} 98 ]& = &
 \{2,3,3,3,2,2,1,1\} & \Rightarrow &  \{2,3,3,3,2,2,1\} & C_{3\,4\,5\,8}&
C_{6\,7\,\mu}\nonumber \\
{d_{7|17}} & = &  {\alpha [} 103 ]& = &
 \{1,2,3,4,2,3,2,1\} & \Rightarrow &  \{1,2,3,4,2,3,2\} &  B_{4\,\mu}& B_{4\,\mu}
 \nonumber \\
{d_{7|18}} & = &  {\alpha [} 104 ]& = &
 \{2,2,3,4,2,3,2,1\} & \Rightarrow &  \{2,2,3,4,2,3,2\} & B_{5\,\mu}&
B_{5\,\mu}\nonumber \\
{d_{7|19}} & = &  {\alpha [} 107 ]& = &
 \{2,3,3,4,2,3,2,1\} & \Rightarrow &  \{2,3,3,4,2,3,2\} & B_{6\,\mu}&
B_{6\,\mu}\nonumber \\
{d_{7|20}} & = &  {\alpha [} 109 ]& = &
 \{2,3,4,4,2,3,2,1\} & \Rightarrow &  \{2,3,4,4,2,3,2\} & B_{7\,\mu}& B_{7\,\mu}
\nonumber \\
{d_{7|21}} & = &  {\alpha [} 111 ]& = &
 \{2,3,4,5,2,3,2,1\} & \Rightarrow &  \{2,3,4,5,2,3,2\} & B_{8\,\mu}&
B_{8\,\mu}\nonumber \\
{d_{7|22}} & = &  {\alpha [} 114 ]& = &
 \{2,3,4,5,2,4,2,1\} & \Rightarrow &  \{2,3,4,5,2,4,2\} & \gamma_\mu{}^{9}& B_{9\mu}
\nonumber \\
{d_{7|23}} & = &  {\alpha [} 8 ]& = &
 \{0,0,0,0,0,0,0,1\} & \Rightarrow &  \{0,0,0,0,0,0,0\} &
\gamma_{3}{}^{4}&\gamma_{3}{}^{4}  \nonumber \\
{d_{7|24}} & = &  {\alpha [} 10 ]& = &
 \{1,0,0,0,0,0,0,1\} & \Rightarrow &  \{1,0,0,0,0,0,0\} & \gamma_{3}{}^{5}&
\gamma_{3}{}^{5} \nonumber \\
{d_{7|25}} & = &  {\alpha [} 17 ]& = &
 \{1,1,0,0,0,0,0,1\} & \Rightarrow &  \{1,1,0,0,0,0,0\}
&\gamma_{3}{}^{6}&\gamma_{3}{}^{6}  \nonumber \\
{d_{7|26}} & = &  {\alpha [} 26 ]& = &
 \{1,1,1,0,0,0,0,1\} & \Rightarrow &  \{1,1,1,0,0,0,0\} & \gamma_{3}{}^{7} &
\gamma_{3}{}^{7} \nonumber \\
{d_{7|27}} & = &  {\alpha [} 35 ]& = &
 \{1,1,1,1,0,0,0,1\} & \Rightarrow &  \{1,1,1,1,0,0,0\} &  \gamma_{3}{}^{8} &
\gamma_{3}{}^{8} \nonumber \\
{d_{7|28}} & = &  {\alpha [} 43 ]& = &
 \{1,1,1,1,0,1,0,1\} & \Rightarrow &  \{1,1,1,1,0,1,0\} & B_{3\,9}&
\gamma_{3}{}^9\nonumber \\
 \end{array}
 \end{eqnarray*}
 \newpage
{{Roots in D[} 7 ]}\nonumber \\
{Magnetic with respect to the electric subgroup $\mathrm{SL(8)} \subset
\mathrm{E_{7(7)}} \subset\mathrm{ E_{8(8)}}$}
\begin{eqnarray*}
\begin{array}{lclclclcc}
{d_{7|29}} & = &  {\alpha [} 112 ]& = &
 \{2,3,4,5,3,3,1,1\} & \Rightarrow &  \{2,3,4,5,3,3,1\} & C_{9\mu} & C_\mu\nonumber \\
{d_{7|30}} & = &  {\alpha [} 75 ]& = &
 \{2,2,2,2,1,1,1,1\} & \Rightarrow &  \{2,2,2,2,1,1,1\} & C_{3\,4} & C_{3\,4\,9}
 \nonumber \\
{d_{7|31}} & = &  {\alpha [} 72 ]& = &
 \{1,2,2,2,1,1,1,1\} & \Rightarrow &  \{1,2,2,2,1,1,1\} &  C_{3\,5} &
C_{3\,5\,9}\nonumber \\
{d_{7|32}} & = &  {\alpha [} 66 ]& = &
 \{1,1,2,2,1,1,1,1\} & \Rightarrow &  \{1,1,2,2,1,1,1\} &  C_{3\,6} &
C_{3\,6\,9}\nonumber \\
{d_{7|33}} & = &  {\alpha [} 62 ]& = &
 \{1,1,1,2,1,1,1,1\} & \Rightarrow &  \{1,1,1,2,1,1,1\} & C_{3\,7} &
C_{3\,7\,9}\nonumber \\
{d_{7|34}} & = &  {\alpha [} 56 ]& = &
 \{1,1,1,1,1,1,1,1\} & \Rightarrow &  \{1,1,1,1,1,1,1\} & C_{3\,8} &
C_{3\,8\,9}\nonumber \\
{d_{7|35}} & = &  {\alpha [} 94 ]& = &
 \{2,3,3,3,1,2,1,1\} & \Rightarrow &  \{2,3,3,3,1,2,1\} & C_{3\,4\,5\,9} &
C_{3\,4\,5}\nonumber \\
{d_{7|36}} & = &  {\alpha [} 90 ]& = &
 \{2,2,3,3,1,2,1,1\} & \Rightarrow &  \{2,2,3,3,1,2,1\} & C_{3\,4\,6\,9} &
C_{3\,4\,6}\nonumber \\
{d_{7|37}} & = &  {\alpha [} 85 ]& = &
 \{2,2,2,3,1,2,1,1\} & \Rightarrow &  \{2,2,2,3,1,2,1\} & C_{3\,4\,7\,9} &
C_{3\,4\,7}\nonumber \\
{d_{7|38}} & = &  {\alpha [} 80 ]& = &
 \{2,2,2,2,1,2,1,1\} & \Rightarrow &  \{2,2,2,2,1,2,1\} & C_{3\,4\,8\,9} &
C_{3\,4\,8}\nonumber \\
{d_{7|39}} & = &  {\alpha [} 78 ]& = &
 \{1,2,2,2,1,2,1,1\} & \Rightarrow &  \{1,2,2,2,1,2,1\} &  C_{3\,5\,8\,9} &
C_{3\,5\,8}\nonumber \\
{d_{7|40}} & = &  {\alpha [} 73 ]& = &
 \{1,1,2,2,1,2,1,1\} & \Rightarrow &  \{1,1,2,2,1,2,1\} & C_{3\,6\,8\,9} &
C_{3\,6\,8}\nonumber \\
{d_{7|41}} & = &  {\alpha [} 68 ]& = &
 \{1,1,1,2,1,2,1,1\} & \Rightarrow &  \{1,1,1,2,1,2,1\} & C_{3\,7\,8\,9} &
C_{3\,7\,8}\nonumber \\
{d_{7|42}} & = &  {\alpha [} 82 ]& = &
 \{1,2,2,3,1,2,1,1\} & \Rightarrow &  \{1,2,2,3,1,2,1\} & C_{3\,5\,7\,9} &
C_{3\,5\,7}\nonumber \\
{d_{7|43}} & = &  {\alpha [} 88 ]& = &
 \{1,2,3,3,1,2,1,1\} & \Rightarrow &  \{1,2,3,3,1,2,1\} & C_{3\,5\,6\,9} &
C_{3\,5\,6}\nonumber \\
{d_{7|44}} & = &  {\alpha [} 79 ]& = &
 \{1,1,2,3,1,2,1,1\} & \Rightarrow &  \{1,1,2,3,1,2,1\} & C_{3\,6\,7\,9} &
C_{3\,6\,7}\nonumber \\
{d_{7|45}} & = &  {\alpha [} 69 ]& = &
 \{2,2,2,2,1,1,0,1\} & \Rightarrow &  \{2,2,2,2,1,1,0\} & B_{3\,4} &
B_{3\,4}\nonumber \\
{d_{7|46}} & = &  {\alpha [} 65 ]& = &
 \{1,2,2,2,1,1,0,1\} & \Rightarrow &  \{1,2,2,2,1,1,0\} & B_{3\,5} &
B_{3\,5}\nonumber \\
{d_{7|47}} & = &  {\alpha [} 61 ]& = &
 \{1,1,2,2,1,1,0,1\} & \Rightarrow &  \{1,1,2,2,1,1,0\} & B_{3\,6} &
B_{3\,6}\nonumber \\
{d_{7|48}} & = &  {\alpha [} 55 ]& = &
 \{1,1,1,2,1,1,0,1\} & \Rightarrow &  \{1,1,1,2,1,1,0\} & B_{3\,7} &
B_{3\,7}\nonumber \\
{d_{7|49}} & = &  {\alpha [} 48 ]& = &
 \{1,1,1,1,1,1,0,1\} & \Rightarrow &  \{1,1,1,1,1,1,0\} & B_{3\,8} &
B_{3\,8}\nonumber \\
{d_{7|50}} & = &  {\alpha [} 42 ]& = &
 \{1,1,1,1,1,0,0,1\} & \Rightarrow &  \{1,1,1,1,1,0,0\} & \gamma_{3}{}^{9} &
B_{39}\nonumber \\
{d_{7|51}} & = &  {\alpha [} 119 ]& = &
 \{3,4,5,6,3,4,2,1\} & \Rightarrow &  \{3,4,5,6,3,4,2\} &  \gamma_{\mu}{}^{4} &
\gamma_{\mu}{}^{4}\nonumber \\
{d_{7|52}} & = &  {\alpha [} 118 ]& = &
 \{2,4,5,6,3,4,2,1\} & \Rightarrow &  \{2,4,5,6,3,4,2\} & \gamma_{\mu}{}^{5}  &
\gamma_{\mu}{}^{5}\nonumber \\
{d_{7|53}} & = &  {\alpha [} 117 ]& = &
 \{2,3,5,6,3,4,2,1\} & \Rightarrow &  \{2,3,5,6,3,4,2\} &  \gamma_{\mu}{}^{6}&
\gamma_{\mu}{}^{6}\nonumber \\
{d_{7|54}} & = &  {\alpha [} 116 ]& = &
 \{2,3,4,6,3,4,2,1\} & \Rightarrow &  \{2,3,4,6,3,4,2\} &  \gamma_{\mu}{}^{7} &
\gamma_{\mu}{}^{7}\nonumber \\
{d_{7|55}} & = &  {\alpha [} 115 ]& = &
 \{2,3,4,5,3,4,2,1\} & \Rightarrow &  \{2,3,4,5,3,4,2\} & \gamma_{\mu}{}^{8} &
\gamma_{\mu}{}^{8}\nonumber \\
{d_{7|56}} & = &  {\alpha [} 113 ]& = &
 \{2,3,4,5,3,3,2,1\} & \Rightarrow &  \{2,3,4,5,3,3,2\} & B_{9\,\mu} & \gamma_{\mu}{}^{9}\nonumber \\
\end{array}
\end{eqnarray*}
{{Roots in D[} 8 ]}
\begin{eqnarray*}
\begin{array}{lcccl|cc}
\hline \mbox{label}& \null & \mbox{root number} &\null &
\mbox{Dynkin label} & \mbox{type II B}&\mbox{type II A} \\\hline
{d_{8|1}} & = &  {\alpha [} 120 ]& = &
 \{3,4,5,6,3,4,2,2\}&  \gamma_\mu{}^3& \gamma_\mu{}^3
 \end{array}
\end{eqnarray*}
\section{Bosonic Field Equations of type II B supergravity}
The bosonic part of the equations can be formally obtained through
variation of the following action \footnote{Note that our $R$ is
equal to $- \ft 1 2 R^{old}$, $R^{old}$ being the normalization of
the scalar curvature usually adopted in General Relativity
textbooks. The difference arises because in the traditional
literature the Riemann tensor is not defined as the components of
the curvature $2$-form $R^{ab}$ rather as $-2$ times such
components.}:
\[
S_{\rm II B} = \frac{1}{2 \kappa^2} \Bigg\{ \int d^{10} x~ \left[
-2  \sqrt{-\det g}~ R \right] - \frac{1}{2} \int \Big[ d \varphi
\wedge \star d \varphi
 \,+\, {\rm e}^{- \varphi} F_{[3]}^{NS}  \wedge \star F_{[3]}^{NS}\,+\, {\rm e}^{2
 \varphi}\, F_{[1]}^{RR} \wedge \star F_{[1]}^{RR}
\]
\begin{equation}
 + \,\,{\rm e}^{\varphi} \,{F}_{[3]}^{RR} \wedge \star
 {F}_{[3]}^{RR} \,+\, \frac{1}{2}\, {F}_{[5]}^{RR}
 \wedge \star {F}_{[5]}^{RR}  \, -\,  C_{[4]} \wedge
 F_{[3]}^{NS}
 \wedge F_{[3]}^{RR} \Big] \Bigg\}
\label{bulkaction}
\end{equation}
where:
\begin{eqnarray}
F^{RR}_{[1]} & = & dC_{[0]} \nonumber\\
F^{NS}_{[3]} & = & dB_{[2]} \nonumber\\
F^{RR}_{[3]}& = & dC_{[2]} -  \, C_{[0]} \,
dB_{[2]}\nonumber\\
F^{RR}_{[5]}& = & dC_{[4]}-  \ft 12 \,\left( B_{[2]} \wedge d C_{[2]}
-  C_{[2]} \wedge d B_{[2]}\right) \label{bosecurve}
\end{eqnarray}
It is important to stress though that the action
(\ref{bulkaction}) is to be considered only a book keeping device
since the $4$-form $C_{[4]}$ is not free, its field strength
$F_{[5]}^{RR}$  being subject to the on-shell self-duality
constraint:
\begin{equation}
F_{[5]}^{RR} = \star F_{[5]}^{RR} \label{selfonshell}
\end{equation}
From the above action the corresponding equations of motion can be
obtained:
\begin{eqnarray}
d \star d \varphi - e^{2\varphi} \, F^{RR}_{[1]} \wedge \star
F^{RR}_{[1]} & = & -\ft 1 2 \, \left( e^{-\varphi} F^{NS}_{[3]}
\wedge \star  F^{NS}_{[3]}-
  e^{\varphi} F^{RR}_{[3]} \wedge \star  F^{RR}_{[3]}\right) \label{NSscalapr}\\
  d\left( e^{2\varphi} \star F^{RR}_{[1]}\right)  & = & - e^{\varphi} \,
F^{NS}_{[3]} \wedge \star  F^{RR}_{[3]}
\label{RRscalapr}\\
d\left( e^{-\varphi} \, \star F_{[3]}^{NS}\right) + e^\varphi \,
F^{RR}_{[1]} \wedge \star F^{RR}_{[3]}
  & = &  - F_{[3]}^{RR} \wedge F^{RR}_{[5]}
\label{3formNS}\\
d\left( e^\varphi \star F_{[3]}^{RR } \right) & = & -F_{[5]}^{RR}
\, \wedge F_{[3]}^{NS}
\label{3formRR}\\
d\star F^{RR}_{[5]} & = & -
  F^{NS}_{[3]} \, \wedge \, F^{RR}_{[3]}
\label{f5RR}\\
-\,2 \,R_{{MN}}&=& \frac{1}{2}\partial_{{M}}\varphi
\partial_{{N}}\varphi+\frac{e^{2\varphi}}{2}
\partial_{{M}} C_{[0]} \partial_{{N}}
C_{[0]}+150
 {F}_{[5]{M}\cdot\cdot\cdot\cdot}
{F}_{[5]{N}}^{\phantom{{M}}\cdot\cdot\cdot\cdot}
\nonumber\\
& &+ 9 \left( e^{-\varphi}F_{[3]{M}\cdot\cdot}^{NS}\,
F_{[3]{N}}^{{NS}\phantom{{M}}\cdot\cdot} +e^{\varphi}{
F}_{[3]{M}\cdot\cdot}^{RR}
{ F}_{[3]{N}}^{RR\phantom{{M}}\cdot\cdot}\right)\nonumber\\
& & -\frac{3}{4}\,
g_{{MN}}\,\left(e^{-\varphi}F_{[3]\cdot\cdot\cdot}^{NS}
F_{[3]}^{NS\cdot\cdot\cdot}+e^{\varphi}{{F}}_{[3]\cdot\cdot\cdot}^{RR}{
F}^{RR\cdot\cdot\cdot}_{[3]}\right) \label{einsteinequa}
\end{eqnarray}
It is not difficult to show, upon suitable identification of the
massless superstring fields, that this is the correct set of
equations which can be consistently obtained from the manifestly
$\mathrm{SU(1,1)}$ covariant formulation of type II B supergravity
\cite{II B}.
\section{A useful integral}
\begin{equation}
  \int \, \exp[a \, x] \, \left ( \cosh[b \, x] \right) ^{1/4} =
  \frac{2^{7/4}}{4a-b}\, \exp[(a-b)x] \, \,  _2 F_1\left( -\ft 18 + \ft
  {a}{2b} , - \ft 14, \ft 78 + \ft {a}{2b} \,; \, - e^{2bx} \right)
\label{star}
\end{equation}


\begin{thebibliography}{99}
\bibitem{experiment} Riess A G {\it et al.} 1998 {\it Astron.\ J.}
 {\bf 116}1009, Perlmutter S {\it et al.} 1999 {\it Astron.\ J.} {\bf 517} 565,
Sievers J L  {\it et al.} 2002 {\it Preprint} astro-ph/0205387
\bibitem{linde90} Linde A D 1990 {\it Particle Physics and Inflationary Cosmology}
 (Switzerland: Harwood Academic)
\bibitem{Kachru:2003sx}
S.~Kachru, R.~Kallosh, A.~Linde, J.~Maldacena, L.~McAllister and S.~P.~Trivedi,
\emph{Towards inflation in string theory},
[arXiv:hep-th/0308055].
\bibitem{Kachru:2003aw}
S.~Kachru, R.~Kallosh, A.~Linde and S.~P.~Trivedi, \emph{De Sitter
vacua in string theory}, Phys.\ Rev.\ D {\bf 68} (2003) 046005
[arXiv:hep-th/0301240].
\bibitem{Burgess:2003ic}
C.~P.~Burgess, R.~Kallosh and F.~Quevedo,
\emph{de Sitter String Vacua from Supersymmetric D-terms},
[arXiv:hep-th/0309187].
\bibitem{Fre:2002pd}
P.~Fre, M.~Trigiante and A.~Van Proeyen, \emph{Stable de Sitter
vacua from N = 2 supergravity}, Class.\ Quant.\ Grav.\  {\bf 19}
(2002) 4167 [arXiv:hep-th/0205119];
M.~de Roo, D.~B.~Westra,
S.~Panda and M.~Trigiante,
 \emph{Potential and mass-matrix in gauged N = 4 supergravity},
JHEP {\bf 0311} (2003) 022 [arXiv:hep-th/0310187].
\bibitem{Gutperle:2002ai}
M.~Gutperle and A.~Strominger, \emph{Spacelike branes}, JHEP {\bf
0204} (2002) 018 [arXiv:hep-th/0202210].
\bibitem{otherSbranes}
V.~D.~Ivashchuk and V.~N.~Melnikov, \emph{Multidimensional
classical and quantum cosmology with intersecting  p-branes}, J.\
Math.\ Phys.\  {\bf 39} (1998) 2866 [arXiv:hep-th/9708157];
L.~Cornalba, M.~S.~Costa and C.~Kounnas, \emph{A resolution of the
cosmological singularity with orientifolds}, Nucl.\ Phys.\ B {\bf
637} (2002) 378 [arXiv:hep-th/0204261];
L.~Cornalba and M.~S.~Costa, \emph{On the classical stability of
orientifold cosmologies}, Class.\ Quant.\ Grav.\  {\bf 20} (2003)
3969 [arXiv:hep-th/0302137].
F.~Leblond and A.~W.~Peet, \emph{A note on the singularity theorem
for supergravity SD-branes}, [arXiv:hep-th/0305059];
M.~Kruczenski, R.~C.~Myers and A.~W.~Peet, \emph{Supergravity
S-branes}, JHEP {\bf 0205} (2002) 039 [arXiv:hep-th/0204144];
N.~Ohta, \emph{Accelerating cosmologies from S-branes}, Phys.\
Rev.\ Lett.\  {\bf 91} (2003) 061303 [arXiv:hep-th/0303238];
R.~Emparan and J.~Garriga, \emph{A note on accelerating
cosmologies from compactifications and S-branes}, JHEP {\bf 0305}
(2003) 028 [arXiv:hep-th/0304124];
A.~Buchel and J.~Walcher, \emph{Comments on supergravity
description of S-branes}, JHEP {\bf 0305} (2003) 069
[arXiv:hep-th/0305055].
\bibitem{Papadopoulos:2002bg}
G.~Papadopoulos, J.~G.~Russo and A.~A.~Tseytlin, \emph{Solvable
model of strings in a time-dependent plane-wave background},
Class.\ Quant.\ Grav.\  {\bf 20} (2003) 969
[arXiv:hep-th/0211289].
\bibitem{que} F.~Quevedo, \emph{Lectures on string / brane cosmology},
hep-th/0210292.
\bibitem{GV} M.~Gasperini and G.~Veneziano, \emph{The pre-big bang scenario
in string cosmology}, hep-th/0207130.
\bibitem{craps} B.~Craps, D.~Kutasov and G.~Rajesh, \emph{String
propagation in the presence of cosmological singularities}, JHEP
{\bf 0206}, 053 (2002) [hep-th/0205101].
\bibitem{ban} T.~Banks and W.~Fischler, \emph{M-theory observables for
cosmological space-times}, hep-th/0102077.
\bibitem{setu} J.~Khoury, B.~A.~Ovrut, N.~Seiberg, P.~J.~Steinhardt and
N.~Turok, \emph{From big crunch to big bang}, Phys.\ Rev.\ D {\bf
65}, 086007 (2002) [hep-th/0108187].
\bibitem{cope}
J.~E.~Lidsey, D.~Wands and E.~J.~Copeland, \emph{Superstring
cosmology}, Phys.\ Rept.\  {\bf 337}, 343 (2000)
[hep-th/9909061].
\bibitem{mart} A.~E.~Lawrence and E.~J.~Martinec, \emph{String field theory
in curved spacetime and the resolution of spacelike
singularities}, Class.\ Quant.\ Grav.\  {\bf 13}, 63 (1996)
[hep-th/9509149].
%
\bibitem{Sen:2002vv}
A.~Sen,
\emph{Time evolution in open string theory},
JHEP {\bf 0210} (2002) 003
[arXiv:hep-th/0207105].
\bibitem{Sen:2002nu}
A.~Sen,
\emph{Rolling tachyon},
JHEP {\bf 0204} (2002) 048
[arXiv:hep-th/0203211].
\bibitem{QXX1f}
B.~de Wit, H.~Samtleben and M.~Trigiante,
\emph{On Lagrangians and gaugings of maximal supergravities},
Nucl.\ Phys.\ B {\bf 655} (2003) 93
[arXiv:hep-th/0212239];
B.~de Wit, H.~Samtleben and M.~Trigiante, \emph{Maximal
supergravity from IIB flux compactifications},
[arXiv:hep-th/0311224];
L.~Andrianopoli, R.~D'Auria, S.~Ferrara and M.~A.~Lledo,
\emph{Gauging of flat groups in four dimensional supergravity},
JHEP {\bf 0207} (2002) 010
[arXiv:hep-th/0203206];
C.~M.~Hull,
\emph{New gauged N = 8, D = 4 supergravities},
[arXiv:hep-th/0204156];
F.~Cordaro, P.~Fre, L.~Gualtieri, P.~Termonia and M.~Trigiante,
\emph{N = 8 gaugings revisited: An exhaustive classification},
Nucl.\ Phys.\ B {\bf 532} (1998) 245 [arXiv:hep-th/9804056].
\bibitem{QXX5}
L.~Andrianopoli, S.~Ferrara and M.~Trigiante,
\emph{Fluxes, supersymmetry breaking and gauged supergravity},
[arXiv:hep-th/0307139];
C.~Angelantonj, S.~Ferrara and M.~Trigiante,
\emph{New D = 4 gauged supergravities from N = 4 orientifolds with fluxes},
[arXiv:hep-th/0306185];
C.~Angelantonj, S.~Ferrara and M.~Trigiante, \emph{Unusual gauged
supergravities from type IIA and type IIB orientifolds},
arXiv:hep-th/0310136;
R.~D'Auria, S.~Ferrara, F.~Gargiulo, M.~Trigiante and S.~Vaula,
\emph{N = 4 supergravity Lagrangian for type II B on
$T^6/\mathbb{Z}_2$ in presence of  fluxes and D3-branes}, JHEP
{\bf 0306} (2003) 045 [arXiv:hep-th/0303049];
R.~D'Auria, S.~Ferrara, M.~A.~Lledo and S.~Vaula,
\emph{No-scale N = 4 supergravity coupled to Yang-Mills: The scalar potential  and
super Higgs effect},
Phys.\ Lett.\ B {\bf 557} (2003) 278
[arXiv:hep-th/0211027];
S.~Ferrara and M.~Porrati, \emph{N = 1 no-scale supergravity from
II B orientifolds}, Phys.\ Lett.\ B {\bf 545} (2002) 411
[arXiv:hep-th/0207135];
R.~D'Auria, S.~Ferrara and S.~Vaula, \emph{N = 4 gauged
supergravity and a II B orientifold with fluxes}, New J.\ Phys.\
{\bf 4} (2002) 71 [arXiv:hep-th/0206241].
\bibitem{bill99}V.~D.~Ivashchuk and V.~N.~Melnikov,\emph{
Billiard representation for multidimensional cosmology with
intersecting p-branes near the singularity}, J.\ Math.\ Phys.\
{\bf 41} (2000) 6341 [arXiv:hep-th/9904077].
\bibitem{dualiza2} T.~Damour, M.~Henneaux and H.~Nicolai,
\emph{Cosmological billiards},
Class.\ Quant.\ Grav.\  {\bf 20} (2003) R145
[arXiv:hep-th/0212256].
\bibitem{Henneaux:2003kk}
M.~Henneaux and B.~Julia,
\emph{Hyperbolic billiards of pure D = 4 supergravities},
JHEP {\bf 0305} (2003) 047
[arXiv:hep-th/0304233].
\bibitem{deBuyl:2003za}
S.~de Buyl, G.~Pinardi and C.~Schomblond,
\emph{Einstein billiards and spatially homogeneous cosmological models},
[arXiv:hep-th/0306280].
\bibitem{Damour:2002tc}
T.~Damour, M.~Henneaux, A.~D.~Rendall and M.~Weaver,
\emph{Kasner-like behaviour for subcritical Einstein-matter systems},
Annales Henri Poincare {\bf 3} (2002) 1049
[arXiv:gr-qc/0202069].
\bibitem{Damour:pq}
T.~Damour and M.~Henneaux,
\emph{Chaos In Superstring Cosmology},
Gen.\ Rel.\ Grav.\  {\bf 32} (2000) 2339.
\bibitem{Damour:2001sa}
T.~Damour, M.~Henneaux, B.~Julia and H.~Nicolai,
\emph{Hyperbolic Kac-Moody algebras and chaos in Kaluza-Klein models},
Phys.\ Lett.\ B {\bf 509} (2001) 323
[arXiv:hep-th/0103094].
\bibitem{Damour:2000hv}
T.~Damour and M.~Henneaux,
\emph{E(10), BE(10) and arithmetical chaos in superstring cosmology},
Phys.\ Rev.\ Lett.\  {\bf 86} (2001) 4749
[arXiv:hep-th/0012172].
\bibitem{Damour:2000th}
T.~Damour and M.~Henneaux,
\emph{Oscillatory behaviour in homogeneous string cosmology models},
Phys.\ Lett.\ B {\bf 488} (2000) 108
[Erratum-ibid.\ B {\bf 491} (2000) 377]
[arXiv:hep-th/0006171].
\bibitem{Damour:2000wm}
T.~Damour and M.~Henneaux,
\emph{Chaos in superstring cosmology},
Phys.\ Rev.\ Lett.\  {\bf 85} (2000) 920
[arXiv:hep-th/0003139].
\bibitem{Demaret:sg}
J.~Demaret, Y.~De Rop and M.~Henneaux,
\emph{Chaos In Nondiagonal Spatially Homogeneous Cosmological Models In Space-Time
Dimensions <= 10},
Phys.\ Lett.\ B {\bf 211} (1988) 37.
\bibitem{julia80}
E.~Cremmer, in \emph{Supergravity '81}, ed. by S.~Ferrara and J.~ G.~ Taylor, pag. 313;
B.~ Julia, in \emph{Superspace and Supergravity}, eds. S.~W.~ Hawking and
M.~Ro\v{c}ek, Cambridge Univ. Press, 1980;
in   \emph{Unified Field Theories and Beyond},
Johns Hopkins Workshop on Current Problems in Particle Physics,
Johns Hopkins University, Baltimore, 1981.
%
\bibitem{Hull:1994ys}
C.~M.~Hull and P.~K.~Townsend,
\emph{Unity of superstring dualities},
Nucl.\ Phys.\ B {\bf 438} (1995) 109
[arXiv:hep-th/9410167].
\bibitem{Julia:1982gx}
B.~Julia,
\emph{Kac-Moody Symmetry Of Gravitation And Supergravity Theories},
{\it Invited talk given at AMS-SIAM Summer Seminar on Applications of Group Theory
in Physics and Mathematical Physics, Chicago, Ill., Jul 6-16,
1982}.
\bibitem{Nicolai:1988jb}
H.~Nicolai and N.~P.~Warner,
\emph{The Structure Of N=16 Supergravity In Two-Dimensions},
Commun.\ Math.\ Phys.\  {\bf 125} (1989) 369.
\bibitem{Nicolai:kx}
H.~Nicolai,
\emph{A Hyperbolic Lie Algebra From Supergravity},
Phys.\ Lett.\ B {\bf 276} (1992) 333.
\bibitem{Mizoguchi:1997si}
S.~Mizoguchi,
\emph{E(10) symmetry in one-dimensional supergravity},
Nucl.\ Phys.\ B {\bf 528} (1998) 238
[arXiv:hep-th/9703160].
\bibitem{e10}
H.~Nicolai and T.~Fischbacher,
\emph{Low level representations for E(10) and E(11)},
[arXiv:hep-th/0301017].
\bibitem{Marcus:1983hb}
N.~Marcus and J.~H.~Schwarz,
\emph{Three-Dimensional Supergravity Theories},
Nucl.\ Phys.\ B {\bf 228} (1983) 145.
\bibitem{nicd3}
B.~de Wit, A.~K.~Tollsten and H.~Nicolai,
\emph{Locally supersymmetric D = 3 nonlinear sigma models},
Nucl.\ Phys.\ B {\bf 392} (1993) 3
[arXiv:hep-th/9208074].
\bibitem{deWit:2003ja}
B.~de Wit, I.~Herger and H.~Samtleben,
\emph{Gauged locally supersymmetric D = 3 nonlinear sigma models},
[arXiv:hep-th/0307006].
\bibitem{Fischbacher:2003yw}
T.~Fischbacher, H.~Nicolai and H.~Samtleben,
\emph{Non-semisimple and complex gaugings of N = 16 supergravity},
[arXiv:hep-th/0306276].
\bibitem{Nicolai:2001sv}
H.~Nicolai and H.~Samtleben,
\emph{Compact and noncompact gauged maximal supergravities in three  dimensions},
JHEP {\bf 0104} (2001) 022
[arXiv:hep-th/0103032].
\bibitem{Nicolai:2000sc}
H.~Nicolai and H.~Samtleben,
\emph{Maximal gauged supergravity in three dimensions},
Phys.\ Rev.\ Lett.\  {\bf 86} (2001) 1686
[arXiv:hep-th/0010076].
\bibitem{Andrianopoli:1996bq}
L.~Andrianopoli, R.~D'Auria, S.~Ferrara, P.~Fre and M.~Trigiante,
\emph{R-R scalars, U-duality and solvable Lie algebras}, Nucl.\
Phys.\ B {\bf 496} (1997) 617 [arXiv:hep-th/9611014].
\bibitem{Andrianopoli:1996zg}
L.~Andrianopoli, R.~D'Auria, S.~Ferrara, P.~Fre, R.~Minasian and
M.~Trigiante, \emph{Solvable Lie algebras in type II A, type II B
and M theories}, Nucl.\ Phys.\ B {\bf 493} (1997) 249
[arXiv:hep-th/9612202].
\bibitem{alekseevskii}
D.~V.~Alekseevskii, \emph{Classification of quaternionic spaces with a transitive
solvable group of motions.}, Izv. Akad. Nauk SSSR Ser. Mat. {\bf 9}, 315-362 (1975);
Math. USSR Izvesstija, {\bf 9}, 297-339 (1975).
\bibitem{Keurentjes:2002xc}
A.~Keurentjes,
\emph{The group theory of oxidation},
Nucl.\ Phys.\ B {\bf 658} (2003) 303
[arXiv:hep-th/0210178].
\bibitem{Cremmer:1999du}
E.~Cremmer, B.~Julia, H.~Lu and C.~N.~Pope,
\emph{Higher-dimensional origin of D = 3 coset symmetries},
[arXiv:hep-th/9909099].
\bibitem{Julia:1980gr}
B.~Julia,
\emph{Group Disintegrations},
LPTENS 80/16
{\it Invited paper presented at Nuffield Gravity Workshop, Cambridge, Eng., Jun 22 -
Jul 12, 1980}.
\bibitem{Campbell:zc}
I.~C.~Campbell and P.~C.~West,
\emph{N=2 D = 10 Nonchiral Supergravity And Its Spontaneous Compactification},
Nucl.\ Phys.\ B {\bf 243} (1984) 112.
\bibitem{II B}
J.~H.~Schwarz, \emph{Covariant Field Equations Of Chiral N=2 D =
10 Supergravity}, Nucl.\ Phys. {\bf B226} (1983) 269;
P.~S.~Howe and P.~C.~West,
\emph{The Complete N=2, D = 10 Supergravity},
Nucl.\ Phys.\ B {\bf 238} (1984) 181;
\emph{The Complete superspace action of chiral D = 10, N=2
supergravity}, Int.\ J.\ Mod.\ Phys. {\bf A8} (1993) 1125.
%
\bibitem{dualiza1}
E.~Cremmer, B.~Julia, H.~Lu and C.~N.~Pope,
\emph{Dualisation of dualities. I},
Nucl.\ Phys.\ B {\bf 523} (1998) 73
[arXiv:hep-th/9710119].
\bibitem{Lu:1996ge}
H.~Lu, C.~N.~Pope and K.~S.~Stelle,
\emph{Weyl Group Invariance and p-brane Multiplets},
Nucl.\ Phys.\ B {\bf 476} (1996) 89
[arXiv:hep-th/9602140].
\bibitem{Bertolini:1999uz}
M.~Bertolini and M.~Trigiante, \emph{Regular R-R and NS-NS BPS black
holes}, Int.\ J.\ Mod.\ Phys.\ A {\bf 15} (2000) 5017
[arXiv:hep-th/9910237].
\bibitem{pope1}H.~Lu, S.~Mukherji, C.~N.~Pope and K.~W.~Xu,
\emph{Cosmological solutions in string theories}, Phys.\ Rev.\ D {\bf
55}, 7926 (1997) [arXiv:hep-th/9610107].
\bibitem{pope2} H.~Lu, S.~Mukherji and C.~N.~Pope,
 \emph{From p-branes to
cosmology}, Int.\ J.\ Mod.\ Phys.\ A {\bf 14}, 4121 (1999)
[arXiv:hep-th/9612224].
\bibitem{lukas1}A.~Lukas, B.~A.~Ovrut and D.~Waldram,
\emph{Cosmological solutions of type II string theory}, Phys.\ Lett.\
B {\bf 393}, 65 (1997) [arXiv:hep-th/9608195].
\bibitem{lukas2}A.~Lukas, B.~A.~Ovrut and D.~Waldram, \emph{String and
M-theory cosmological solutions with Ramond forms}, Nucl.\ Phys.\
B {\bf 495}, 365 (1997) [arXiv:hep-th/9610238].
\bibitem{quevedo}C.~Grojean, F.~Quevedo, G.~Tasinato and  I. Zavala, \emph{Branes
 on Charged Dilatonic Backgrounds: Self-Tuning, Lorentz Violations and
Cosmology}, JHEP {\bf 0108}, 005 (2001) [arXiv:hep-th/0106120].
\bibitem{andythermo}
A.~Maloney, A.~Strominger and X.~Yin, \emph{S-brane
thermodynamics}, [arXiv:hep-th/0302146].
\bibitem{stro2}
A.~Strominger, \emph{Open string creation by S-branes},
[arXiv:hep-th/0209090].
\bibitem{stro1}
M.~Gutperle and A.~Strominger, \emph{Timelike boundary Liouville
theory}, [arXiv:hep-th/0301038].
\end{thebibliography}
\end{document}